\documentclass[12pt,doublespace]{JHEP3}
 
\usepackage{amsmath,amssymb,amsfonts,xspace} 
\usepackage{epsfig,supertabular,longtable}

\renewcommand{\Im}{\imag}

\newcommand{\be}{\begin{equation}}
\newcommand{\ee}{\end{equation}}
\newcommand{\beq}{\begin{eqnarray}}
\newcommand{\eeq}{\end{eqnarray}}
\newcommand{\ben}{\begin{eqnarray}\displaystyle}
\newcommand{\een}{\end{eqnarray}}
\newcommand{\bea}[2]{\be\label{#2}\begin{array}{#1}}
\newcommand{\eea}{\end{array}\ee}

\def\Tr{\,{\rm Tr}\, }
\def\det{\,{\rm det}\, }

\def\sign{{\rm sign}}

\def\Im{\,{\rm Im}\, }

\def\({\left(}
\def\){\right)}
\def\[{\left[}
\def\]{\right]}

\def\11{1\!\! 1}

\def\eps{\varepsilon}

\newcommand{\bZ}{\mathbb{Z}}

\renewcommand{\L}{\cal{L}}

\newcommand{\de}{\mathrm{d}}

\newcommand{\I}{\mathrm{i}}
\newcommand{\cA}{\mathcal{A}}
\newcommand{\cB}{\mathcal{B}}
\newcommand{\cL}{\mathcal{L}}
\newcommand{\cQ}{\mathcal{Q}}

\newcommand{\cP}{\mathcal{P}}

\newcommand{\cK}{\mathcal{K}}
\newcommand{\cM}{\mathcal{M}}

\newcommand{\cN}{\mathcal{N}}

\newcommand{\cX}{\mathcal{X}}

\newcommand{\cO}{\mathcal{O}}
\newcommand{\cR}{\mathcal{R}}

\newcommand{\cJ}{\mathcal{J}}

\newcommand{\IR}{\mathbb{R}}
\newcommand{\IC}{\mathbb{C}}
\newcommand{\IZ}{\mathbb{Z}}

\def\varpi{t}

\def\pa{\partial}

\newcommand{\kahler}{{K\"ahler}\xspace}

\def\bse{\begin{subequations}}
\def\ese{\end{subequations}}

\def\qli2{{\bf E}}

\newcommand{\tot}{\Omega}

\newcommand{\ttz}{t}

\newcommand\bOm{\bar\Omega}

\def\ea#1\ea{\begin{align}#1\end{align}}

\usepackage{graphicx}
\usepackage[arrow,curve,matrix,arc]{xy}

\numberwithin{equation}{section}
\newcommand{\one}{S}
\newcommand{\coll}{coll}

\newcommand{\coulombf}{eclassq}

\title{A fixed point formula for the index of \\ multi-centered  $\cN=2$ black holes}

\author{Jan Manschot$^{1}$, Boris Pioline$^{2}$, Ashoke Sen$^{3}$
\\
$^1$ {\it Institut de Physique Th\'eorique, CEA Saclay, CNRS-URA 2306,\\
91191 Gif sur Yvette, France}
\\
$^2$ {\it Laboratoire de Physique Th\'eorique et Hautes
Energies, CNRS UMR 7589, \\
Universit\'e Pierre et Marie Curie,
4 place Jussieu, 75252 Paris cedex 05, France} \\
$^3$ Harish-Chandra Research Institute,
Chhatnag
Road, Jhusi, Allahabad 211019, India
\\

\vspace*{2mm} {\tt e-mail: \email{
jan.manschot@cea.fr, pioline@lpthe.jussieu.fr,
sen@hri.res.in}
} \vspace*{-3mm}

}

\abstract{
We propose a formula for computing the 
(moduli-dependent) contribution of multi-centered solutions to the total 
BPS index in terms of the  
(moduli-independent) indices associated
to single-centered solutions. 
The main tool
in our analysis is the 
computation of 
the refined index $\Tr(-y)^{2J_3}$ of 
configurational degrees
of freedom of multi-centered BPS black hole 
solutions in $\cN=2$ supergravity
by localization methods. When the 
charges carried by the centers do not  
allow for scaling solutions (i.e. solutions 
where a subset of the centers 
can come arbitrarily close to each other), 
the phase space of classical BPS solutions 
is compact and the refined index
localizes to a finite set of isolated fixed points 
under rotations, corresponding
to collinear solutions. When the charges allow for scaling solutions, 
the phase space is non-compact but appears to admit a 
compactification with finite volume and additional
non-isolated fixed points. We give a 
prescription for determining 
the contributions of these fixed submanifolds 
by means of a `minimal modification 
hypothesis', which we prove 
in the special case of dipole halo 
configurations. 
}

\begin{document}

\section{Introduction and summary}  \label{sintro}

In $\cN=2$ supersymmetric string vacua,  BPS
states in suitable large charge limits can be represented 
as multi-centered black hole solutions of $\cN=2, D=4$ 
supergravity \cite{Denef:2000nb, Denef:2002ru,Bates:2003vx}. 
To compute the moduli-dependent index $\Omega(\gamma;t)$ associated with such configurations,
one needs to combine two independent sets of
data. 

The first  part of the data are the indices $\Omega^{\one}(\gamma)$ 
associated with single centered
BPS black holes carrying electromagnetic charges $\gamma$. 
In the supergravity approximation, the index 
is given by the exponential of the Bekenstein-Hawking 
entropy \cite{Bekenstein:1973ur,Hawking:1974sw}, and is 
independent of the asymptotic values of the scalar fields $t$ (within a
given basin of attraction) 
by virtue of the attractor phenomenon \cite{Ferrara:1995ih,Ferrara:1996um,
Ferrara:1997tw}.
Effects of classical higher derivative corrections to the low energy effective 
action can be incorporated 
by using Wald's modification of the Bekenstein-Hawking
formula \cite{Wald:1993nt,Behrndt:1996jn,
LopesCardoso:1998wt,Cardoso:1999za}, while 
quantum corrections to the index can
in principle be computed using the quantum entropy
function formalism \cite{Sen:2008yk, Gupta:2008ki,Sen:2008vm}. 

The second part of the data
is the  index of the supersymmetric quantum mechanics 
describing multi-centered black hole configurations.
In this description, the centers are treated as pointlike, 
entropy-less particles carrying (in general, mutually non-local)
electromagnetic charges 
$\alpha_1,\dots ,\alpha_n$, and kept in equilibrium 
by balance of forces  \cite{Denef:2000nb}.
The space of solutions of this mechanical problem
is a $2(n-1)$-dimensional symplectic space $\cM_n$
with an hamiltonian action of the rotation 
group $SO(3)$, which  can be quantized 
by the standard procedure of  geometric quantization
 \cite{Denef:2002ru,deBoer:2008zn}.
Unlike the first part of the data, the index of these configurational
degrees of freedom, which we denote by $g(\{\alpha_i\};t)$, depends sensitively 
on the asymptotic values of the scalar fields $t$. 
In our previous work \cite{ Manschot:2010qz}, reviewed in \cite{Pioline:2011gf},
we showed how to compute
 the jump of this index across a wall of marginal 
stability by localization with respect to a $U(1)$ 
subgroup of $SO(3)$ corresponding to rotations 
along the $z$-axis. 
In this work, we extend the techniques of
\cite{ Manschot:2010qz} to compute the configurational index away from the 
walls of marginal stability, and propose a formula to combine this result
with the indices associated to single-centered black holes in order to 
compute the total BPS index.

The approach of  \cite{ Manschot:2010qz} was based on several simplifying facts. 
First it was shown that since identical centers do not interact, one 
can  replace the Bose-Fermi statistics carried by the centers by Boltzmann statistics, 
provided the index $\Omega(\alpha_i)$ carried by the centers
-- or more generally the `refined index'\footnote{Here
$\Tr'$ denotes the trace over states carrying a fixed set
of charges after removing the contribution
from the fermion zero modes associated with
broken supersymmetry generators.
While $\Tr'(-y)^{2J_3}$ is in general not a protected index away from $y=1$ (except in rigid $\cN=2$ field theories), 
it is essential to allow for $y\neq 1$, at least as the intermediate steps, in order for
localization methods to apply. It would be interesting to understand 
the dependence of $\Tr'(-y)^{2J_3}$  on the string coupling 
and other hypermultiplet fields.}
$\Omega_{\rm ref}(\alpha_i,y)=\Tr'(-y)^{2J_3}$,  where $J_3$ denotes the generator of rotations
along the $z$ axis and $y$ is a real parameter  -- is replaced 
by an effective rational index $\bar\Omega_{\rm ref}(\alpha_i,y)$ \cite{ks,Joyce:2008pc,
Manschot:2010qz, Manschot:2010nc}. Thus, even 
when some of the $\alpha_i$'s are identical,  
one may still treat the centers as distinguishable. 
Second, it was assumed that the `refined index' of this quantum 
mechanical problem is related
to the equivariant volume  (i.e. the  integral of $y^{2J_3}$ over $\cM_n$)
by a simple overall multiplicative renormalization required 
for consistency with angular momentum quantization. 
Third, it was important that, in the case of loosely bound constituents relevant for 
wall-crossing, the phase space  $\cM_n$ was compact and the action of $J_3$ 
had only  isolated fixed points, corresponding to collinear 
multi-centered solutions along the $z$-axis. 
Under these circumstances, the 
localization theorem of  Duistermaat and Heckman
 \cite{Duistermaat:1982vw} can be used to express 
 the configurational index $g(\{\alpha_i\},t)$, or rather its
 refined generalization $g(\{\alpha_i\};y,t)$, as a finite sum over 
the fixed points of $J_3$. 
This result was found to match the prediction of the known
wall crossing formulae from supergravity \cite{Denef:2007vg, Andriyash:2010qv} 
and mathematics \cite{ks, Joyce:2008pc} in all cases where it was tested. 
As we shall show in this work, provided $\cM_n$ is 
compact and the fixed points are isolated, the 
multiplicative renormalisation postulated in  \cite{ Manschot:2010qz} 
can in fact be derived from
the Atiyah-Bott Lefschetz fixed point theorem \cite{MR0212836,MR0232406,MR1365745,MR1215720}, 
a quantum-mechanical 
version of the Duistermaat-Heckman formula.

In this paper, we show that  the same approach can be used to find
the spectrum of multi-centered black hole solutions at
a generic point in the moduli space,  given the indices associated
to single-centered black holes. There are however
some important differences:
\begin{enumerate}

\item The set of collinear multi-centered
solutions must not only  satisfy the BPS equilibrium  conditions of 
\cite{Denef:2000nb}, but also lead to a regular metric. 
This condition was automatically satisfied for loosely bound
states near a wall of marginal stability, but needs to be 
checked when computing the index at a generic point in moduli space
and for generic charges. This condition is expected to rule out all
but finitely many decompositions of the total charge $\gamma$ into a sum
$\sum_{i=1}^n \alpha_i$  \cite{Denef:2007vg}. It  can also rule out certain connected 
components of $\cM_n$ even when the decomposition $\gamma=
\sum_{i=1}^n \alpha_i$ is allowed, see 
\S\ref{d6d6d0} for an example. A necessary (but not sufficient) 
condition is that the collinear configuration be regular along the $z$-axis.

\item For some range of charges carried by the centers,
the phase space includes `scaling solutions', i.e. 
regions where the relative distances
between a subset (or all) the centers  can become arbitrarily
small \cite{Denef:2002ru,Bena:2006kb,
Bena:2007qc,Denef:2007vg,deBoer:2009un}.  
As a result, the space $\cM_n$ is non-compact, and the 
sum over regular collinear configurations fails to produce a 
sensible answer. In particular, it does not have a finite limit 
as $y\to1$,  and cannot be interpreted as a 
sum of  characters of $SO(3)$ (nor of its double cover $SU(2)$). 
However, despite being non-compact, $\cM_n$ turns out to have
a finite symplectic volume, suggesting that it
may admit a  compactification.
In the case of `dipole halo' configurations, introduced in 
\cite{deBoer:2008zn,deBoer:2009un}, it is straightforward to
construct the compactification explicitly. 
The resulting space  
$\hat \cM_n$ still admits an hamiltonian action of
$SO(3)$, but the fixed points of $J_3$ are no longer isolated, 
in particular there is a codimension 4 submanifold of fixed points
where the total angular momentum vanishes, and 
which parametrizes scaling solutions.
We shall assume that $\cM_n$ always admits a compactification $\hat \cM_n$, 
although we shall not require the details of its construction. 

\item Due to the fact that the action of $J_3$  on $\hat \cM_n$ has non-isolated 
fixed points, the equivariant volume and equivariant index 
are no longer related by a simple multiplicative
renormalization. 
While the `refined index' could  still be in principle computed
by localization using the Atiyah-Bott theorem, this would require  a complete understanding
of the compactification $\hat\cM_n$ which we have not achieved so far. Instead, 
we propose a `minimal modification 
hypothesis' which determines the contribution of these scaling regions 
from that of the regular,  well-separated
collinear fixed points. Our prescription amounts to requiring that 
scaling solutions contribute with the smallest possible 
angular momentum compatible with the final result being 
a character of $SU(2)$.  
This prescription is motivated by the fact that classically 
(ignoring angular momentum 
quantization), scaling solutions 
carry zero total angular momentum. 
While we do not have a proof of this hypothesis, we shall demonstrate that it is
consistent with wall-crossing and with the split 
attractor flow conjecture \cite{Denef:2007vg}.
For a special class of `dipole halo' configurations,
where the moduli space of 
multi-centered solutions is fully understood, 
we shall verify that our prescription agrees with the
geometric quantization of $\cM_n$ performed 
in~\cite{deBoer:2008zn,deBoer:2009un}, for an 
arbitrary number of centers.

\end{enumerate}

We shall now summarize our proposal.
We denote by $\Omega^{\one}_{\rm ref}(\alpha, y)$ the index 
$\Tr'(-y)^{2J_3}$ carried by a single centered black hole
with charge $\alpha$ (here, $\Tr'$ denotes the
trace after factoring out the bosonic and fermionic zero
modes). Since single centered black holes
carry zero angular momentum\cite{Sen:2009vz,Dabholkar:2010rm}, we expect 
that $\Omega^{\one}_{\rm ref}(\alpha,y)$ is independent
of $y$. However we shall not make use of this
information in our analysis, and proceed with general
$\Omega^{\one}_{\rm ref}(\alpha, y)$. 
Let us denote by 
$\tot_{\rm ref}(\gamma,y)\equiv 
\Tr'(-y)^{2 J_3}$ the total contribution to the
index from single and multi-centered black hole
solutions carrying total charge 
$\gamma$, and by
\be \label{edefsbar}
\bar \tot_{\rm ref}(\gamma, y) = \sum_{m|\gamma}\, 
m^{-1} {y-y^{-1}\over y^m - y^{-m}} \, \tot_{\rm ref}(\gamma/m, y^m)\, .
\ee
First consider the case when
there are no scaling solutions. 
Then our proposal for $\bar \tot_{\rm ref}(\gamma, y)$ is:
\be \label{enn1int}
\bar \tot_{\rm ref}(\gamma, y)= 
\sum_{\{\alpha_i\in \Gamma\}\atop \sum_i\alpha_i=\gamma} 
{1\over {\rm Aut}(\{\alpha_i\})}
\, g_{\rm ref} (\alpha_1, \dots , \alpha_n;y) \, 
\bOm_{\rm ref}^{\one}(\alpha_1;y)
\cdots \bOm^{\one}_{\rm ref}(\alpha_n;y)\, .
\ee
Here $\Gamma$ is the charge lattice,
${\rm Aut}(\{\alpha_i\})$ is the symmetry factor appropriate
 for Maxwell-Boltzmann 
 statistics,\footnote{${\rm Aut}(\{\alpha_i\})$ 
 is defined as 
 the order of the subgroup of the 
 permutation group of $n$ elements which preserves 
 the ordered set $(\alpha_1,\dots, \alpha_n)$, for a fixed 
 (arbitrary) choice of  ordering.
 Thus if the set  
 $\{\alpha_i\}$ consists of $r_1$ copies of $\beta_1$,
 $r_2$ copies of $\beta_2$ etc. then
 $|{\rm Aut}(\{\alpha_i\})|=\prod_k r_k!.$ \label{fo1}} 
 and $\bOm^{\one}_{\rm ref}(\alpha,y)$ is the 
 `rational refined index',
related to the refined index by
\be \label{enn2int}
\bOm^{\one}_{\rm ref}(\alpha, y) = \sum_{m|\alpha}
m^{-1} {y-y^{-1}\over y^m - y^{-m}} \Omega^{\one}_{\rm ref}
(\alpha/m,y^m)\, .
\ee
The coefficient $g_{\rm ref}$ is the refined  index of the configurational degrees
of freedom of $n$-centered BPS black hole solutions. By localization, it evaluates
to
\be \label{eclassint}
g_{\rm ref} (\alpha_1, \dots , \alpha_n;y)
=(-1)^{\sum_{i<j} \alpha_{ij} +n-1}
 \left[
(y-y^{-1})^{1-n} \, \sum_{p} \,  
s(p)\, 
y^{\sum_{i<j} \alpha_{ij}\, \sign[z_j-z_i]}\right]\, ,
\ee
where  $\alpha_{ij}\equiv \langle \alpha_i,\alpha_j\rangle$
is the (Dirac-Schwinger-Zwanziger)
symplectic inner product between $\alpha_i$ and
$\alpha_j$, the sum over $p$ represents sum over all
collinear solutions to 
 the BPS equilibrium conditions \eqref{denef1d}, \eqref{eregu},
 and 
$s(p)$ takes value $\pm 1$ 
as determined from eq.\eqref{esigneq}.
As mentioned above, the set of possible
decompositions of $\gamma$ that contributes to the sum
\eqref{enn1int}
is expected to be finite \cite{Denef:2007vg}.
We can recover the integer invariant $\tot_{\rm ref}$ from the rational invariant $\bar \tot_{\rm ref}$
using the inverse
formula~\cite{Manschot:2010qz}
\be \label{sfrombars}
\tot_{\rm ref}(\gamma, y)
= \sum_{m|\gamma} \mu(m) \, 
m^{-1} (y - y^{-1}) (y^m - y^{-m})^{-1} 
\bar \tot_{\rm ref}(\gamma/m, y^m)
\ee
where $\mu(m)$ is the M\"obius
function.  Expressing this in terms of 
$\Omega^{\one}_{\rm ref}$ using \eqref{enn1int}, 
\eqref{enn2int} we can arrive at an expression of the form
\be \label{eform22}
\tot_{\rm ref}(\gamma, y) =
\sum_{\{\beta_i\in \Gamma\}, \{m_i\in\bZ \}\atop
m_i\geq 1, \sum_i m_i\beta_i =\gamma}
G_{}(\{\beta_i\}, \{m_i\};y) \, \prod_i 
\Omega^{\one}_{\rm ref}(\beta_i, y^{m_i})\, ,
\ee
for some function $G_{}$. This expresses the
total index  in terms of the indices
of single centered solutions. Of course, the sum in
\eqref{eform22} also includes the contribution from single
centered black holes given by $\Omega^{\one}_{\rm ref}
(\gamma,y)$.

In the presence of scaling solutions, \eqref{eform22} cannot be the
full answer for the following reason.
If we follow the procedure outlined above ignoring the 
presence of scaling solutions, and denote the corresponding
functions $G$ by $G_{\rm \coll}$, we shall find that
the functions $G_{\rm \coll}(\{\beta_i\}, \{m_i\};y)$ 
are not Laurent polynomials in $y$, i.e. finite linear combinations
of $y^{\pm m}$ with integer $m$. 
Hence the corresponding
expression \eqref{eform22}
cannot be interpreted as the generating function of
the spectrum of a quantum mechanical system with
quantized angular momentum.
Our prescription for taking into account the effect of
scaling solutions is to 
modify \eqref{eform22} to
\be \label{eform22modint}
\tot_{\rm ref}(\gamma, y) =
\sum_{\{\beta_i\in \Gamma\}, \{m_i\in\bZ\}\atop
m_i\ge 1, \, 
\sum_i m_i\beta_i =\gamma}
G_{\rm \coll}(\{\beta_i\}, \{m_i\};y) \, \prod_i 
\left( \Omega^{\one}_{\rm ref}(\beta_i, y^{m_i})
+ \Omega_{\rm scaling}(\beta_i, y^{m_i})\right)
\, ,
\ee
where $\Omega_{\rm scaling}(\alpha, y)$ is given by
\be \label{edefHint}
\Omega_{\rm scaling}(\alpha, y) =
\sum_{\{\beta_i\in \Gamma\}, \{m_i\in\bZ\}\atop
m_i\ge 1, \, \sum_i m_i\beta_i =\alpha}
H(\{\beta_i\}, \{m_i\};y) \, \prod_i 
\Omega^{\one}_{\rm ref}(\beta_i, y^{m_i})
\, ,
\ee
for some function $H(\{\beta_i\}, \{m_i\};y)$. 
To determine $H$ we
substitute 
\eqref{edefHint} into \eqref{eform22modint}
to express the latter equation as
\be \label{eModint}
\tot_{\rm ref}(\gamma, y) =
\sum_{\{\beta_i\in \Gamma\}, \{m_i\in\bZ\}\atop
m_i\ge 1, \, \sum_i m_i\beta_i =\gamma}
G(\{\beta_i\}, \{m_i\};y) \, \prod_i 
\Omega^{\one}_{\rm ref}(\beta_i, y^{m_i})
\, ,
\ee
for some functions $G$. We fix $H$
by requiring that  $G(\{\beta_i\}, \{m_i\};y)$ be a Laurent polynomial in $y$. 
The ambiguity of adding to $H$ a
Laurent polynomial is resolved by using the minimal
modification hypothesis, which requires that $H$ must 
be symmetric under $y\to y^{-1}$ and vanish
as $y\to\infty$. An iterative procedure for determining the
functions $H$ and hence $G$ has been described in
\S\ref{scaling}.

One advantage of our construction compared to the 
split attractor flow conjecture of
\cite{Denef:2007vg} is that it allows us to compute 
the contributions from scaling solutions. Since the latter 
are usually stable across walls of marginal stability (as
illustrated for three centers in \S\ref{secfinite}
and Fig.\ref{figcplane} below), their
contributions cannot be obtained from attractor flow trees.
This advantage is however mitigated by the fact that 
we do not know how to determine {\it a priori} which
decompositions $\gamma=\sum \alpha_i$  lead
to regular multi-centered solutions, except by 
checking \eqref{eregu} numerically. 
It should also be emphasized  that in cases where
the BPS spectrum is described by some 
quiver quantum mechanics with non-zero
superpotential, the 
configurational degrees of freedom 
carried by scaling solutions do not exhaust, 
by far \cite{deBoer:2009un}, the  exponentially large 
number of states present on the 
Higgs branch \cite{Denef:2007vg}. These Higgs branch
states have macroscopic entropy and should 
be included as part of the single-centered configurations
counted by $\Omega^{\one}_{\rm ref}(\gamma)$.

The rest of the paper is organized as follows. In \S\ref{secphase},
we review the structure of the phase space of multi-centered 
BPS solutions in $\cN=2$ supergravity, establish the
finiteness of  its symplectic volume, and
compute the classical and quantum refined index (also known as 
equivariant volume and equivariant index) by localization.
In \S\ref{sexplicit}, we investigate several examples of three-centered
solutions in a simple one-modulus supergravity, paying attention
to the regularity condition \eqref{eregu} and to the contributions
of scaling solutions.
In \S\ref{snon} we describe our proposal for the
index of multi-centered black hole configuration 
when there are no scaling solutions, and 
show the consistency of this
formula with wall crossing 
and with the split attractor flow conjecture.
 In \S\ref{scaling} we describe our proposal for modifying the result of
\S\ref{snon} in the presence of scaling solutions.
In \S\ref{sdipole} we prove the validity of our prescription in the 
solvable case of dipole halo configurations. Further technical
details are relegated to appendices: in appendix \ref{ssignrule}
we provide a detailed analysis of certain sign rules which govern the
contributions of collinear fixed points. 
In appendix \ref{sinteg} we prove that in the absence of
scaling solutions the right hand side of \eqref{eclassint}
can be expressed as a Laurent polynomial in $y$. 
Explicit computations of the equivariant  volume and index for 
dipole halos with 4 and 5 centers can be found in appendix
\ref{sc}.

\section{The phase space of multi-centered 
configurations\label{secphase}}

We begin by reviewing some relevant
properties of supersymmetric multi-centered black hole 
solutions in $\cN=2$ supergravity.  Such solutions
fall into the stationary metric ansatz 
\be
\de s^2= - 
e^{2U} \, (\de t+\cA )^2 + e^{-2U} \de \vec r^2 
\ee
where the scale function $U$, the Kaluza-Klein 
one-form $\cA$ and the vector multiplet scalars $\ttz^a,
a=1, \dots , n_v$ depend on the coordinate $\vec r$ on 
$\IR^3$. 

\subsection{Equilibrium and regularity conditions}  

Let $\Gamma$ denote the charge lattice. Locally in the
moduli space of the theory a charge vector $\alpha\in
\Gamma$ may be split into its electric and magnetic
components $(p^\Lambda, q_\Lambda)$. Given two such
vectors $\alpha$ and $\alpha'$ we can define a
symplectic product 
\be \label{edefsymp}
\langle \alpha, \alpha'\rangle = q_\Lambda p^{\prime \Lambda}
- q'_\Lambda p^\Lambda\, .
\ee
For $n$ centers located at $\vec r_1,\dots, 
\vec r_n$, carrying electromagnetic charges $\alpha_1,\dots ,
\alpha_n$ in the charge lattice $\Gamma$, the values of the vector multiplet
scalars and of the scale factor $U$
are obtained by solving the ``attractor equations"  \cite{Denef:2000nb}.
\be
\label{eqatt}
-2\, e^{-U(\vec r)} \Im\left[ e^{-\I\phi} Y\left((\ttz^a(\vec r)
\right) \right]=  \beta +
\sum_{i=1}^n {\alpha_i\over |\vec r - \vec r_i|},
\qquad \phi=\arg Z_\gamma,
\ee
where $Y(\ttz^a)=-e^{\cK/2}(X^\Lambda(\ttz),F_\Lambda(\ttz))$ 
is the
symplectic section afforded by the special geometry of the
vector multiplet moduli space, 
$\cK=-\ln i(F_\Lambda \bar X^\Lambda - \bar F_\Lambda
X^\Lambda)$, and $Z_\gamma$
is the central charge 
\be \label{edefcentral}
Z_\gamma = \langle \gamma, Y(\ttz_\infty)\rangle\, 
\ee
associated to the total charge
$\gamma=\alpha_1+\cdots+\alpha_n$. 
$Y(\ttz_\infty)$ denotes the value of $Y$ at infinity.
The constant vector $\beta$
on the right-hand side of \eqref{eqatt} is determined in 
terms of  the asymptotic values of the moduli at infinity $\ttz_\infty^a$
by 
\be \label{edefbeta}
\beta = -2 \, {\rm Im} \left[e^{-\I\phi}\, Y(\ttz_\infty) \right]\, .
\ee
The locations $\vec r_i$ are subject 
to the equilibrium conditions
(also known as integrability equations) \cite{Denef:2000nb}
\be \label{denef3d}
\sum_{j=1\atop j\ne i}^n \frac{\alpha_{ij}}{r_{ij}}
=  c_i\, ,
\ee
where $r_{ij} = |\vec r_i - \vec r_j|$, $\alpha_{ij}
\equiv \langle \alpha_i, \alpha_j\rangle$, 
and the real constants 
\be
\label{eci}
c_i \equiv 2 \, {\rm Im}\, (e^{-\I\phi} Z_{\alpha_i})
\ee
depend on the the asymptotic values of the moduli. Since $\phi=\arg Z_\gamma$,
these constants satisfy $\sum_{i=1}^n c_i=0$.
Finally,
$\cA$ is given by
\be \label{ecaexp}
*_3d\cA = \left\langle d \sum_{i=1}^n {\alpha_i\over 
|\vec r - \vec r_i|}, \beta +
\sum_{i=1}^n {\alpha_i\over |\vec r - \vec r_i|}
\right\rangle \, ,
\ee
where $*_3$ denotes Hodge dual in 3 flat dimensions.
The conditions  \eqref{denef3d} 
guarantee the existence of a Kaluza-Klein connection
$\cA$ such that the above configuration is a supersymmetric solution
of the equations of motion.

It follows from \eqref{eqatt} that the scale factor $U$ 
is given by evaluating the Bekenstein-Hawking entropy function 
$S(\gamma)$ on the harmonic function appearing on the right-hand side
of \eqref{eqatt} \cite{Bates:2003vx},
\be
\label{esk2aa}
e^{-2U(\vec r)}  = \frac{1}{\pi} S\left( \beta +\sum_{i=1}^n {\alpha_i\over |\vec r - \vec r_i|}
\right)\ .
\ee
In order for the solution to be physical 
one must 
require that the scale factor be everywhere real and
positive  \cite{Denef:2007vg}\footnote{This condition is somewhat too strong, since 
it rules out the 'empty hole' configurations representing e.g. the BPS states which 
become massless at the conifold point 
\cite{Denef:2000nb}. A more accurate requirement would be that regions with 
$S\leq 0$ are shielded by a shell where the moduli lie at singular conifold-type 
points, such that the regions inside the shell can be replaced by flat regions 
with constant values of the scalars. To avoid this complication we shall restrict 
to charges $\alpha_i$ which satisfy $D(\alpha_i)\geq 0$, where $D$ is the 
quartic polynomial such that $S=\sqrt{D}/\pi$ in the large volume limit.
},
\be \label{eregu}
S\left( \beta +
\sum_{i=1}^n {\alpha_i\over |\vec r - \vec r_i|}
\right) > 0\, , \qquad \forall \ \vec r\in \IR^3\, ,
\ee
where $\vec r_i$ is the location of the $i$-th
center. This is necessary to ensure that there exists 
a regular solution
to the attractor equations \eqref{eqatt} at all points in $\IR^3$.

\subsection{Symplectic structure and equivariant volume}

Leaving aside the regularity condition \eqref{eregu} for now, 
let us denote by
 $\cM_{n}(\{\alpha_{ij}\}; \{c_i\})$ the space of
solutions $\{\vec r_1,\dots , \vec r_n\}$
to the equilibrium conditions \eqref{denef3d}, 
modulo overall translations of the centers. 
$\cM_n$ is a (possibly disconnected)
$2n-2$-dimensional submanifold of $\IR^{3n-3}\backslash \Delta$,
where $\Delta$ is the locus in $\IR^{3n-3}$ where two or more 
of the centers $\vec r_i$ coincide. $\IR^{3n-3}\backslash \Delta$
is equipped with the closed two-form\footnote{Our normalisation differs by 
a factor of two from the one used in \cite{deBoer:2008zn}.
This ensures that $\omega/2\pi$ has integer periods, see \S\ref{secqindex}.}
\be \label{edefomega}
\omega =\frac14 \sum_{i,j\atop i<j} \alpha_{ij}\, 
\frac{\epsilon^{abc}\, \de  r^a_{ij} \wedge  \de r^b_{ij}  \, r^c_{ij} 
}{|r_{ij}|^3} = \frac12 \sum_{i,j\atop i<j} \alpha_{ij}\, \sin\theta_{ij} \, \de\theta_{\ij}
\wedge \de \phi_{ij}\ ,
\ee 
where $\theta_{ij},\phi_{ij}$ are the polar angles parametrizing the direction of the vector 
$\vec r_{ij}$ with respect to a fixed unit vector $\vec u$, for example $\vec u=(0,0,1)$. For generic values of $c_i$, the 
restriction of $\omega$ to $\cM_n$ is 
non-degenerate\footnote{A notable exception 
occurs when $c_i=0$, where the space of solutions admits an exact 
dilation symmetry $r_{ij}\to \varepsilon\, r_{ij}$, along which the 
two-form $\omega$ becomes degenerate. 
This case corresponds to multi-centered solutions asymptotic to 
$AdS_2\times S^2$, and will become relevant in the analysis of scaling solutions below.
\label{notablefoo} }, 
and provides $\cM_n$ with a
symplectic structure  \cite{deBoer:2008zn}. 

By construction, the symplectic form $\omega$ is invariant under $SO(3)$ rotations. The 
moment map associated to infinitesimal rotations is the  angular momentum
\be \label{ejexp}
\vec J= \frac12 \sum_{i,j\atop i<j} \alpha_{ij}\, 
\frac{\vec r_{ij}}{|r_{ij}|} = \frac12 \sum_i c_i \, \vec r_i\ ,
\ee 
where the second equality follows by using \eqref{denef3d}. 
After some further algebraic manipulations, the norm 
of $\vec J$ can be written as  \cite{deBoer:2008zn}
\be
\label{defj}
j \equiv \sqrt{\vec J^2} = \sqrt{-\frac14 \sum_{i,j\atop 
i\neq j} 
c_i \, c_j \, r_{ij}^2}\ .
\ee
For $n\geq 3$ centers, the orbits of the $SO(3)$ action are generically 3-dimensional,
except on the two-dimensional subspace $\cM_{n;{\rm \coll}}\subset \cM_n$ corresponding 
to collinear configurations. Removing this locus, the quotient of
$\cM_n\backslash \cM_{n;{\rm \coll}}$ 
by the $SO(3)$ action is a $(2n-5)$-dimensional Poisson manifold  $\tilde \cM_n$.
Let $\cJ=j(\cM_n)\subset\IR^+$ be the range\footnote{In 
general, $\cJ$ is bounded, and consists of a set of intervals in $\IR^+$. The regularity
condition \eqref{eregu} may rule out certain intervals in $\cJ$.} 
 spanned by the total angular momentum
$j$ on the space of configurations $\cM_n$.
The symplectic leaves of $\tilde \cM_n$ are the
hypersurfaces $\tilde \cM_n(j)$ with fixed total angular momentum $j\in \cJ$. If one so wishes,
one may parametrize $\tilde \cM_n$ by $2n-5$  relative distances $r_{ij}$
(suitably chosen among the $n(n-1)/2$ radii $r_{ij}$, and subject to triangle inequalities), 
and the leaves $\tilde\cM_n(j)$ by their projectivization.
Using Euler angle coordinates $\theta,\phi,\sigma$ on $SO(3)$, 
the symplectic form on $\cM_n$ may then be decomposed into
\be
\label{omj}
\omega =  j\, \sin\theta \, \de\theta \wedge \de\phi
- \de j\wedge (\de\sigma+\cos\theta\de\phi) 
+ \tilde\omega \ ,
\ee
where $\tilde\omega$ is the symplectic form on the symplectic leaf $\tilde \cM(j)$. 
Eq. \eqref{omj} follows by requiring the invariance of $\omega$ under the vector fields
$\pa_\sigma, \pa_\phi, (\pa_\sigma-\cos\theta \pa_\phi)/\sin\theta$, 
generalizing the argument  in \cite{deBoer:2008zn} to an
arbitrary number of centers.

Our goal in this work will be to determine the `refined index' $\Tr '(-y)^{2J_3}$ of the 
supersymmetric quantum mechanics of $n$-centered BPS configurations, where $J_3$ 
is the angular momentum operator along the 
$z$-axis.\footnote{We shall use the
symbol $J_3$ to denote both the quantum angular
momentum operator as well as the classical angular momentum \eqref{ejexp}.
The correct interpretation should be clear from the context.}
We defer to \S\ref{secqindex} a detailed discussion
of this quantum mechanics, and focus for now on 
the classical
version of the refined index, 
the phase space integral 
\cite{Manschot:2010qz}
\be \label{ephaseint}
g_{\rm classical} (\{\alpha_i\};y) \equiv  
{(-1)^{\sum_{i<j}
\alpha_{ij} -n+1} \over (2\pi)^{n-1} (n-1)!}\,  \int_{\cM_n} 
\, e^{2\nu\, J_3}\, \omega^{n-1}\, ,
\ee
where throughout this paper, we denote
\be
 \nu\equiv \ln y\ .
\ee
Such integrals over  a symplectic manifold
of the exponential of the moment map of some Hamiltonian
action  are well studied in the mathematical literature 
under the name of `equivariant volume' (see e.g. \cite{MR2334206} for a survey). 
The convergence of 
the integral \eqref{ephaseint} will be addressed in \S\ref{secfinite}.
The sign $(-1)^{\sum_{i<j}
\alpha_{ij} -n+1}$ in \eqref{ephaseint} is inserted for reasons which will become 
clear in \S\ref{secqindex}.
Leaving aside this sign, the equivariant volume 
 \eqref{ephaseint} is expected to be
a good approximation to $\Tr'  y^{2 J_3}$ in
the classical limit, where all symplectic
products $\alpha_{ij}$ are scaled to infinity and $y\to 1$ 
(this last condition
ensuring that the function $y^{2J_3}$ 
varies slowly on the phase space). 

Using the description of $\cM_n$ as 
$SO(3)\times \tilde\cM_n$ we can rewrite the integral 
\eqref{ephaseint} as 
\be \label{ephaseint2}
g_{\rm classical} (\{\alpha_i\};y) \equiv  
{(-1)^{\sum_{i<j} \alpha_{ij} -n+1}
\over (2\pi)^{n-1} (n-3)!}\,  \int_{\cJ} j \, \de j \int_{SO(3)}  
\sin\theta \,\de \theta\, \de\phi\, \de \sigma  
\, e^{2\nu\, j\cos\theta} \int_{\tilde\cM_n(j)} 
\, \tilde\omega^{n-3}\ .
\ee
Carrying out the angular integral, we arrive at 
\be \label{ephaseint3}
g_{\rm classical} (\{\alpha_i\};y) \equiv  (-1)^{\sum_{i<j}
\alpha_{ij} -n+1}
 \int_{\cJ} \de j\, \frac{\sinh(2\nu j)}{\nu}
\, \tilde g_{\rm classical} (\{\alpha_i\},j)  
\ee
where 
\be
\tilde g_{\rm classical} (\{\alpha_i\},j) ={1\over (2\pi)^{n-3}
(n-3)!}\,  \int_{\tilde\cM_n(j)} 
\, \tilde\omega^{n-3}
\ee
is the symplectic volume of the leaf $\tilde\cM_n(j)$. 

\subsection{Scaling solutions and finiteness of the equivariant volume\label{secfinite}}

In order to assess the convergence of the equivariant volume \eqref{ephaseint}, it
is useful to start with the simplest three-body case, which was discussed in detail 
in \cite{deBoer:2008zn}. Recall that the solutions to the 
equilibrium conditions
\be 
\label{esa6}
{\alpha_{12}\over r_{12}} +
{\alpha_{13}\over r_{13}}  = c_1\, ,
\quad 
-{\alpha_{12}\over r_{12}} +
{\alpha_{23}\over r_{23}}  = c_2\, , \quad
-{\alpha_{13}\over r_{13}} -
{\alpha_{23}\over r_{23}}  =  c_3 = - c_1 -c_2
\ee
can be parametrized by \cite{deBoer:2008zn}
\be \label{eintersolution}
r_{12} = {\alpha_{12}\over \rho}, \quad r_{23}
= {\alpha_{23}\over \rho + c_2}, \quad
r_{13} = {\alpha_{31}\over \rho - c_1}\, ,
\ee
where $\rho$ runs over the subset of $\IR$ 
satisfying $r_{ij}>0$ and the triangular inequalities 
$r_{12}\leq r_{13}+r_{23}$, $r_{23}\leq r_{12}+r_{13}$, 
$r_{13}\leq r_{12}+r_{23}$. In general, 
the allowed range of $\rho$ consists of at most two intervals, possibly reaching 
$\pm \infty$, whose finite endpoints correspond to collinear 
configurations (see Fig. \ref{figcplane} for a pictorial determination
of the allowed range of $\rho$ as a function of the constants $c_i$). 
Trading $\rho$ for the total angular momentum $j$ using \eqref{defj}, one finds that 
the range $\cJ$ of the latter is always a bounded interval $[j_-,j_+]$.

When $\alpha_{12}$,
$\alpha_{23}$ and $\alpha_{31}$ are positive and 
satisfy the same triangular inequalities, the region $\rho\to \infty$
is included in the allowed range, and corresponds to 
scaling solutions where the three centers come arbitrarily 
close to each other\cite{deBoer:2008zn}. From the second equality
in  \eqref{ejexp} it follows that  such configurations have
vanishing total angular momentum $j$ in the strict limit $\rho\to\infty$, hence $j_-=0$.

Even though the space $\cM_3$ is non-compact when such solutions are allowed,
its equivariant   volume $g_{\rm classical} (\{\alpha_i\};y)$ (in particular, 
its symplectic volume) is in fact finite. 
To see this, note that the reduced symplectic space $\tilde\cM_3(j)$ 
consists of a single point when $j\in  [j_-,j_+]$, or is empty otherwise.
Thus,
 \be \label{ephaseint33}
g_{\rm classical} (\alpha_1,\alpha_2,\alpha_3;y) 
= \frac12
(-1)^{\alpha_{12}+\alpha_{23}+\alpha_{13}}  
\frac{\cosh(2\nu j_+)-\cosh(2\nu j_-)}{\nu^2}
\ee
In the presence of scaling solutions, $j_{-}=0$, but the 
integral is still convergent. In fact the space $\cM_3$ may be compactified
by adding one point , corresponding to scaling solutions in the strict $\rho=\infty$ limit.
Since $j=0$ in this limit, this point is fixed under the action of $SO(3)$.

\FIGURE{
\centerline{\includegraphics[totalheight=5.5cm]{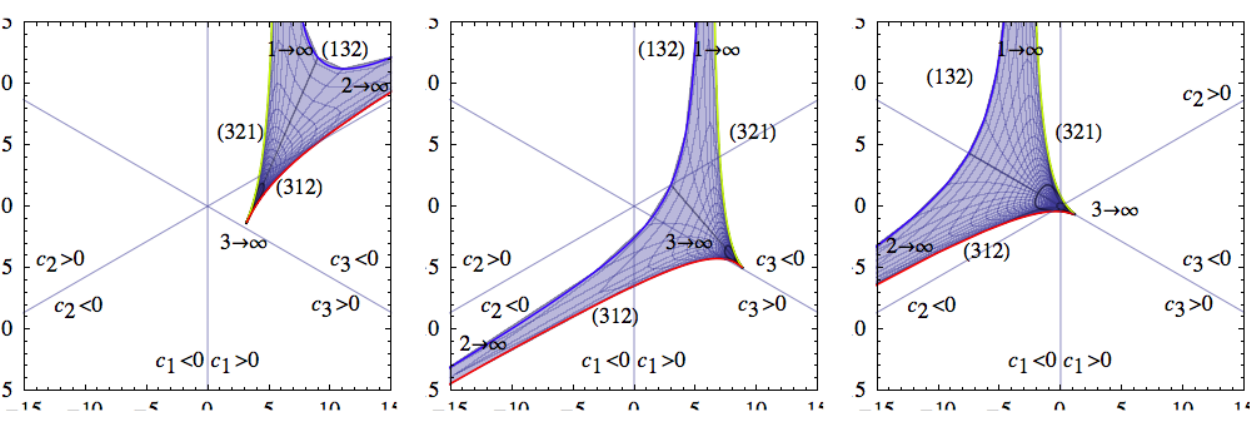}}
\caption{Phase diagram for 3-center configurations, in the plane 
$(x,y)=(-\frac32(c_2+c_3),\frac{\sqrt{3}}{2}(c_2-c_3))$. 
The shaded area corresponds to the values of $c_i\, r_{12}$ 
reached by varying the position of the third center, keeping the position of the centers 
1 and 2 and $\alpha_{ij}$ fixed. The range of $r_{12}$ (and therefore, of the parameter $\rho$ 
in \eqref{eintersolution}) at fixed values of $c_i$ can be determined by 
intersecting the shaded 
area with a radial line extending from the origin through the point $c_i$
(Thus if the radial line passing through 0 and
$\vec c$ meets the boundary of the shaded region at  
$\vec c^{(1)}$ and $\vec c^{(2)}$, then the range of $r_{12}$
is from $|\vec c^{(1)}|/|\vec c|$ to $|\vec c^{(2)}|/|\vec c|$.
On the other hand if the origin falls inside the shaded region
then the range of $r_{12}$ varies from $0$ to  
$|\vec c^{(1)}|/|\vec c|$).
The blue, red and yellow lines
correspond to collinear configurations in order 132, 213, 321 (or their reverse), 
respectively. 
The lines with $\arg(x+iy)=\pi/2, -5\pi/6,-\pi/6$ denote the locus where $c_1=0, 
c_2=0, c_3=0$, respectively.
i) $\alpha_{12}>0,
\alpha_{23}>0, \alpha_{13}>0$: solutions exist in the sector $c_1>0,c_3>0$
ii) $\alpha_{12}>0,
\alpha_{23}>0, \alpha_{13}<0$, $\alpha_{12}>\alpha_{31}+\alpha_{23}$
(non-scaling regime):
solutions exist in the sectors $c_1>0$ or $c_2<0$
iii)  $\alpha_{12}>
\alpha_{23}>0, \alpha_{13}<0$, $\alpha_{12}<\alpha_{31}+\alpha_{23}$
(scaling regime): solutions exist for all values of $c_i$. The figures were 
produced using  $(\alpha_{12},\alpha_{23},\alpha_{13})=(1,3,2),(6,3,-2),(1,3,-3)$
in cases i), ii), iii), respectively. 
 \label{figcplane}}
}

We now turn to solutions with arbitrary number of centers.
In the special case where the centers can be ordered via a permutation $\sigma$ 
of $\{1,\dots, n\}$ such that $\alpha_{\sigma(i)\sigma(j)}\geq 0$ 
whenever $i>j$, \footnote{This 
is equivalent to the condition that the associated quiver (with $n$ nodes
and  $|\alpha_{ij}|$ arrows from $i$ to $j$ if $\alpha_{ij}>0$,
or from $j$ to $i$ if $\alpha_{ij}<0$) contains no oriented 
closed loop. This condition is sufficient for the absence of scaling
solutions, but not necessary, as illustrated by the 3-center case.
It is in particular obeyed when all charges $\alpha_i$ belong to 
a positive cone in a two-dimensional 
charge lattice, the case studied in \cite{Manschot:2010qz}.} 
it is straightforward to see that the $r_{ij}$'s are bounded
from below by a non-zero $r_{\rm min}$, therefore scaling solutions 
do not exist. Away from walls of marginal stability
or threshold stability\footnote{Walls of marginal  or threshold stability 
correspond to loci where $\sum_{i\in A} c_i=0$ for a proper subset $A$
of $\{1,\dots, n\}$, such that the centers in the subset $A$ can move off
to infinity. Threshold stability corresponds to the special case 
where the charge $\gamma_A=\sum_{i\in A} \alpha_i$ of the subset
$A$ and the charge $\gamma-\gamma_A$ of its complement are mutually
local \cite{deBoer:2008fk}.}, 
the $r_{ij}$'s are also bounded
from above, and the phase space $\cM_n$ is therefore compact.
In such cases, the equivariant volume \eqref{ephaseint} is manifestly
finite. 

If on the other hand, the centers cannot be ordered such that 
$\alpha_{\sigma(i),\sigma(j)}\geq 0$ 
whenever $i>j$, then there may exist a 
subset $A$ of $(1,\dots , n)$ for which
there exist vectors $\vec r_i\in \IR^3$ ($i\in A$)
satisfying
\be \label{escond0}
\sum_{j\in A\atop j\ne i} \, {\alpha_{ij} \over \left|\vec r_{ij}
\right|} = 0, \qquad \forall \quad i\in A\, .
\ee 
In this case 
the centers in the subset $A$ may form a scaling
solution, i.e. may come arbitrarily close to each other.
In this case, the space $\cM_n$ is non-compact. 
However, we shall now outline a proof that the equivariant volume \eqref{ephaseint}
is in fact finite even in this case.

We start by considering potential divergences from `maximally scaling solutions',
where all centers come arbitrarily close to each other. 
Since from eq.\eqref{edefomega} the symplectic form $\omega$ 
is invariant under $r_{ij}\to\varepsilon r_{ij}$, the volume element 
$\omega^{n-1}$ is superficially expected to behave as 
$\de\varepsilon / \varepsilon$ in the region $\varepsilon\to 0$,  
suggesting  a logarithmic divergence of the equivariant volume. 
However, notice that in the $\varepsilon\to 0$ limit, 
the symplectic form $\omega$ becomes degenerate 
in the direction corresponding to overall dilations of the 
configuration (as noted in footnote \ref{notablefoo} above).  
This dilation is an exact symmetry of the equilibrium equations
\eqref{denef3d} at $c_i=0$, but is broken for finite $\varepsilon$ 
at non-zero $c_i$. Thus, the naive estimate $\omega^{n-1}\sim \de\varepsilon / \varepsilon$
must be modified by an integer power\footnote{In fact, using the decomposition
\eqref{ephaseint3} of the equivariant volume and the second relation \eqref{ejexp},
one finds that the volume element $j\de j$ along the  angular momentum
direction scales as $c^2 \varepsilon \de \varepsilon$. 
If  the symplectic volume of the 
reduced phase space $\tilde\cM_n(j)$ has a finite non-zero value at $j=0$, which is the case
for general 3-centered configurations or for dipole halo configurations with 
an arbitrary number of centers, it follows that the volume
element goes as $\eps d\eps$ near $\eps=0$.} 
$k\geq 1$ 
of the dimensionless parameter $c\varepsilon$
(where $c$ denotes a function of the $c_i$ homogeneous of degree one). In this case 
the volume element $\omega^{n-1}\sim (c\, \varepsilon)^k \de\varepsilon / \varepsilon$ 
vanishes at $c_i=0$. Thus, the contribution of the region  $\varepsilon\to 0$
 to the equivariant volume is finite, and 
 there is no divergence coming from `maximally scaling solutions'. 

We should also consider situations where $m$ of the $n$
centers come arbitrarily close to each other. The
phase space near this configuration is locally a product
of the phase space of an $(n-m+1)$ centered configuration,
in which all the $m$ close-by centers are regarded as
a single center, and that of a $m$-centered scaling solution.
The symplectic volume of the former  is manifestly finite, 
while the symplectic volume of the latter is
finite by the argument given earlier. This can be easily
generalized to the cases where there are several 
subsets, each containing a set of centers which
come close to each other. Thus, this reasoning indicates that the  
equivariant volume 
\eqref{ephaseint} is finite even in the presence of scaling solutions.
In \S\ref{sdipole},
we shall verify this claim
in the case of dipole halo configurations, where the 
phase space admits a natural compactification into a toric 
symplectic manifold, whose volume can be 
evaluated explicitly. It is an important problem to  
construct a compactification of $\cM_n$ in the  general case
involving  scaling solutions.

\subsection{A fixed point formula for the equivariant volume \label{sec_equivol}}  
  
Having established that the equivariant volume \eqref{ephaseint} is finite, we 
shall now evaluate it by localization.  For this purpose, we need to determine 
the fixed points of the action of $J_3$ on $\cM_n$. In general, the fixed points 
correspond either to collinear solutions along the $z$ axis, or to the coinciding
limit of scaling solutions  approaching a point on this axis. First we restrict to the
case where the charges $\alpha_i$ do not allow for scaling solutions. In this
case, $\cM_n$ is compact and all fixed points are isolated.

Collinear solutions  lying  along the $z$ axis satisfy 
a one-dimensional version of the equilibrium 
conditions 
\eqref{denef3d},
\be \label{denef1d}
\sum_{j=1\atop j\ne i}^n \frac{\alpha_{ij}}{|z_i- z_j|} \, 
= c_{i}\,  \ ,\qquad \sum_i z_i=0\ ,
\ee
where the second equation  fixes the
translational zero-mode.
These equations then select the critical points 
of  the `superpotential' \cite{Manschot:2010qz}
\be
\label{defhatW}
\hat W(\lambda, \{z_i\}) = -\sum_{i<j} \alpha_{ij}\,   {\rm sign}[z_j-z_i]\, 
\ln| z_j - z_i| - \sum_i (c_i - \lambda/n)  z_i 
\ee
as a function of $n+1$ variables $\lambda, z_1,\cdots z_n$. 
In the vicinity of one such fixed point $p$, 
the angular momentum $J_3$
and the symplectic form $\omega$ take the form
\be \label{ej3omega}
J_3 = {1\over 2} \sum_{i<j} \alpha_{ij}\, \sign[z_j-z_i]
- {1\over 4} M_{ij}(p)\,  (x_i x_j + y_i y_j) + \cdots,
\qquad \omega = 
\frac12 \, M_{ij}(p)\, \de x_i\wedge \de y_j +\cdots \, ,
\ee
where $M_{ij}(p)$ is the Hessian matrix of $\hat W(\lambda, \{z_i\})$ with 
respect to $z_1,\dots , z_n$, and $(x_i,y_i)$ are coordinates in the plane
transverse to the $z$-axis at the center $i$, subject to the condition 
$\sum_i x_i=\sum y_i=0$. Except for the overall translational zero-mode,
the matrix $M_{ij}$ is non-degenerate, and the critical points are isolated.
Since $\cM_n$ is compact, we have a finite set of
collinear fixed points $p$. 
It will be useful to note that 
$\widehat\det M=-\det \hat M$, where 
$\widehat\det M$ denotes the determinant of the $(n-1)\times (n-1)$ 
matrix $M$ with the first column and first row removed (i.e. the Hessian
of $\hat W$ as a function of $z_2, \dots , z_n$), and 
$\hat M$ is the Hessian of $\hat W$ as a function of the $n+1$ variables
$\lambda, z_1,\dots , z_n$ \cite{Manschot:2010qz}.

Since $\cM_n$ is compact and the fixed points are isolated, 
the Duistermaat-Heckman formula \cite{Duistermaat:1982vw} 
allows to express the equivariant volume \eqref{ephaseint} as 
a sum over fixed points. The contribution of each fixed 
point $p$ is
given by the  formal Gaussian integral\footnote{This Gaussian integral is ill-defined
since the quadratic form is in general not positive definite, but the 
Duistermaat-Heckman formula guarantees that formal Gaussian
integration leads to the correct result.}
 obtained by replacing 
the moment map by its quadratic approximation around $p$, 
and the symplectic form by its value at the same point. Thus
the net contribution is given by
\be \label{eclass0} 
\begin{split}
g_{\rm classical} (\{\alpha_i\};y) = & \frac{(-1)^{\sum_{i<j} \alpha_{ij} +n-1}}
{(2\pi)^{n-1}} \, \sum_{p} \,  
\int \prod_{i=2}^{n} \de x_i\, \de y_i\, 
\frac{(\widehat\det  M(p))}{2^{n-1}} \\
&  \exp\Big\{\nu 
\sum_{i,j=1\atop i<j}^n \alpha_{ij} \, \sign[z_j - z_i]
- {\nu\over 2}  \sum_{i,j=2\atop i<j}^n M_{ij}(p)\,  (x_i x_j + y_i y_j)\Big\}
\, .
\end{split}
\ee
After carrying out the Gaussian integral over $x_i, y_i$, 
we arrive at \cite{Manschot:2010qz}
\be \label{eclass} 
g_{\rm classical} (\{\alpha_i\};y) = \frac{(-1)^{\sum_{i<j} \alpha_{ij} +n-1}}
{(2\ln y)^{n-1}} \, \sum_{ p} \,  
s(p)\, 
y^{\sum_{i<j} \alpha_{ij}\, \sign[z_j-z_i]}
\, ,
\ee
where the coefficient $s(p)$ is given by
\be \label{esigneq}
s(p) =\sign \widehat \det M(p) = - \sign \det \hat M(p)\, .
\ee
$-s(p)$ is 
recognized as Morse 
index of the critical point $p$ of $\hat W$. 
It is worthwhile noting that the prefactor $(2\nu)^{n-1}$ in the denominator
of \eqref{eclass} originates from the determinant of the quadratic form in the $2n-2$
transverse directions to the fixed point.

In the presence of scaling solutions,  $\cM_n$ is non-compact and the
Duistermaat-Heckman formula does not apply directly. In the examples
that we shall study in \S\ref{sdipole}, it appears that $\cM_n$ admits
a natural compactification $\hat \cM_n$, which introduces additional, non-isolated fixed
points. Their contribution can in principle be evaluated  using a 
suitable generalization of the  Duistermaat-Heckman formula \cite{MR1215720}. 
For lack of a complete understanding of the 
compactification $\hat \cM_n$, we shall not pursue this approach. Instead, one may 
give a `minimal modification prescription' which determines  
the contribution of the non-isolated fixed points by requiring that 
the Fourier transform of the equivariant volume $\int_{\IR} g_{\rm classical} 
(\{\alpha_i\},\nu) e^{2\I\nu m} \de\nu$
becomes a piecewise polynomial function of $m$ with compact support, as
must be the case for any compact symplectic manifold with a 
Hamiltonian action \cite{Duistermaat:1982vw}. Since our interest lies
eventually in the equivariant index rather than the equivariant volume, we
shall not explain this prescription here, referring to \S\ref{sdipole} and \S\ref{sc}
for direct computations of the equivariant volume for dipole halo configurations,
and identifications of the corresponding  fixed points.

Finally, we return to the regularity condition \eqref{eregu}. In the case considered in 
\cite{Manschot:2010qz}, namely multi-centered solutions
whose charges all lie in a positive cone of a two-dimensional lattice, it appears 
that this condition is automatically satisfied near the 
wall. 
In general, this need not be so.
On physical grounds, we do
not expect that this condition should cut out regions of a given connected component of $\cM_n$.
On the other hand, it is quite possible that this condition could forbid certain connected components of $\cM_n$ altogether. In this case, we simply need to omit 
the corresponding fixed points from the sum \eqref{eclass}. We shall see an example
of this phenomenon in \S\ref{d6d6d6} below.  
 
\subsection{A fixed point formula for the refined index  \label{secqindex}}
 
We now turn to the problem of quantizing the configurational degrees of freedom
of $n$-centered black hole solutions  and the computation of the refined 
index $\Tr' (-y)^{2J_3}$, of which \eqref{eclass}
is supposed to be the classical approximation.
 
 We start by recalling and expanding upon some general aspects of 
 the quantization of BPS 
 multi-centered black hole solutions, first laid out 
 in \cite{Denef:2002ru, deBoer:2008zn}. 
 As in these references, we 
 assume that the Hilbert space of BPS states, in principle obtained by
 first quantizing the full configuration space of $\cN=2$ supergravity
 (or more generally, string theory) and then restricting 
 to the subspace annihilated by 4 supercharges, can be alternatively
 obtained by first restricting to classical solutions annihilated by
 the  same number of supercharges, and then quantizing. 
   
 Since each BPS solution of the type \eqref{eqatt} preserves 4 supercharges, 
 one expects that the locations $\vec r_i$ should be quantized into operators
 in a quantum mechanics with $\cN=4$ supersymmetries, such that the 
 classical supersymmetric vacua
 are in one-to-one correspondence with solutions to the equilibrium 
 conditions \eqref{denef3d}. 
 The final result should not be sensitive to the details
 of the potential in this
 quantum mechanics as long as the 
 space of classical supersymmetric vacua
 is given by \eqref{denef3d}.
 One way to obtain the Lagrangian is to consider
 the Coulomb branch of an $\cN=4$ supersymmetric quiver quantum
 mechanics with $n$ vector multiplets (each including 3 
 real scalars $\vec r_i$,  a $U(1)$ gauge field $A_i$ and 
 four fermions $\psi_i$) and, for each pair of centers $i,j$,
 $|\alpha_{ij}|$ chiral multiplets with charge $\sign(\alpha_{ij})$ under $A_i$
 and $-\sign(\alpha_{ij})$ under $A_j$. On the 
 Coulomb branch\footnote{The quiver quantum 
 mechanics also has a Higgs branch. 
 In some cases, including the wall-crossing problem
 considered in \cite{Manschot:2010qz}, the BPS spectra on the Coulomb branch
 and Higgs branch are identical, but this is not so in general, and  the Higgs branch 
 states cannot always be viewed as multi-centered black holes.}, 
 the chiral multiplets are massive and can be integrated out at one-loop.
 The result is an $\cN=4$ supersymmetric Lagrangian for the positions $\vec r_i$
 and their fermionic partners $\psi_i$,  with a potential which vanishes when
  the $3n$ scalars satisfy the equilibrium conditions \eqref{denef3d}. In addition, 
  there is a first order coupling 
 $\int \lambda$ to  the one-form\cite{Denef:2002ru}
 \be \label{edeflambda}
\lambda = \frac12 \sum_{i,j\atop i<j} \alpha_{ij}\, 
\frac{\epsilon^{abc}\, r^a_{ij} \, \de r^b_{ij}  \, u^c 
}{|\vec r_{ij}| (|\vec r_{ij}| +\vec r_{ij}\cdot \vec u)} =
\frac12 \sum_{i,j\atop i<j} \alpha_{ij}\, (1-\cos\theta_{ij}) \, \de \phi_{ij}\ ,
\ee 
where $\vec u$ is the same unit vector used in \eqref{edefomega}.
The degrees of freedom associated to the center of mass
motion are decoupled and can be removed by setting 
$\sum_i \vec r_i = \sum_i \psi_i=0$, so that the 
bosonic configuration space consists of 
$(\IR^{3n-3}\backslash\Delta)$, where 
$\Delta$ is the coinciding
locus.  Quantum mechanically, the Hilbert space consists
of square integrable sections of $\tilde S\otimes\tilde\cL$ 
over $\IR^{3n-3}\backslash \Delta$, where $\tilde S$ 
is the  trivial bundle of rank $2^{2n-2}$ over $\IR^{3n-3}$
obtaining by quantizing the $4(n-1)$ fermionic modes 
$\psi^i$, and $\tilde\cL$
is a complex line bundle with  first Chern 
class\footnote{The normalisation chosen in 
\eqref{edefomega}
ensures that $\omega\in H^2(\IR^{3n-3}\backslash \Delta,\IZ)$, moreover
$\IR^{3n-3}\backslash \Delta$ is simply connected so $\tilde\cL$ is uniquely
defined. The unit vector $\vec u$ in \eqref{edeflambda} correspond to 
a choice of Dirac strings for the line bundle $\tilde\cL$.} 
$c_1(\tilde\cL)=\omega=\de\lambda$. 
BPS states correspond to supersymmetric ground states
of this $\cN=4$ quantum mechanics.

For the purposes of studying these ground states, one may return to the 
classical Coulomb branch Lagrangian, integrate 
out the fluctuations transverse to the zero-energy 
submanifold $\cM_n\subset \IR^{3n-3}$, and then quantize the
system. While we have not studied this problem in detail,
we assume, in the spirit of \cite{deBoer:2008zn}, 
that BPS states correspond to harmonic spinors on $\cM_n$, i.e.
sections of $S\otimes \cL$ annihilated by $D$, where
$S$ is the total spin bundle\footnote{We assume that
$\cM_n$ admits a spin structure, unlike a generic
symplectic manifold which only admits a spin$_c$ 
structure \cite{MR1365745}.
By integrating out the massive fermions around the supersymmetric vacua, it should 
be possible to construct the spin bundle $S$ as a  rank $2^{n-1}$ subbundle of 
$\tilde S$, but we have not investigated this in detail.
}
 of $\cM_n$, $\cL$ is the restriction of $\tilde \cL$ to $\cM_n$ 
 and  $D$ is the Dirac operator 
for the metric $g_{\mu\nu}$ induced from the flat metric on $\IR^{3n-3}\backslash\Delta$, 
 twisted by  the line bundle $\cL$.
The Dirac operator $D$ decomposes as $D=D_+ + D_-$, where $D_\pm$ maps 
$S_\pm$ to $S_\mp$ where $S=S_+\oplus S_-$ is the decomposition into spinors 
with positive and negative chirality. 
Moreover, the action of $SO(3)$ on $\cM_n$ 
lifts to an action of $SU(2)$ 
on $S_\pm \otimes \cL$. 
The refined index $\Tr'(-y)^{2J_3}$ is then given by
\be
g_{\rm ref} (\{\alpha_i\};y) = 
\Tr_{{\rm Ker} D_+}(-y)^{2J_3} +  \Tr_{{\rm Ker} D_-}(-y)^{2J_3}\ . 
\ee
We shall assume that ${\rm Ker} D_-=0$, so that the refined index $g_{\rm ref} (\{\alpha_i\};y)$
reduces to 
\be
\label{equivi}
g_{\rm ref} (\{\alpha_i\};y) = \Tr_{{\rm Ker} D_+} (-y)^{2J_3} -  \Tr_{{\rm Ker} D_-} (-y)^{2J_3}\ . 
\ee

When $\cM_n$ is \kahler, as is the case for $n=2,3$ or for the dipole halo
configurations studied in \S\ref{sdipole}, and $\alpha_{ij}$ large enough, the 
fact that ${\rm Ker} D_-=0$ can be proved as follows.
In this case
$S_+$ and $S_-$ are isomorphic to 
$\Lambda^{\rm even} (T^{(0,1)}_\IC \cM_n) \otimes K^{1/2}$ 
and $\Lambda^{\rm odd} (T^{(0,1)}_\IC \cM_n) \otimes K^{1/2}$,
respectively, where $K$ is  the canonical bundle of $\cM_n$, i.e. the 
complex line bundle of $(n-1,0)$ forms (the square root exists since $\cM_n$
is assumed to be spin). 
In this case 
the Kodaira vanishing theorem
states that the cohomology groups 
$H^{(0,q)}(\cM_n,\cL\otimes K^{1/2})$ all vanish 
except for $q=0$. This shows that ${\rm Ker} D_-=0$.
In general, we do not how to prove this assumption, but it is supported by the fact that it 
leads to results in agreement with wall-crossing.  Since the refined index $\Tr'(-y)^{2J_3}$ 
is not protected\footnote{In 
$\cN=2$ gauge theories at low energies, it  may be possible to compute the protected spin character (PSC) \cite{Gaiotto:2010be} by quantizing 
the space of classical multi-centered Abelian dyons
along the lines of \cite{Lee:2011ph}. Since the PSC is protected, it is plausible that the configurational index will be directly equal to the 
equivariant index \eqref{equivi}, without the need to assume 
that ${\rm Ker} D_-$ vanishes. The vanishing of ${\rm Ker} D_-=0$
in this context is closely related to the weak positivity conjecture of \cite{Gaiotto:2010be}.} 
in $\cN=2$ supergravity, for the purposes of computing the index 
$\Tr'(-1)^{2J_3}$ it suffices to make the weaker assumption 
that $\Tr_{{\rm Ker} D_-} (-1)^{2J_3}=0$. 

Under this assumption, $g_{\rm ref} (\{\alpha_i\};-y)$
reduces to the equivariant index of the Dirac operator $D$.  
Using the  Lefschetz fixed point formula established
in \cite{MR0212836,MR0232406,MR805808,MR1215720}\footnote{We are grateful 
to M. Vergne for guidance into the math literature.} ,
assuming that $\cM_n$ is compact, the equivariant index can be written as the integral
\be
\label{ephaseq}
g_{\rm ref} (\{\alpha_i\}; - y) = 
\int_{\cM_n} {\rm Ch}(\cL,\nu)\,  \hat A(\cM_n,\nu)\vert_{n-1}
\ee
where ${\rm Ch}(\cL,\nu)$ is the equivariant Chern character of the line bundle $\cL$,
and $\hat A(\cM_n,\nu)$ is the equivariant $\hat A$-genus of $\cM_n$, defined by 
\be
{\rm Ch}(\cL,\nu) =\exp\left( 2\nu J_3 + \frac{1}{2\pi}  \omega\right) \ ,\quad 
\hat A(\cM_n, \nu) = \det \left(\frac{2\nu L +\frac{1}{2\pi} \cR}
{2\sinh\frac12 \left(2\nu L + \frac{1}{2\pi} \cR\right)}
\right)\ .
\ee
Here, $J_3$ is the moment map of the action of rotations around the $z$ axis, 
$L$  is the endomorphism of the holomorphic  tangent bundle $T^{(1,0)} 
\cM_n$ induced by the same action, $\cR$ is the curvature two-form on 
$T^{(1,0)} \cM_n$,and $A\vert_{p}$ denotes the degree $2p$ part
of a multi-form $A$. The integral \eqref{ephaseq} can be evaluated
by localization \cite{MR805808}, leading to 
\be
\label{ephaseqloc}
g_{\rm ref} (\{\alpha_i\}; - y) = \int_{\cM_n^{\rm fixed}} 
\frac{ {\rm Ch}(\cL,\nu)\,  \hat A(\cM_n,\nu)}
{{\rm Eu}(N\cM_n^{\rm fixed})}\Big\vert_p
\ee
where $\cM_n^{\rm fixed)}\subset \cM_n$ denotes the  fixed point locus, of  
dimension $2p$, and 
\be
{\rm Eu}(N\cM_n^{\rm fixed}) = \det \left(2\nu L +\frac{1}{2\pi} \cR\right)
\ee
is the equivariant Euler character of the normal bundle of $\cM_n^{\rm fixed}$. 

In the absence of scaling solutions, $\cM_n$ is compact and all fixed points are isolated, so
$N\cM_n^{\rm fixed}=T\cM_n$ and the Euler character cancels the 
numerator in the $\hat A$-genus. Moreover, $J_3=\frac12 \sum_{i<j} \alpha_{ij}\sign[z_j-z_i]$
and the operator $L$, representing the action of $J_3$
on $T^{(1,0)}(\cM_n)$, has eigenvalues $\pm 1$, 
with $\det L =  s(p)$. This
leads to 
the following explicit formula for
the quantum refined index:\footnote{The same localization techniques show that
the index of the untwisted Dirac operator 
$\int_{\cM_n} \hat A = {(y-1/y)^{n-1}} \, \sum_{ p} \,  s(p)$ vanishes, in particular
is integer, consistently with the fact that $\cM_n$ is spin. }
\be
\begin{split}
g_{\rm ref} (\{\alpha_i\};-y) = &
{(y-1/y)^{n-1}} \, \sum_{ p} \,  
s(p)\, 
y^{\sum_{i<j} \alpha_{ij}\sign[z_j-z_i]}
\, . \end{split}
\ee
Hence, after changing\footnote{The fact that $y\to -y$ changes the refined index by an overall
sign shows that for fixed 
$\alpha_{ij}$ and $n$, all states carry the same parity of $2J_3$.
}
 $y\to -y$,
\be 
 \label{eclassq} 
g_{\rm ref} (\{\alpha_i\};y) = \frac{(-1)^{\sum_{i<j} \alpha_{ij}+n
-1}}
{(y-1/y)^{n-1}} \, \sum_{ p} \,  
s(p)\, 
y^{\sum_{i<j} \alpha_{ij}\sign[z_j-z_i]}
\, .
\ee
In particular, the refined index \eqref{eclassq} is related to the equivariant
volume \eqref{eclass} by an overall rescaling by a factor of $(\nu/\sinh\nu)^{n-1}$.
This multiplicative renormalization was postulated in \cite{Manschot:2010qz} 
on the basis of angular momentum quantization.  As the present derivation 
shows, this multiplicative renormalization in fact follows from the Atiyah-Bott Lefschetz 
fixed point formula for the equivariant index of the Dirac operator on $\cM_n$.

Eq. \eqref{eclassq} will be our main tool
for computing the index of multi-centered black holes
in the absence of scaling solutions. Although it was
derived under various assumptions, we shall carry out
various consistency checks which build our confidence
in this formula.
In the presence of scaling solutions, the space $\cM_n$ is non-compact
and its compactification includes non-isolated fixed points for the action
of $J_3$. Our limited understanding of the geometry of the fixed submanifold
${\rm \cM}_n^{\rm fixed}$ prevents us from computing the index
using \eqref{ephaseqloc}, nevertheless in \S\ref{scaling}
we shall give
a prescription for computing the contributions from these non-isolated
fixed points, from the knowledge of the isolated ones.

\section{Case studies in a one-modulus supergravity model}
\label{sexplicit}

In this section, we analyze several examples of three-centered solutions 
in the context of a simple $\cN=2$ supergravity model with a single modulus.
This model, introduced in \S\ref{sexam}, 
arises  by compactifying type IIA string 
theory on a Calabi-Yau 3-fold 
$\cX$ with $b_2=1$ in the large volume limit. The first example in  \S\ref{d6d6d6} 
illustrates the importance of the regularity condition
\eqref{eregu}, while the second example in 
\S\ref{d6d6d0} shows the necessity of including 
contributions from 
fixed points associated with scaling  solutions. 
This example is a particular 
case of a general class of dipole halo configurations which 
will be discussed
further in \S\ref{sdipole}. The example in \S\ref{sunphys}
illustrates the role of the regularity condition 
\eqref{eregu} in deciding whether a given scaling solution
is physical or not.

\subsection{One-modulus model} \label{sexam}
We consider a simple supergravity model with one vector multiplet, governed
by the prepotential 
\be \label{eex1}
F(X^0, X^1) = - \kappa \, (X^1)^3 / 6 X^0 + {1\over 2}
\cA (X^1)^2 + {1\over 24} \cB X^0 X^1\, .
\ee
This model arises in the large volume limit of type IIA string theory 
compactified on a Calabi-Yau 3-fold $\cX$ with $b_2=1$. In this case
$\kappa=\int_\cX \omega^3$ is the cubic self-intersection of the unique
generator $\omega$ of of $H^2(\cX, \bZ)$, $\cB=\int \omega\wedge c_2$ 
is an integer such that $\cB=-2\kappa \mod 12$  
and $\cA$ is an half-integer constant such 
that $\cA=\kappa/2 \mod 1$.\footnote{These congruence conditions
are special cases of the conditions $\frac12 \kappa_{abc} p^b p^c -
\cA_{ab} p^b=0 \mod 1$, $\frac16  \kappa_{abc} p^a p^b p^c + \frac{\cB_a}{12} p^a
\mod 1$ for all integer vector $p^a$, see  \cite{Alexandrov:2010ca}.
For the quintic,  we have
$\kappa=5$, $\cA=-11/2$, $\cB=50$.}
The ratio $X^1/X^0$ is the complexified K\"ahler modulus $t$:
\be \label{eex2}
 t=B+\I J=\frac{X^1}{X^0}\, ,
\ee
with $B$ the NS 2-form potential and $J$ the real K\"ahler modulus.
The central charge of a BPS state with electromagnetic charges 
$(p^\Lambda,q_\Lambda)$ is given by
\be \label{eex3}
\begin{split}
Z &= e^{\cK/2} (p^\Lambda 
F_\Lambda-q_\Lambda X^\Lambda ) \cr
& =
e^{\cK/2} X^0 \left[  {1\over 6} \kappa\, t^3 p^0 -{1\over 2} 
\kappa\, t^2 p^1 - t \tilde q_1 - \tilde q_0
\right] \, ,
\end{split}
\ee
where $e^{-\cK}=\I(\bar X^\Lambda F_\Lambda-X^\Lambda \bar F_\Lambda)
={4 \over 3}\kappa J^3\, |X^0|^2$ is the K\"ahler potential, and
\be \label{deftildeq}
\tilde q_1= q_1 - \frac{\cB}{24} p^0 - \cA p^1, \qquad
\tilde q_0 =q_0 - \frac{\cB}{24} p^1 \, .
\ee
With the above conditions on $\cA,\cB$,  the
charges $p^0$, $p^1$, $q_1$ and $q_0$ are
quantized in integer 
units\cite{Alexandrov:2010ca} and represent (up to
a sign) $D6$, $D4$, $D2$ and $D0$-brane charges. 
We shall denote 
\be
 \label{edeftildeal}
\alpha=(p^0,p^1, q_1,q_0)\ ,\qquad
\tilde\alpha=(p^0,p^1, \tilde q_1,\tilde q_0)\, .
\ee
Note that given a pair of vectors $\alpha$, $\alpha'$ we
have $\langle \alpha, \alpha' \rangle = \langle \tilde \alpha,
\tilde \alpha'\rangle$. Solving the attractor equations gives for the Bekenstein-Hawking
entropy $\pi |Z_\alpha(X^I_{\mathrm{attr.}})|^2$ of a large black hole
 in this model \cite{Shmakova:1996nz}:
\be \label{eex4}
S =\pi \,   
\sqrt{D(p^0,p^1,  \tilde q_1,  \tilde q_0)},
\ee
where $D$ is the quartic polynomial
\be
D(p^0,p^1,  \tilde q_1,  \tilde q_0)
={\kappa^2\over 9}\, 
\left[ 3\frac{(  \tilde q_1p^1)^2}{\kappa^2} - 
18 \frac{  \tilde q_0 \, p^0 \,  \tilde  q_1 \, p^1}{\kappa^2}
- 9 \frac{  \tilde q_0^2 \,(p^0)^2}{ \kappa^2} - 6 
\frac{(p^1)^3 \,   \tilde q_0}{\kappa} 
+ 8  \frac{p^0 \, (  \tilde q_1)^3}{\kappa^3}\
\right]\, .
\ee

For a multi-centered black hole solution with charges 
$\alpha_i$, 
the consistency condition \eqref{eregu} can now be expressed as
\be\label{eex4.5}
D\left(  \tilde \beta +
\sum_{i=1}^n { \tilde \alpha_i\over |\vec r - \vec r_i|}
\right) > 0\, , \qquad \forall \quad \vec r\in \IR^3\, .
\ee
For single centered supersymmetric black holes $D(p^0,p^1, 
\tilde q_1,  \tilde q_0)>0$ implies 
\eqref{eex4.5} for $J\gg 0$. For
multi-center black holes \eqref{eex4.5} does not follow from
regularity of the near-horizon regions but must be checked
independently.  

We shall conclude this subsection with some comments on
the effect of $\alpha'$ corrections to the supergravity action. 
One source of corrections originates from world-sheet instantons
which contribute to the prepotential $F$. It is 
straightforward in principle  to incorporate such corrections in the analysis
of multi-centered solutions. In addition, there are also 
higher derivative corrections 
to the four-dimensional effective action 
which greatly complicate the construction of exact multi-centered
solutions \cite{LopesCardoso:1998wt,LopesCardoso:2000fp}.
While these corrections are essential in computing 
the index $\Omega^{\one}(\alpha)$ of  certain single centered black 
holes \cite{Cardoso:1999za},
their effect on the functions $g_{\rm ref}(\{\alpha_i\};y)$ 
governing the index of multi-centered black 
holes is mild, since the
detailed information about the action and the solution
is needed only to determine whether a given collinear solution
has a non-singular metric or not; but given such a solution
the contribution to $g_{\rm ref}(\{\alpha_i\};y)$ given in
\eqref{eclassq}, and its generalization to scaling solutions
which we shall discuss later, is independent of the
action. Thus in our analysis we shall  ignore the
effect of these higher derivative corrections. For explicit
computation we shall also set 
$\cA$ and $\cB$ to zero for convenience.

\subsection{Non-scaling solutions: $D6-D6-\overline{D6}$}
\label{d6d6d6}

We now consider  a 3-centered configuration
of $D6$-branes with fluxes, carrying charges
$e^{U\omega}$, $e^{V\omega}$ and $-e^{(U+V)\omega}$ where $U,V$
are positive integers. This example was studied in detail in
\cite{Denef:2007vg}, section 5.2.2.
To compute the 
corresponding charge vectors, 
recall that the central charge associated with a
$D6$-brane carrying charge $e^{U\omega}$ is proportional to
\be \label{eex5b}
-\int_{\cX} e^{-t\omega }\wedge \left[e^{U\omega}
\left(1 + {c_2\over 24}\right)\right]
\ .
\ee
Comparing this with \eqref{eex3}, and using the fact
that $\int c_2\wedge\omega=\cB$,
$\int \omega\wedge\omega\wedge \omega=\kappa$,
we see that the first
center carries charges
\be \label{eex5}
\tilde  \alpha_1 = \left( 1, U, 
- \frac12 \kappa\, U^2- \frac{1}{24}\cB, 
\frac16\kappa\, U^3+ \frac{1}{24}\cB U\right)\, ,
\ee
where we have determined the overall normalization of
$ \tilde\alpha_1$ by using the 
fact that for a single $D6$-brane
$p^0=1$. Similarly the second and the third
centers carry charges
\be \label{eex6}
\begin{split}
\tilde  \alpha_2 &=\left( 1, V, 
- \frac12\kappa\,V^2- \frac{1}{24}\cB, 
\frac16 \kappa\,V^3 + \frac{1}{24}\cB V\right) \, , \\
\quad  \tilde \alpha_3 &= \left(-1, -(U+V),  
\frac12\kappa\, (U+V)^2 + \frac{1}{24}\cB, -\frac16 \kappa\,(U+V)^3 
-  \frac{1}{24}\cB (U+V) \right)\, ,
\end{split}
\ee
so that the total charge is given by
\be \label{eex7}
 \tilde \gamma =  \tilde \alpha_1+ \tilde \alpha_2+ 
 \tilde \alpha_3
= \left(1, 0, \kappa\, UV-\frac{1}{24}\cB, -\frac12 
\kappa\, UV(U+V) \right)\, .
\ee
It is easy to see that for integer $U$ and $V$, the charges
$\tilde q_i$ are not integers in general, but $q_i$
computed via \eqref{deftildeq} are.
Using $\alpha_{ij}=\tilde \alpha_{ij}$,
the integer symplectic products are then given by
\be \label{eex7.5}
\alpha_{21} = \frac16 \kappa\, (V-U)^3 + \frac{1}{12}\cB (V-U)\ , 
\quad \alpha_{23}= \frac16\kappa\, U^3 + \frac{1}{12}\cB U\ ,
\quad \alpha_{13} =  \frac16\kappa\, V^3 + \frac{1}{12}\cB V\ .
\ee

Our goal is to find the index of supersymmetric bound
states in the large volume limit. We choose as a concrete example 
\be \label{eex7a}
\kappa=6\ ,
\quad \cB=0, \qquad 
U=1, \quad V=2, \quad B+iJ=3\, i\, .
\ee
This choice differ from that of \cite{Denef:2007vg} in
the choice of $\kappa$, but this is a simple normalization
factor and still allows us to
compare our findings with the results of
\cite{Denef:2007vg}.  The value of $J$ is chosen 
to be large enough to lie
in the large volume chamber.
The precise value of $J$ 
affects the numerical values of the
solutions to eq.\eqref{denef1d}, but is otherwise 
irrelevant. We
use the projective symmetry of special geometry to fix $X^0=1$.
For the values of charges and moduli corresponding to \eqref{eex7a},
this leads to
\be
\begin{split}
&\alpha_{12}=-1\ ,\quad \alpha_{23}=1\ ,\quad \alpha_{13}=8\ ,
\\
&\textstyle c_1={73 \over 477}\sqrt{318}, \quad c_2={170 \over 477}\sqrt{318}\ ,\quad 
\beta =\sqrt{53\over 6}\, ({7\over 159}, {2\over 53},
{63\over 53},-{18\over 53})\ .
\end{split}
\ee
Numerical analysis of  \eqref{denef1d} leads to 
the following collinear solutions:\footnote{We have
presented the solution in the $z_1=0$ gauge instead of
$\sum_i z_i=0$ gauge.}
\be
 \label{eex8}
\begin{array}{|r|r|r|r|r|r|r|}
\hline
\sigma & z_{\sigma(1)} & z_{\sigma(2)} & z_{\sigma(3)} & \rho & s(\sigma) & \mbox{reg.} \\
\hline
132 & 0 & 2.59 & 2.75& -0.363 & -1 & \surd \\
231 & -2.75 & -2.59 & 0& -0.363 & -1 & \surd\\
123 & 0 & 2.37 & 2.54 & -0.422 & 1 & \surd \\
321 & -2.54& -2.37 & 0 & -0.422 & 1 & \surd\\
123 & 0 & 0.195 & 1.02  & -5.14 & -1 & \times \\ 
321 & -1.02 & -0.195 & 0 &-5.14 & -1 & \times\\
213 & -0.182 & 0 & 0.974 & -5.49 &   1 & \times \\
312 & -0.974 & 0 & 0.182 &-5.49& 1 & \times \\
\hline
\end{array}
\ee
In this table, we have displayed the permutation $\sigma$
of 123
specifying the order of the $z_i$'s for a given solution
according to $z_{\sigma(1)}<z_{\sigma(2)}<
z_{\sigma(3)}$, the 
location of the centers $z_{\sigma(i)}$, the value of the parameter
$\rho$ in the parametrization given in \eqref{eintersolution}, the 
value of the sign $s(p)$, and in 
the last column, we indicated by
$\surd$ solutions obeying the regularity condition \eqref{eregu}, and 
by $\times$ those which do not (it turns out that the region where
the discriminant $D$ becomes negative intersects the $z$-axis, 
and therefore can be found by plotting $D$ as a function of $z$). 
Note that the parameter $\rho$ takes the same value for 
two configurations related by a reversal of the $z$-axis.
More generally,  solutions of \eqref{denef3d}
satisfying the triangular inequalities arise from the disconnected 
intervals $-.422<\rho< -.363$ and $-5.49< \rho< -5.14$ along the $\rho$ axis, 
and correspond to general non-collinear solutions with angular momentum 
in the intervals $0.444<j^2<0.694$ and 
$0.25<j^2<0.444$, respectively. 
Only the first 
interval leads to solutions which satisfy the regularity 
solution \eqref{eregu}.

According to our prescription only the solutions marked by $\surd$ must be
included in the sum in \eqref{eclassq}, and we get
\be \label{eex9}
g_{\rm ref}(\alpha_1,\alpha_2, \alpha_3;y) = 
(-1)^{\alpha_{21}+\alpha_{23}+\alpha_{13}}\, 
{1\over \sinh^2 \nu} \sinh(\alpha_{23}\nu) \,
\sinh((\alpha_{13}-\alpha_{21}) \nu)\, , \quad
\nu\equiv \ln y,
\ee
in agreement with \cite{Denef:2007vg}.
Had we ignored the constraint 
\eqref{eex4.5} and included  contributions from
all the solutions given in \eqref{eex8}, we would have got
\be \label{ediff}
(-1)^{\alpha_{21}
+\alpha_{23}+\alpha_{13}}\, 
{1\over \sinh^2 \nu} \sinh(\alpha_{13}\nu) \,
\sinh((\alpha_{23}+\alpha_{21}) \nu)\, .
\ee 
This last expression is indeed the
result produced by quiver quantum mechanics,
but as emphasized in \cite{Denef:2007vg}, 
quiver quantum mechanics in this case fails to give the correct
result. We see that the difference between the correct
result and quiver quantum mechanics is accounted for
by the additional constraint \eqref{eex4.5} that must
be imposed on the solutions besides
\eqref{denef1d}.

\subsection{Scaling solutions} 
\label{spres}
 
Consider a three centered black hole
solution with the centers carrying charges $\alpha_1$,
$\alpha_2$ and $\alpha_3$. Suppose further that
$\alpha_{12}=3$, $\alpha_{23}=4$ and $\alpha_{31}=5$.
In this case $\alpha_{12}$, $\alpha_{23}$ and $\alpha_{31}$
satisfy a triangle inequality, showing that the scaling solution
can exist. Thus this provides us with a 
laboratory for studying the
role of scaling solutions in the computation of the index
of multi-centered black hole solutions.

For now we focus on the constraints imposed by
eqs.\eqref{denef1d} without worrying about \eqref{eregu};
we shall return to this later.
Since we have two moduli $J$ and $B$ at our disposal
we can adjust them to set $c_1$ and $c_2$ as we like
(within an appropriate range), $c_3$ is then fixed to be
$-c_1-c_2$.
Let us use this freedom to choose a point in the
moduli space where
$c_i=-\Lambda \sum_j \alpha_{ij}$ for some positive
constant $\Lambda$. We can now look for solutions to
\eqref{denef1d}. Numerical
analysis shows that there are only two collinear
configurations which contribute, corresponding to the
alignments 123 and 321. Furthermore both contribute 
with positive sign. As a result the net contribution
is given by $(y-y^{-1})^{-2} \left(y^{\alpha_{12}+\alpha_{13}
+\alpha_{23}} + y^{-\alpha_{12}-\alpha_{13}
-\alpha_{23}}\right)$.

This causes a puzzle since this does not have a finite
$y\to 1$ limit. We must however recall that there are also
scaling solutions to Denef's constraints which
correspond to all three points approaching each other.
Clearly this is also  a fixed point of $J_3$ if we choose
the point of approach to lie on the $z$ axis. These however
will not show up in the numerical determination of the fixed
points which assumes from the beginning that the centers
have finite separation.

The fixed point associated with the scaling solution
has $J_3=0$ and hence it contributes a constant to the index.
There may be additional factors from the integration measure
near the scaling solution. 
These can in principle be determined from
a detailed analysis of the scaling solution.
However we can try to guess the contribution from the
scaling solution by requiring that the total contribution has
a finite $y\to 1$ limit. Our first guess would be
\be \label{esec1}
g_{\rm ref}(\alpha_1,\alpha_2,\alpha_3;y)
= (y - y^{-1})^{-2} \, \left(y^{\alpha_{12}+\alpha_{13}
+\alpha_{23}} + y^{-\alpha_{12}-\alpha_{13}
-\alpha_{23}}-2
\right)\, .
\ee
For 
$\alpha_{12}+\alpha_{13}
+\alpha_{23}$ even, the numerator has a factor 
of $(y-y^{-1})^2$, and after canceling this against the
denominator we are left with a Laurent polynomial
in $y$. This is a sensible result,
with the coefficient of $y^m$ counting $(-1)^m$ times
the number of states
with $J_3=m/2$. But \eqref{esec1} 
 does not lead to
a sensible spectrum for odd values of
$\alpha_{12}+\alpha_{13}
+\alpha_{23}$, since the factor of $(y-y^{-1})^2$ in the
denominator is not cancelled.
Our proposal is to replace the subtraction
constant 2 by $y + y^{-1}$ in this case so that 
$(y-y^{-1})^{-2}$
factor in the denominator is cancelled and we again get a 
Laurent polynomial in $y$.

The above analysis 
shows that we can predict the existence of
and the contribution from
a scaling solution from the results on the
non-scaling solutions which are simpler to 
find.
This is the general procedure we shall follow, \i.e.\
assume that the role of the fixed points associated
with the scaling solution is to essentially make the final
result satisfy desirable properties. This by itself will not
fix the contribution uniquely. For example in \eqref{esec1}
we could have taken the subtraction term to be 
$y^2 + y^{-2}$ instead of 2. To fix this ambiguity we shall
make a further assumption, that the extra contribution due
to the scaling solutions vanish as $y\to\infty$. This
fixes the correction terms uniquely. 
We shall call this the `minimal modification hypothesis'.
We shall
describe the general rule for such replacements
in \S\ref{scaling}.
While we have no
{\it a priori} justification for this assumption, it seems to work
in all known cases.

For specific choices of the prepotential we also need to
verify that the solutions we consider satisfy the constraint
\eqref{eex4.5}. This can be easily implemented by testing
\eqref{eex4.5} for each of the two collinear solutions
described above and if we find that \eqref{eex4.5}
fails for these solutions then we would conclude that the
subtraction terms associated with the scaling solutions
must also vanish. Thus there will be no contribution to
the index from this configuration. 
For implementing this condition we need to work with
specific charge vectors. We shall now discuss two 
examples, -- one where this condition is satisfied, and
another where it fails.

\subsubsection{Example 1: 
$\overline{D6}-D6-D0$} \label{d6d6d0}

Let us take a $D6$-$\overline{D6}$-brane pair
with flux as in the last subsection and a 
pure $D0$-charge. 
In particular we choose
\be \label{esc1}
\begin{split}
&  \tilde\alpha_1 = (1, -U, -\kappa\, 
U^2/2-\cB/24, -\kappa\, U^3/6-\cB U/24), \\
&  \tilde\alpha_2 = (-1, -U,  \kappa\, 
U^2/2 + \cB/24, -\kappa\, U^3/6-\cB U/24), \\
&  \tilde\alpha_3 = 
(0,0, 0, n), \qquad  \, 
\end{split}
\ee
for some integer $U$. In this case we have
\be \label{esc2}
\alpha_{12}=4\kappa U^3/3+\cB U/6, \quad \alpha_{23}
= \alpha_{31} = n\, .
\ee
We define $\alpha\equiv\alpha_{23}=\alpha_{31}$, and assume that 
$\alpha_{12}>0, \alpha>0$. Scaling solutions are expected when
the triangular inequalities are satisfied, i.e. when 
$2\alpha \geq \alpha_{12}$. 

We choose the following explicit values of the parameters: 
\be \label{evalue2}
\kappa=6, \quad \cB=0, \qquad U=1, \quad
t=B + iJ = 1 + 3\, i\, .
\ee 
As example of a system which does not allow for scaling
solutions we take $n=3$, while we take $n=6$ for an 
example of a system that does. For those values, we have
\begin{eqnarray}
&n={3}:& \qquad c_1=\textstyle\frac{135}{2}\sqrt{{6\over 3697}}, \qquad
c_2=-\frac{147}{2}\sqrt{{6\over 3697}},\\
&n=6:& \qquad c_1=\textstyle\frac{33}{10}\sqrt{{3\over 5}}, \qquad \qquad
c_2=-\frac{39}{10}\sqrt{{3\over 5}}\, .
\end{eqnarray}
For $n={3}$ one finds collinear fixed points for
 $\rho=3.41$ with  alignment $312$ (or its reverse)
 and $s(p)=1$, and for $\rho=11.37$ with 
 alignment $132$ (or its reverse)
and $s(p)=-1$. A detailed study 
of the full supergravity solutions shows
that they satisfy the regularity condition \eqref{eex4.5}
 everywhere outside the centers. The contribution of these fixed points  gives
\be \label{extra1}
g_\mathrm{ref}(\{   \alpha_i\};y)=\frac{y^8+y^{-8}-y^2-y^{-2}}{(y-y^{-1})^2}\ ,
\ee
which  is finite for $y\to 1$ as in \S\ref{d6d6d6}.

For $n=6$ one finds a single value of $\rho=3.99$ corresponding
to collinear solutions with alignment $312$ and $s(p)=1$ 
(and its reverse).
We have checked that this solutions satisfies \eqref{eex4.5} everywhere.
Adding
the contributions of these fixed points gives
$(y-y^{-1})^{-2}(y^8+y^{-8})$,  which does
not have a finite $y\to 1$ limit. 
Following the procedure described earlier we now
take the total contribution to be
\be \label{esec1special}
g_{\rm ref}(\{  \alpha_i\};y)
= (y - y^{-1})^{-2} \, \left(y^{8} + y^{-8}-2\, \right)\, ,
\ee
which does give a sensible result for the
spectrum.

We close this subsection with a comment on the 
structure of the solution in the regime
$0\leq B \ll 1$, $J\gg 1$. For this discussion 
we parametrize $c_1$ and $c_2$ by
$c_1=\mu-\eta$ and $c_2=-\mu-\eta$. In the 
limit described above, $\mu>0$ and
$0 \leq \eta \ll 1$. For $2\alpha<\alpha_{12}$,
there are  two 
collinear configurations (up to reversal of the $z$ axis)
with alignments $132$ and $213$, 
parametrized by 
\be
\label{rhoeta}
\rho_{(132)}=\frac{\alpha_{12}\mu}{\alpha_{12}-2\alpha}+\mathcal{O}(\eta^2) \quad
\mbox{and}\quad \rho_{(213)}=
\mu+\sqrt{\frac{2\mu\alpha \eta}
{\alpha_{12}}}+\mathcal{O}(\eta)\ ,
\ee
respectively. For $2\alpha\geq \alpha_{12}$, only the second solution
leads to physical solutions. 
In the  $\eta\to 0$ limit
the first solution in \eqref{rhoeta} has 
a smooth limit, with the center 3 sitting at the middle point
between the centers 1 and 2, separated by $r_{12}\sim (\alpha_{12}-2\alpha)/\mu$.
For the second solution, we have instead 
$r_{23}\simeq 
r_{13}= \sqrt{\frac{\alpha\alpha_{12}}{2\mu \eta}}+\mathcal{O}(\eta^0)$ with
$r_{12}\simeq \alpha_{12}/\mu$. 
Therefore the point 3 moves 
off to infinity along the $z$ axis, similarly to what happens at a wall
of marginal stability. However, since $\langle \alpha_1+\alpha_2, \alpha_3\rangle=0$,
the index does not change across the locus $\eta=0$. This is an example of 
a  wall of threshold stability \cite{deBoer:2008fk}. While a non-compact 
direction opens up in $\cM_3$ at $\eta=0$, the equivariant volume \eqref{eclass}
stays unchanged. In  \S\ref{sdipole} we shall see that the locus $\eta=0$ 
(or its analogue for more centers) is very convenient to analyze the phase space, 
even though some of the collinear fixed points sit at infinity. Such 'infrared divergent' configurations should however not be confused with the 'ultraviolet divergent'
scaling solutions, which cannot be removed by moving away from $\eta=0$.

\subsubsection{Example 2:
Unphysical scaling solutions} \label{sunphys}

In the example in \S\ref{d6d6d0} the
entropy associated with individual centers vanish as can
be easily seen using eq.\eqref{eex4} and \eqref{esc1}.
We now consider another example where each center
describes a regular black hole with finite entropy.
Again for simplicity we take $\kappa=6$ and $\cB=0$ and
choose
\be \label{scn1}
\alpha_1 = (0,7,48,0), \quad 
\alpha_2=(0,9,42,0),
\quad \alpha_3= (0,-8,-48,6)\, .
\ee
In this case we have
\be \label{escn2}
\alpha_{12}=138, \quad \alpha_{23}=96, \quad \alpha_{31}
= 48\, .
\ee
Thus the triangle inequality is satisfied and we can look for
scaling solution. Taking $t=B+iJ=3i$, we find that collinear
solutions to \eqref{denef1d} exist for  the
alignment $132$ (or its reverse),
with $s(p)=1$ and $\rho=62.68$. Naively,
to the contributions
from these collinear fixed points we should add 
 the contribution $-2$ from scaling solutions,
 so as to obtain an admissible $SU(2)$ character.
 Numerical analysis shows
however that the collinear solutions  fail to satisfy 
\eqref{eex4.5} in some region of the $z$-axis and hence
are not valid solutions. Thus the scaling solutions must
also be absent. This can be verified explicitly
by  examining the full  solution
space parametrized by \eqref{eintersolution}: in this case
the values of $\rho$ consistent with the triangular inequalities
range from $\rho=62.68$ (corresponding the above collinear solution)
to $\rho=\infty$ (corresponding to the scaling regime). 
We find that throughout this range,
the left hand side of \eqref{eex4.5} fails to be positive
somewhere in the three dimensional space. Thus, the regularity
condition \eqref{eex4.5} rules out the entire phase space $\cM_3$
in this case.

\section{Index from non-scaling multi-centered configurations} \label{snon}

In $\cN=2$ supersymmetric string theory, typically the
 index in the sector of charge $\gamma$ receives 
 contributions  both from single centered black holes carrying
charge $\gamma$, and from multi-centered black hole 
solutions, with individual centers
carrying charges $\alpha_1,\dots , \alpha_n$ such that
$\sum_i\alpha_i=\gamma$.
In this section, we shall consider contributions from those
multi-centered configurations which
do not allow for scaling solutions, \i.e.\ 
solutions where three or more centers can come
arbitrarily close. As explained in \S\ref{secphase}, 
this requires  that there is no subset $A$ of 
$\{1,\dots , n\}$ for which
we can find vectors $\vec r_i$ ($i\in A$)
satisfying \eqref{escond0}.
We work at some fixed point in the
moduli space, and denote by $Z_\alpha$ the central
charge associated with the charge $\alpha$. 
Using the same logic as in \cite{Manschot:2010qz}
we arrive at the following expression for the
(refined) index
$\tot_{\rm ref}(\gamma;y)\equiv 
\Tr'(-y)^{2 J_3}$ from multi-centered black hole
solutions carrying total charge 
$\gamma$ (assuming that $\gamma$ is 
a primitive vector of the charge lattice):
\be \label{enn1}
\sum_{\{\alpha_i\}\atop \sum_i\alpha_i=\gamma} 
{1\over {\rm Aut}(\{\alpha_i\})}
\, g_{\rm ref} (\alpha_1, \dots , \alpha_n;y) \, 
\bOm_{\rm ref}^{\one}(\alpha_1;y)
\cdots \bOm^{\one}_{\rm ref}(\alpha_n;y)\, .
\ee
Here, ${\rm Aut}(\{\alpha_i\})$ is the symmetry factor appropriate
 for Maxwell-Boltzmann 
 statistics (see footnote \ref{fo1}), while
  $g_{\rm ref} (\alpha_1, \dots , \alpha_n;y)$ given in
  \eqref{eclassq}  is the
contribution to $\Tr'(-y)^{2 J_3}$ from
$n$-centered black hole configuration with the
centers regarded as distinguishable particles with unit index.
$\bOm^{\one}_{\rm ref}(\alpha;y)$ is 
the `rational refined index',
related to the refined index by
\be \label{enn2}
\bOm^{\one}_{\rm ref}(\alpha, y) = \sum_{m|\alpha}
m^{-1} {y-y^{-1}\over y^m - y^{-m}} \Omega^{\one}_{\rm ref}
(\alpha/m,y^m)\, .
\ee
It is worthwhile noting that 
\eqref{enn1} and \eqref{enn2} exhibit a manifest `charge conservation property', 
whereby each power
of $\bOm^{\one}_{\rm ref}(\alpha, y)$ 
carries charge $\alpha$ and each power
of $\Omega_{\rm ref}(\alpha,y^m)$ carries charge $m\alpha$. 
The usual protected index is obtained by taking the
$y\to 1$ limit of \eqref{enn1}, at the cost of obscuring manifest 
`charge conservation'. If $\gamma$ is not primitive, then
\eqref{enn1} represents $\bar \tot_{\rm ref}(\gamma, y)$ 
defined in \eqref{edefsbar}
rather than $\tot_{\rm ref}(\gamma, y)$.
  
Using \eqref{enn1}, \eqref{enn2} and the relationship
between $\tot_{\rm ref}$ and $\bar \tot_{\rm ref}$ 
we can express $\tot_{\rm ref}(\gamma,y)$
as
\be \label{eform22new}
\tot_{\rm ref}(\gamma, y) =
\sum_{\{\beta_i\in \Gamma\}, \{m_i\in\bZ\}\atop
m_i\ge 1, \, \sum_i m_i\beta_i =\gamma}
G(\{\beta_i\}, \{m_i\};y) \, \prod_i 
\Omega^{\one}_{\rm ref}(\beta_i, y^{m_i})\, ,
\ee
for some function $G$. 
The $i$-th term in the sum represents a contribution from
configurations with a total of $\sum_i m_i$ centers, with
$m_i$ centers of charge $\beta_i$. If two  or
more of the $\beta_i$'s are identical,  the
total number of centers carrying a given charge $\beta_i$ is
the sum of the corresponding $m_i$'s.
Finally the sum also contains the contribution from a single
centered black hole of charge $\gamma$, represented by
the term $\Omega^{\one}_{\rm ref}(\gamma,y)$.
In order for  the right hand side of \eqref{eform22new}
to be a bona fide $SU(2)$ character whenever 
the $\Omega^{\one}_{\rm ref}(\beta_i, y)$'s are,  
the functions
$G$ which appear in \eqref{eform22new} must be Laurent polynomials
in $y$. The reader is referred to appendix \ref{sinteg} for a proof of
this property  in the absence of scaling solutions.

Eq.\eqref{enn1} (or equivalently \eqref{eform22new}) gives
the net contribution to the total index from all possible
single and multi-centered solutions carrying a fixed
charge $\gamma$. However it is also useful to
identify which terms in \eqref{enn1} represent the contribution
from a specific multi-centered solution. We shall assume
for simplicity that $\gamma$ is primitive so that
$\bar \tot_{\rm ref}(\gamma, y)=\tot_{\rm ref}(\gamma, y)$.
When all $\alpha_i$'s are different then the summand in 
\eqref{enn1} with $\bOm^{\one}_{\rm ref}$ replaced by 
$\Omega^{\one}_{\rm ref}$ 
represents the contribution to the index from multi-centered
black holes carrying charges $\alpha_1, \dots , \alpha_n$.
However when some of the $\alpha_i$'s are equal
(say $r_1$ copies of $\beta_1$, $r_2$ copies of
$\beta_2$ etc.), then
there is additional contribution to this index due to the fact
that the 
$\Omega^{\one}_{\rm ref}
(\alpha/m,y^m)$ term in \eqref{enn2} represents
the contribution from $m$ identical centers, each of charge
$\alpha/m$. Thus, 
the contribution to the index from a multi-centered
configuration with $r_1$ centers of charge $\beta_1$, $r_2$
centers of charge $\beta_2$ etc. is given by
\be \label{eomexp}
\sum_{\{n_k\ge 1\}, \{s^{(a)}_k \geq 1 \} \atop
\sum_{a=1}^{n_k} s^{(a)}_k = r_k}
\hat g_{\rm ref}(\{ s^{(a)}_k \beta_k\};y) \, 
\prod_k \bigg[\prod_{a=1}^{n_k} \bigg\{ {1\over s^{(a)}_k} {y - y^{-1}\over
y^{s^{(a)}_k} - y^{- s^{(a)}_k}} 
\Omega^{\one}_{\rm ref}(\beta_k,y^{s^{(a)}_k})
\bigg\}
\bigg] \, ,
\ee
where 
\be \label{edefghat}
\hat g_{\rm ref} (\alpha_1, \dots , \alpha_n;y)
= {1\over {\rm Aut}(\{\alpha_i\})}
\, g_{\rm ref} (\alpha_1, \dots , \alpha_n;y)\, .
\ee
After summing \eqref{eomexp} over all possible choices
of $\{\beta_k\}$, $\{r_k\}$ satisfying $\sum_k r_k \beta_k
=\gamma$, we recover \eqref{enn1}.
As explained in \S\ref{secphase}, 
the index $g_{\rm ref}(\alpha_1,\dots , \alpha_n;y)$ can be computed
by localization. Since we assume that the 
phase space contains no scaling solutions,
all fixed points are isolated and the quantum index is 
given by \eqref{eclassq}.

In the remainder of this section we shall carry out various consistency
tests of our proposal for the index associated with the
multi-centered black hole solutions.

\subsection{Consistency with wall crossing} \label{swall}

Let us examine whether our proposal is compatible with wall crossing.
For this purpose, assume that the moduli are chosen near a wall of
marginal stability where the state carrying total charge 
$\gamma\equiv \alpha_1+\cdots +\alpha_n$ becomes
marginally unstable against decay into states carrying
charges $\gamma_A=\sum_{i\in A} \alpha_i$ and 
$\gamma_B=\sum_{i\in B}\alpha_i$ where $A$ and $B$ are
two complementary subsets of $\{1,\dots, n\}$. At the wall, 
the phases of the central charges $Z_{\gamma_A}$ and 
$Z_{\gamma_B}$ align, and we have
\be \label{enn5.5}
\sum_{i\in A} c_i = - \sum_{j\in B} c_j \to 0\, .
\ee
We shall assume that the moduli are chosen on the side
of the wall where 
\be \label{enn5}
\langle \gamma_A,\gamma_B\rangle
\sum_{i\in A} c_i  > 0\, .
\ee
In this region of the moduli space, a class of 
solutions to \eqref{denef1d} can be constructed 
by joining two solutions involving the centers in
the set $A$ and those in the set $B$  as follows. 
Let  $n_A$ and $n_B$ be the cardinality
of the sets $A$ and $B$.
Now, choose the relative distances 
$z_{i}-z_{j}$ for $i,j\in A$
according to a particular  collinear solution
involving the charges $\alpha_{i\in A}$, and 
the relative distances 
$z_{i}-z_{j}$ for $i,j\in B$
according to the particular  collinear solution
involving the charges $\alpha_{i\in B}$.  
Finally, choose  the relative separation
between the $z_i$'s in the 
set $A$ and the $z_i$'s in the 
set $B$ such that
\be \label{enn6}
{
\gamma_{AB}
\over |z_A-z_{B}|} = \sum_{i\in A} c_i\, , \qquad
\gamma_{AB}\equiv \langle\gamma_A,\gamma_B\rangle\, ,
\ee
where $z_A=(\sum_{i\in A} z_i)/n_A$ and 
$z_B=(\sum_{i\in B} z_i)/n_B$
are the average positions of the centers in the 
sets $A$ and $B$ 
respectively. 
Near the wall of marginal stability $\sum_{i\in A} c_i
\to 0$, and the separation $|z_A-z_B|$ 
becomes large.
We claim that this configuration satisfies 
eq. \eqref{denef1d} in the limit \eqref{enn5.5}, and can be systematically
corrected to an exact solution of eq. \eqref{denef1d} in the vicinity of the wall.

To see this,  we note that if 
$i$ belong to the set $A$ then eq.\eqref{denef1d}
receives significant contribution only when 
$j$ also belongs to the set $A$ since the
distance $|z_A-z_B|$ computed from
\eqref{enn6} is large in the limit \eqref{enn5.5}. 
Thus the equations
reduce approximately to the equations for
collinear multi-centered solutions involving the set
$A$ only. By assumption our solution satisfies the
latter equations. A similar argument holds when
$i$ belongs to the set $B$. There is however
a small caveat stemming from the fact that the solutions
in the set $A$ are labelled by $n_A-1$ relative distances
while solutions in the set $B$ are labelled by $(n_B-1)=
(n-n_A-1)$
relative distances. This gives $(n-2)$ parameters, but
\eqref{denef1d} contains $(n-1)$ independent equations
(the sum of the equations over all $i$ being trivially
satisfied). Since we cannot adjust $(n-2)$ parameters to
solve $(n-1)$ independent equations we must have missed
some equation that determines the relative distance
between the points in the set $A$ and those in the
set $B$. To find the missing equation we sum \eqref{denef1d} over
all $i\in A$ to obtain
\be \label{enn7}
\sum_{i\in A} 
\sum_{j=1\atop j\ne i}^n 
\frac{\alpha_{ij}}{|z_{j} - z_{i}|} \, 
= \sum_{i\in A} c_{i}\,  .
\ee
Dividing the sum over $j$ on the left hand side into
those for which $j\in A$ and those for which
$j\in B$, and noting that the first term vanishes
by $i\leftrightarrow j$ symmetry, we get
\be \label{enn8}
\sum_{i\in A} 
\sum_{j\in B} \frac{\alpha_{ij}}
{|z_{j} - z_{i}|} \, 
= \sum_{i\in A} c_{i}\,  .
\ee
Approximating $z_j-z_i\simeq z_{B}-z_A$ for each term 
in the sum
we recognize \eqref{enn6}. It is then  clear that this
approximate solution can be extended to an exact
solution in the vicinity of the wall, by correcting 
the locations $z_i$ by a Taylor series in $1/|z_A-z_B|$.

Eq.\eqref{enn5} shows
that the solution to \eqref{enn6} exists as we
approach the wall of marginal stability but ceases to
exist as we cross the wall since the left hand side of
\eqref{enn5} and hence $|z_A-z_B|$ computed from
\eqref{enn6} now becomes negative.
Thus the jump in the index will be given by the contribution
 to \eqref{\coulombf} from this class of fixed points.
 To evaluate this jump, we first note that 
\be \label{enk1}
y^{\sum_{i<j} \alpha_{ij}\, \sign[z_j-z_i]}
= y^{\sum_{i,j\in A, i<j} \alpha_{ij}\, \sign[z_j-z_i]}\, 
y^{\sum_{i,j\in B, i<j} \alpha_{ij}\, \sign[z_j-z_i]}\, 
y^{\gamma_{AB}\, \sign[z_B-z_A]}
\, .
\ee
Similarly,
\be \label{efactorsign}
(-1)^{\sum_{i<j} \alpha_{ij} +n-1}
= (-1)^{n_A+n_B+\gamma_{AB}-1}\, 
(-1)^{\sum_{i,j\in A, i<j} \alpha_{ij}}\,
(-1)^{\sum_{i,j\in B, i<j} \alpha_{ij}}\,
\, .
\ee
Finally we need to compute $s(p)$ near the wall
of marginal stability. This has been analyzed in
detail in appendix \ref{sd}. The net result is that for the
configuration of the type we are considering, 
$s(p)$ is given by
\be \label{esigndiv}
s(p) = s_A\, s_B\, \sign (\gamma_{AB})\, \sign[z_B-z_A]\, .
\ee
where $s_A$ is the sign which appears in
the contribution to $n_A$ centered collinear
configurations with centers at $\{z_i, i\in A\}$,
and $s_B$ is the sign which appears in
the contribution to $n_B$ centered collinear
configurations with centers at $\{z_i, i\in B\}$.

Putting \eqref{enk1}, \eqref{efactorsign} and
\eqref{esigndiv}  into  \eqref{\coulombf}  and summing
over all $n_A$-centered solutions involving the charges
in the set $A$ and all $n_B$-centered solutions
involving charges in the set $B$ we see that the net
contribution from solutions of this type is given by
\be \label{enetwall}
-  \sign (\gamma_{AB}) \, \sign[z_B-z_A]\, (y-y^{-1})^{-1}\, 
(-y)^{\gamma_{AB}\, \sign[z_B-z_A]} \, 
g^{}_{\rm ref}(\{\alpha_i, i\in A\};y)\,
g^{}_{\rm ref}(\{\alpha_j, j\in B\};y)\, .
\ee
Adding up the contributions from the configurations for
which $z_B-z_A>0$ and those for which $z_B-z_A<0$,
we get
the net contribution to the
index from the solutions which decay across the
wall of marginal stability:
\be \label{efinal}
  - \sign (\gamma_{AB}) \, (y-y^{-1})^{-1}
\left[(-y)^{\gamma_{AB}} 
- (-y)^{-\gamma_{AB}}
\right]  \, g^{}_{\rm ref}(\{\alpha_i, i\in A\};y)\, 
 g^{}_{\rm ref}(\{\alpha_j, j\in B\};y)\, .
\ee
This is the correct wall crossing formula for
primitive decays. Besides the
solutions considered here, \eqref{denef1d} can have
other solutions for which the relative distances between
the centers remain finite as we approach the wall of
marginal stability. They will continue to exist on the other side
of the wall and must decay at other walls of marginal
stability before we reach the attractor point.

The analysis can be easily generalized to the case of
general non-primitive decays. 
We shall sketch the derivation below.
Suppose that we
are near a wall of marginal stability where the total charge
$\gamma$ can decay into $L$ states carrying charges
$\gamma_{m} = \sum_{i\in A_m}\alpha_i$ for
$m=1,\dots , L$. For this we need the charges 
$\gamma_{m}$
for different $m$ to lie in a two dimensional plane and
have $\sum_{i\in A_m} c_i\to 0$ for each $m$.
We approach the wall from the side where
\be \label{ejj1}
\langle \gamma_{m}, \gamma_{n} \rangle 
\left(\sum_{i\in A_m} c_i - \sum_{i\in A_n} c_i\right)
> 0\, ,
\ee
for any pair $(A_m, A_n)$. In this chamber there exists a
class of solutions in which  the elements of the set
$A_m$ are bunched together for each $m$ within a finite
distance and the relative separation between the elements
of the set $A_m$ and the elements of the set $A_n$
go to infinity for every pair $(m,n)$.
These are the solutions which disappear
across the wall of marginal stability; hence the
change in the index is given by the index associated
with these configurations. We shall order the sets $A_m$
such that $\langle \gamma_m,
\gamma_n\rangle>0$ for $m<n$.
Let us assume that for a given collinear solution $p$
of this type, 
the sets $A_m$ are arranged along
the $z$-axis as $A_{\sigma(1)}, \dots , A_{\sigma(L)}$
for some permutation $\sigma(1),\dots , \sigma(L)$
of $1,\dots , L$.
Then the contribution to the summand in
\eqref{\coulombf} from such a configuration will be given by
\be \label{ejj2}
s(p)\prod_{m=1}^L \left( y^{\sum_{i,j\in A_m}
\alpha_{ij}\, \sign[z_j-z_i]}\right) 
\prod_{m<n} \left( y^{\langle\gamma_{\sigma(m)},\gamma_{\sigma(n)} \rangle}
\right)\, .
\ee
The sum over collinear fixed points $p$ 
which respect the bunching of
the centers into the sets $A_1,\dots , A_L$ 
will involve independent sum
over collinear fixed points inside each set
$A_m$, generating the contribution 
$g_{\rm ref}(\{i\in A_m\};y)$
and the sum over permutations of the 
sets $A_m$,
generating a contribution 
$g_{\rm ref}(\{ \gamma_{m}\};y)$.
The final result, multiplied by the
$\prod_i \bOm^{\one}_{\rm ref}(\{\alpha_i\}, y)$, will be given by
\be \label{ejj3}
g_{\rm ref}(\{ \gamma_{m}\};y) \prod_m \left[
g_{\rm ref}(\{\alpha_i, i\in A_m\};y)\prod_{i\in A_m}
\bOm^{\one}_{\rm ref}(\{\alpha_i\}, y)\right]\, .
\ee
This is precisely what is needed to produce the wall
crossing formula given in \cite{Manschot:2010qz}.
The $g_{\rm ref}(\{\alpha_i, i\in A_m\};y)
\prod_{i\in A_m}
\bOm^{\one}_{\rm ref}(\{\alpha_i\}, y)$ factor
contribute to the index of the $m$-th `black hole
molecule' introduced in \cite{Manschot:2010qz}, 
while $g_{\rm ref}(\{ \gamma_{m}\};y) $ is the function
multiplying the indices of black hole molecules in the
wall crossing formula of \cite{Manschot:2010qz}.

\subsection{Consistency with split attractor flow conjecture} \label{ssplit}

The above analysis also shows that our proposal is
consistent with the split attractor flow conjecture
under certain assumptions.
To see this let us consider the system of black
holes carrying charges $\alpha_1,\dots , \alpha_n$ and
let the moduli flow all the way to the attractor point.
On the way we may cross several walls of marginal
stability. 
If at the attractor point there are no multi-centered
solutions, the result for the index at the
original point in the moduli space can be computed by adding
the contributions from the jumps across
different walls.
Suppose on the $s$-th wall the system is marginally
unstable against decay into 
a pair of states carrying charges $\sum_{i\in A_s} \alpha_i$
and $\sum_{i\in B_s}\alpha_i$. Then the index associated
with the states which decay across the wall is given by
\eqref{efinal} with $(A,B)$ replaced by $(A_s, B_s)$.
Thus the net contribution to the index at the original point
from this multi-centered configuration is given by
\ben \label{esplit1}
g^{}_{\rm ref}(\{\alpha_i\};y) 
&=&
-\sum_s \sign (\gamma_{A_sB_s}) 
{(-y)^{\gamma_{A_sB_s}} 
- (-y)^{-\gamma_{A_sB_s}}
\over y-y^{-1}
} \nonumber \\
&&\qquad  \, g^{}_{\rm ref}(\{\alpha_i, i\in A_s\};y)\, 
 g^{}_{\rm ref}(\{\alpha_j, j\in B_s\};y)\, .
\een
In the $y\to 1$ limit this gives
\be \label{esplit2}
g^{}(\{\alpha_i\})
=\sum_s (-1)^{\gamma_{A_sB_s}+1}
|\gamma_{A_sB_s}|
\, g^{}(\{\alpha_i, i\in A_s\})\, 
 g^{}(\{\alpha_j, j\in B_s\})\, .
\ee
We can now take the multi-centered configurations
carrying charges 
$\{\alpha_i, i\in A_s\}$, $\{\alpha_j, j\in B_s\}$ and calculate
their indices by flowing along their attractor flow
lines in the same way. Continuing this process till
we are left with only single centered black holes, we arrive
at the split attractor flow conjecture.

Note however that the above analysis relies on an assumption:
that the only jumps in $g^{}_{\rm ref}$ (and in $g^{}$)
take place at the walls of marginal stability. In particular
the solutions should not disappear away from
the walls of marginal stability 
(or if they do then they disappear in pairs
so that there is no net change in the index).

\section{Index from scaling multi-centered solutions}  \label{scaling}

The goal of this section 
is to determine the contribution of scaling solutions
to  the index of a multi-centered black hole
configuration using the Coulomb branch analysis. 
As shown in appendix \ref{sinteg},   in the
absence of scaling solutions 
the functions $G$ defined in \eqref{eform22new} are
automatically Laurent polynomials, as is required in 
order for the result to be a bona fide character
of $SU(2)$. This property however does not hold
in general  when scaling solutions are present, and
\eqref{eform22new} has to be corrected.
We shall denote by $G^{\rm \coll}_{\rm ref}$ the
contribution due to regular collinear fixed points only,
and determine the corrections to \eqref{eform22new}
by requiring that after adding these corrections
the final expression must be a proper 
$SU(2)$ character, whenever the single-centered
refined indices $\Omega^{\one}(\alpha, y)$
are $SU(2)$ characters.

Specifically, we propose
to modify \eqref{eform22new} into
\be \label{eform22mod}
\tot_{\rm ref}(\gamma, y) =
\sum_{\{\beta_i\in \Gamma\}, \{m_i\in\bZ\}\atop
m_i\ge 1, \, 
\sum_i m_i\beta_i =\gamma}
G_{\rm \coll}(\{\beta_i\}, \{m_i\};y) \, \prod_i 
\left( \Omega^{\one}_{\rm ref}(\beta_i, y^{m_i})
+ \Omega_{\rm scaling}(\beta_i, y^{m_i})\right)
\, ,
\ee
where $\Omega_{\rm scaling}(\alpha, y)$ is given by
\be \label{edefH}
\Omega_{\rm scaling}(\alpha, y) =
\sum_{\{\beta_i\in \Gamma\}, \{m_i\in\bZ\}\atop
m_i\ge 1, \, \sum_i m_i\beta_i =\alpha}
H(\{\beta_i\}, \{m_i\};y) \, \prod_i 
\Omega^{\one}_{\rm ref}(\beta_i, y^{m_i})
\, ,
\ee
for some function $H(\{\beta_i\}, \{m_i\};y)$
to be determined. To determine $H$ we
substitute 
\eqref{edefH} into \eqref{eform22mod}
to express the latter equation as
\be \label{eMod}
\tot_{\rm ref}(\gamma, y) =
\sum_{\{\beta_i\in \Gamma\}, \{m_i\in\bZ\}\atop
m_i\ge 1, \, \sum_i m_i\beta_i =\gamma}
G(\{\beta_i\}, \{m_i\};y) \, \prod_i 
\Omega^{\one}_{\rm ref}(\beta_i, y^{m_i})
\, ,
\ee
for some functions $G$. We fix $H$
by requiring that  $G(\{\beta_i\}, \{m_i\};y)$
are given by Laurent polynomials in $y$. 
This leaves open the possibility of adding
Laurent polynomials to $H$.
This ambiguity is resolved using the minimal
modification hypothesis, which says that $H$ must 
be symmetric under $y\to y^{-1}$ and vanish
as $y\to\infty$.

In practice we solve for the functions $H$ using an
iterative scheme involving the number of centers.
For this suppose we know  $H(\{\beta_i\}, \{m_i\};y)$
in all cases for $\sum_i m_i \le (n-1)$. 
Now we can
substitute \eqref{edefH} 
into eq.\eqref{eform22mod} and compute
the coefficient of 
$\prod_i 
\Omega^{\one}_{\rm ref}(\beta_i, y^{m_i})$ for all terms
with $\sum_i m_i \le n$.
The only unknown term is 
$\Omega_{\rm scaling} (\gamma;y)$, originating from
the replacement of $\Omega^{\one}(\gamma,y)$
by $\Omega^{\one}(\gamma,y)
+\Omega_{\rm scaling}(\gamma, y)$ in the right hand
side of \eqref{eform22mod}.
This gives
\be
\sum_{\{\beta_i\in \Gamma\}, \{m_i\in\bZ\}\atop
m_i\ge 1, \, \sum_i m_i\beta_i =\gamma}
H(\{\beta_i\}, \{m_i\};y) \, \prod_i 
\Omega^{\one}_{\rm ref}(\beta_i, y^{m_i})\, .
\ee
 Thus requiring
the coefficient of $\prod_i 
\Omega^{\one}_{\rm ref}(\beta_i, y^{m_i})$ to be a 
Laurent polynomial in $y$ we can determine
$H(\{\beta_i\}, \{m_i\};y)$ for $\sum_i m_i = n$. This
procedure can then be repeated to find
$H(\{\beta_i\}, \{m_i\};y)$ for $\sum_i m_i = n+1$ and
so on.
 
The algorithm given above gives a prescription for
finding the net contribution from the scaling solutions
for a fixed total charge $\gamma$.
In the rest of the section we shall see how this prescription
can be used to
determine the contribution of the scaling solutions
to a configuration containing a fixed
set of centers.

\subsection{Correction to
$g_{\rm ref}$ for non-identical centers} \label{snonid}

We shall now show that if the centers carry non-identical
charges then the minimal modification hypothesis translates
to a simple rule for correcting the function 
$g_{\rm ref}(\alpha_1,\dots ,\alpha_n;y)$.
First suppose that $\alpha_1,\dots , \alpha_n$ have been
chosen such that there is a scaling solution where all
the centers come together, but no scaling
solution where a subset of the centers come together.
In this case our proposal for the contribution of the 
scaling solution to $g_{\rm ref}(\alpha_1,\dots , \alpha_n;y)$
gives
\be \label{escalprop}
g_{\rm scaling}(\alpha_1,\dots , \alpha_n;y)
= (-1)^{\sum_{i<j} \alpha_{ij}+n-1}
(y-y^{-1})^{1-n} \,
\sum_{0\le k\le (n-2)\atop k -\sum_{i<j}\alpha_{ij}\in 2\IZ}
a_k \left\{y^k - (-1)^n y^{-k}\right\}\, ,
\ee
where $a_k$'s are constants to be adjusted so that, after
adding \eqref{escalprop} to the contribution \eqref{eclassq}
of the collinear fixed points, the result has a finite 
limit as $y\to 1$.
It is easy to see that the number of $a_k$'s is precisely
equal to the number of divergent terms in the $y\to 1$
limit, so that requiring finiteness as $y\to 1$ uniquely
fixes all the $a_k$'s. For example for even $n$ the 
possible values of $k$ range from $1$ to $(n-2)$, with 
$k$ taking either only even or only odd values
depending on
the parity of $\sum_{i<j}\alpha_{ij}$. The number of
$a_k$'s is then $(n-2)/2$. On the other hand using
$y\leftrightarrow y^{-1}$ symmetry we see that the 
divergent terms are of the form $(y-y^{-1})^{-2s}$
for $s=1,2,\dots , (n-2)/2$, giving precisely $(n-2)/2$
possible divergent terms. For $n$ odd the allowed
values of $k$ are in the range 0 to $(n-2)$ and again
$k$ takes either only even values or only odd values.
This gives $(n-1)/2$ possible $a_k$'s.  On the other
hand the possible divergent terms are of the form
$(y-y^{-1})^{-2s}$ for $1\le s \le (n-1)/2$, giving 
precisely $(n-1)/2$ possible divergent terms.
It is also worthwhile to note that the condition $k\leq n-2$
on the powers of $y$ appearing in the numerator of 
\eqref{escalprop} is equivalent to the requirement that 
the correction \eqref{escalprop} vanishes in the 
limit $y\to\infty$.

Now consider the more general case where there are scaling
solutions in which a subset of the centers come
together.
As mentioned earlier,
we shall proceed by  induction, \i.e.\ assume that
all the fixed points (including scaling solutions)
and their contributions have been determined
for any number of centers less or equal to $n-1$ 
and then show how this can be used to infer  
the result for $n$ centered black hole solution.
Let us consider an $n$-centered black hole
configuration with centers carrying charges $\alpha_1,
\dots , \alpha_n$. 
Now such configurations  will include a set of `regular'
fixed points where all the centers are separated
along the $z$-axis. We can determine them using 
numerical methods. To those we need to add the
contribution from fixed points where a (subset of) the
centers lie on top of each other. A generic fixed point
of this type will have the charges $\{\alpha_{{k^{(l)}_1}},
\alpha_{{k^{(l)}_2}}, 
\dots , \alpha_{{k^{(l)}_{n_l}}}\}$ lying on top of each other 
for $1\le l\le s$, $n_l\ge 1$
such that $\sum_l n_l = n$.
To determine its contribution
we proceed as follows. We first consider an $l$-centered
configuration with individual centers carrying charges
$\alpha_{{k^{(l)}_1}}+\cdots \alpha_{{k^{(l)}_{n_l}}}$ 
with $1\le l\le s$
and determine all its regular fixed points. 
In this computation the effective constant $c_l$ for the $l$-th
charge will be given by $2 \, {\rm Im}(e^{-i\alpha}
Z_{\alpha_{{k^{(l)}_1}}+\cdots \alpha_{{k^{(l)}_{n_l}}}})$.
Now we multiply
this contribution by the product of the weight factors of
the $s$ scaling solutions, with the $l$'th solution containing
centers $\{\alpha_{{k^{(l)}_1}},
\alpha_{{k^{(l)}_2}}, 
\dots , \alpha_{{k^{(l)}_{n_l}}}\}$. These are known 
by induction
except for the case $s=1$. We now add these
to the contribution from the regular fixed points of
the $n$-centered solution.
This procedure leaves out the $s=1$ term, corresponding to 
$l=1$, $n_1=n$: this is the maximally scaling configuration,
where all the centers are on top of each other.
Our proposal for the contribution of this term 
to $g_{\rm ref}(\alpha_1,\dots , \alpha_n;y)$ is 
again given by
\eqref{escalprop},
where $a_k$'s are constants to be adjusted so that after
adding \eqref{escalprop} to the other contributions (including the non-maximally
scaling ones), the result can be expressed as
a Laurent polynomial in $y$. 

The procedure just described  generalizes the one  introduced in \S\ref{spres}
in the absence of scaling solutions. It relies on the assumption that the
effect of the scaling solutions is to correct $g_{\rm ref}$ into a proper $SU(2)$ 
character, with the smallest possible powers of $y$ 
and $y^{-1}$ in the numerator
(reflecting the classical fact that scaling solutions carry zero angular momentum).
Eq.\eqref{escalprop} together with \eqref{\coulombf}
gives a complete prescription for computing the
functions $g_{\rm ref} (\{\alpha_i\};y)$. 
This construction guarantees that
$g_{\rm ref} (\{\alpha_i\};y)$ is given by a Laurent
polynomial in $y$.
As long as the charges $\alpha_i$ are all different,
the factor $g_{\rm ref}$ just described multiplied by
$\prod_{i=1}^n \Omega^S(\alpha_i; y)$
determine the contribution to the index from
multi-centered black hole solutions carrying charges
$\alpha_1,\dots , \alpha_n$. 
The resulting expression is given by a Laurent
polynomial in $y$.
As discussed in \S\ref{snon}, even when some of the $\alpha_i$'s
are equal but provided there are no scaling solutions, Eqs.
\eqref{eomexp}, \eqref{edefghat}  give the
correct result for the index of multi-centered black hole
solutions.
However when some of
the $\alpha_i$'s are equal and  scaling solutions are allowed, 
some additional
corrections to \eqref{eomexp} 
are necessary, which we shall now determine.

\subsection{Effect of identical particles} \label{sidentical}

We shall now consider the case where some of the centers
carry identical charges, {\it e.g.} we have
$r_1$ centers
of charge $\beta_1$, $r_2$ centers of charge $\beta_2$
etc. We shall assume for simplicity that the only allowed
scaling solutions involve all the centers coming together,
-- more general cases may be dealt with using the method
of induction as before. In this case our proposal for the
index associated with this configuration is the following
generalization of \eqref{eomexp}:
\be \label{enn1full}
\begin{split}
&\sum_{\substack{\{n_k\}, \{s^{(a)}_k\} \\
n_k,s^{(a)}_k\in\bZ; n_k,s^{(a)}_k\ge 1, \\
\sum_{a=1}^{n_k} 
s^{(a)}_k = r_k}}
\hat g_{\rm ref}(\{ s^{(a)}_k \beta_k\};y) \, 
\prod_l \bigg[\prod_{b=1}^{n_l} 
\bigg\{ {1\over s^{(b)}_l} {y - y^{-1}\over
y^{s^{(b)}_l} - y^{- s^{(b)}_l}} 
\Omega^{\one}_{\rm ref}(\beta_l,y^{s^{(b)}_l})
\bigg\}
\bigg] \\ &
\qquad \qquad \qquad \qquad
+ I_{\rm cor}(\{\beta_k\}, \{ r_k\}; y)\, ,
\end{split}
\ee
where the function $I_{\rm cor}$ vanishes if all the
$r_i$'s are 1, 
but can be non-zero if some
of the $r_i$'s are larger than 1. 
The function $\hat g_{\rm ref}$ is defined as in
\eqref{edefghat} with $g_{\rm ref}$ including 
the corrections due to scaling solutions described
in \S\ref{snonid}.
The need for the additional correction terms 
$I_{\rm cor}$ can be seen by
noting that due to the
presence of the $(y^{s^{(b)}_l} - y^{-s^{(b)}_l})$ 
factor in the denominator
it is not guaranteed that the first term in 
\eqref{enn1full} can be expressed as a Laurent
polynomial in $y$. $I_{\rm cor}$ is adjusted to
compensate for this. 
We choose
\be \label{efullp2}
 I_{\rm cor}(\{\beta_k\}, \{ r_k\}; y)= 
\sum_{\substack{\{n_k\}, \{s^{(a)}_k\} \\
n_k,s^{(a)}_k\in\bZ; n_k,s^{(a)}_k\ge 1, \\
\sum_{a=1}^{n_k} 
s^{(a)}_k = r_k}}
h(\{\beta_k\}; \{s^{(a)}_k\}; y)
\prod_l \prod_{b=1}^{n_l} 
\Omega^{\one}_{\rm ref}(\beta_l,y^{s^{(b)}_l})
\, ,
\ee
where $h(\{\beta_k\}; \{s^{(a)}_k\}; y)$
is chosen so that
\begin{enumerate}
\item It is invariant under $y\to y^{-1}$.
\item 
$ \lim_{y\to \infty}  h(\{\beta_k\}; \{s^{(a)}_k\}; y)=0$.
\item \eqref{enn1full} has an expansion of the form
$\sum_{m\in \bZ} a_m y^m$ 
with a finite number
of terms whenever the $\Omega^{\one}_{\rm ref}(\beta_i, y)$'s
have this property.
\end{enumerate}
Note that the second condition above is another
manifestation of the `minimal modification hypothesis'.
These three requirements fix the function $h$ completely
in any given situation. Thus this prescription gives a
complete algorithm for computing the spectrum of
multi-centered black holes given the collinear fixed
point solutions to \eqref{denef1d} satisfying the
requirement \eqref{eregu}.

The term in which all the $s^{(a)}_k$'s in the argument
of $h$ are 1
would combine with the term where each $s^{(a)}_k=1$
in the first term in \eqref{enn1full}. Since
the latter terms do not
have any unwanted denominators and are automatically
given by Laurent
polynomials in $y$, 
there
is no need for any correction terms.
Thus
$h(\beta_1,\beta_2,\cdots; \{s^{(a)}_k\}; y)$ vanishes if all the 
$s^{(a)}_k$'s are 1.
This shows that 
if all the
$r_i$'s are 1, \i.e.\ the $\alpha_i$'s are all different, then
$I_{\rm cor}$ vanishes.

\subsection{Illustration} \label{sillustrate}

Since the above discussion has been somewhat abstract
we shall now demonstrate this procedure by a
hypothetical example (which will map on to a
real example in \S\ref{sidcen}). Suppose we
have a four centered solution with charges
$\alpha_1$, $\alpha_2$, $\alpha_3=r_1\beta$,
$\alpha_4=r_2\beta$ such that $\langle\alpha_1,
\beta\rangle = -\langle\alpha_2,
\beta\rangle$. Suppose further that in some region of the
moduli space we find that the only collinear configurations
involve the order $1,2,3,4$ and its mirror 
$4,3,2,1$ with
$s(p)=\pm 1$. Then the
net contribution to $g_{\rm ref}(\alpha_1,\dots , \alpha_4;y)$
from the collinear fixed points is
\be \label{enet1}
(-1)^{\alpha_{12} +3} (y-y^{-1})^{-3} (y^{\alpha_{12}}
- y^{-\alpha_{12}})\, .
\ee 
Let us suppose further that there are no 
solutions where
a proper 
subset of the centers are in the scaling configurations,
--  the only additional contribution comes from the
scaling solution where all the centers approach each
other. 
We can determine this term
according to the prescription \eqref{escalprop}. This gives
the net contribution to $g_{\rm ref}(\alpha_1,\dots , 
\alpha_4;y)$ to be
\be \label{enet2}
\begin{split}
g_{\rm ref}(\alpha_1,\dots , 
\alpha_4;y)=& (-1)^{\alpha_{12} +3} ( y-y^{-1})^{-3} 
\left\{y^{\alpha_{12}}
- y^{-\alpha_{12}} - {1\over 2}\,
\alpha_{12}(y^2 - y^{-2})
\right\}  \quad \hbox{for} \quad \alpha_{12}\in 2\bZ,
\\
=& (-1)^{\alpha_{12} +3} ( y-y^{-1})^{-3} 
\left\{y^{\alpha_{12}}
- y^{-\alpha_{12}} -
\alpha_{12}(y - y^{-1})
\right\}  \quad \hbox{for} \quad \alpha_{12}\in 2\bZ+1\, .
\end{split}
\ee
After taking out a factor of $y^{-\alpha_{12}}$ we see
that the term inside $\{~\}$ is a polynomial in $y^2$
and has triple zero at $y=1$. Thus it must have
a factor of $(1-y^2)^3$ cancelling the $(1-y^2)^3$ factor
in the denominator, and the expressions for
$g_{\rm ref}(\alpha_1,\dots , 
\alpha_4;y)$ given in \eqref{enet2} can be
expressed as a Laurent
polynomial in $y$. 

As long as all $\alpha_i$'s are different this ends the 
discussion
for the contribution from this four centered terms. Suppose
however $r_1=r_2=r$ so that $\alpha_3=\alpha_4$. In this
case the net contribution to the index from this
four centered configuration will be given by
\be \label{enet3}
\begin{split}
& {1\over 2}
g_{\rm ref} (\alpha_1, \alpha_2, r \beta, r\beta; y)\, 
\Omega^{\one}_{\rm ref}(\alpha_1;y) \, \Omega^{\one}_{\rm ref}(\alpha_2,y)\,
\Omega^{\one}_{\rm ref}(r\beta, y)^2 \\
& + {1\over 2}\, 
g_{\rm ref}(\alpha_1, \alpha_2, 2 r \beta;y)
{y-y^{-1}\over y^2 - y^{-2}}\, 
\Omega^{\one}_{\rm ref}(\alpha_1,y) \, \Omega^{\one}_{\rm ref}(\alpha_2,y)\, 
\Omega^{\one}_{\rm ref}(r\beta, y^2)\\
&
+ h(\alpha_1, \alpha_2, r\beta;  s^{(1)}_1=1,
s^{(1)}_2=1,s^{(1)}_3=2; y)\,
\Omega^{\one}_{\rm ref}(\alpha_1,y) \,
\Omega^{\one}_{\rm ref}(\alpha_2,y)\,
\Omega^{\one}_{\rm ref}(r\beta, y^2)\, .
\end{split}
\ee
The second term comes
from the term
$g_{\rm ref}(\alpha_1, \alpha_2, 2 r \beta;y)
\Omega^{\one}_{\rm ref}(\alpha_1,y) \Omega^{\one}_{\rm ref}(\alpha_2,y)
\bOm^{\one}_{\rm ref}(2r \beta, y)$ and the third term 
is the correction
term given in \eqref{enn1full}, \eqref{efullp2}.
To proceed we need to know 
$g(\alpha_1, \alpha_2, 2 r \beta;y)$. Suppose at the
same point in the moduli space the collinear
three centered configurations carrying charges
$\alpha_1$, $\alpha_2$, $\alpha_3=2r\beta$ are of the
form $123$ and $321$. In this case we have
\be \label{enet4}
\begin{split}
g_{\rm ref}(\alpha_1, \alpha_2, 2 r \beta;y)
= & (-1)^{\alpha_{12} +2} ( y-y^{-1})^{-2} 
\left\{y^{\alpha_{12}}
+ y^{-\alpha_{12}} - 
2
\right\}  \quad \hbox{for} \quad \alpha_{12}\in 2\bZ,
\\
= & (-1)^{\alpha_{12} +2} ( y-y^{-1})^{-2} 
\left\{y^{\alpha_{12}}
+ y^{-\alpha_{12}} -
(y + y^{-1})
\right\}  \quad \hbox{for} \quad \alpha_{12}\in 2\bZ+1\, .
\end{split}
\ee
Note that we have added the correction terms due
to the scaling solutions according to \eqref{escalprop}.
We can now substitute \eqref{enet2} and \eqref{enet4}
into \eqref{enet3} and determine $h$ by requiring
that the resulting expression is given by a Laurent
polynomial in $y$ for any choice of 
$\Omega^{\one}_{\rm ref}$ satisfying similar properties.
Now the first term clearly has this property. In
the second term $g_{\rm ref}(\alpha_1, \alpha_2, 
2 r \beta;y)$ has this property, but the factor
of $(y-y^{-1})/(y^2 - y^{-2}) = y/(1+y^2)$ has 
a factor of $(1+y^2)$ in the denominator which
could potentially spoil this property unless
the numerator has a factor of $(1+y^2)$. In this case
we need a non-vanishing $h$ to cancel the
unwanted terms. To proceed we note that if
$\alpha_{12}$ is odd ($\alpha_{12}
=(2k+1)$ with $k\in\bZ$)
then we have
\be \label{egref1}
{1\over 2}\,{y\over 1+y^2}\,
g_{\rm ref}(\alpha_1, \alpha_2, 2 r \beta;y)
= -{1\over 2}\, y^{-2k+2} (1+y^2)^{-1}
( 1-y^2)^{-2} 
(1-y^{2k}) (1 - y^{2k+2})\, .
\ee
Each of the two factors $(1-y^{2k})$ and $(1 - y^{2k+2})$
has a factor of $(1-y^2)$ canceling the $(1-y^2)^{-2}$ factor.
Furthermore $(1-y^{2k})$ for even $k$ and 
$(1 - y^{2k+2})$ for
odd $k$ also has a factor of $(1+y^2)$ that 
cancels the $(1+y^2)^{-1}$. 
Thus in this case \eqref{enet3}
gives  a Laurent
polynomial in $y$
without any $h$ term and we can set $h=0$. On the
other hand for even $\alpha_{12}$ 
($\alpha_{12}
=2k$ with $k\in\bZ$) we have
\be \label{egref2}
{1\over 2}\,
{y\over 1+y^2}\,g_{\rm ref}(\alpha_1, \alpha_2, 2 r \beta;y)
=  {1\over 2}\,
 y^{-2k+3} (1+y^2)^{-1}
( 1-y^2)^{-2} 
(1-y^{2k})^2 \, .
\ee
$(1-y^{2k})^2$ 
has two factors of $(1-y^2)$ cancelling the $(1-y^2)^{-2}$
factor. Furthermore for even $k$ it
also has two factors of $(1+y^2)$ killing the factor
of $(1+y^2)^{-1}$. Thus again in this case there is no
need for any correction term and we can set $h$
to 0. Finally for $k$ odd we can express the right hand
side of \eqref{egref2} as
\be \label{egref3}
{1\over 2}\,
y^{-2k+3} (1+y^2)^{-1}
\{1 + y^2 +\cdots y^{2k-2}
\}^2\, .
\ee
Now as $y^2\to -1$ this term approaches
${1\over 2}\,
y (1+y^2)^{-1}$. Thus
from \eqref{enet3} we see that the unwanted terms
may be cancelled by choosing
$h$ to be negative of this term. This gives
\be \label{egref4}
\begin{split}
h(\alpha_1, \alpha_2, r\beta;  s^{(1)}_1=1,
s^{(1)}_2=1,s^{(1)}_3=2; y)
& = 0 \quad \hbox{for} \quad \alpha_{12}\in 2\bZ+1\\
& = 0 \quad \hbox{for} \quad \alpha_{12}\in 4\bZ\\
& =- {1\over 2}\,
 (y+y^{-1})^{-1}
\quad \hbox{for} \quad \alpha_{12}\in 4\bZ+2\, .
\end{split}
\ee
In \S\ref{sidcen} we shall realize this example in the case of dipole halo
configurations. 

\subsection{Wall crossing re-examined} \label{srevisit}

Given the modifications due to the scaling
solutions, we need to re-examine the analysis in
\S\ref{swall} on the compatibility of our
prescription with the wall crossing formula. 
Rather than doing a detailed analysis, we note 
that our prescription for
including contributions from scaling
solutions affects the factors $g_{\rm ref}(\{\alpha_i,
i\in A_m\};y) \prod_{i\in A_m} \bOm^{\one}_{\rm ref}(\alpha_i, y)$
in the square bracket of \eqref{ejj3}, as each set $A_m$ may
allow for scaling solutions. Thus, the index associated 
with individual black hole molecules will change.
However, the factor $g_{\rm ref}(\{ \gamma_{m}\};y)$ in front, which 
determines the jump of the index under wall-crossing,
will not be modified since the elements belonging to different 
sets $A_m$ always remain separated. Thus,
the contributions of scaling solutions do not affect the
consistency with the  wall-crossing formula.

\section{Dipole halo configurations\label{sdipole}}

In this section, we verify our prescriptions for a class of multi-centered configurations
whose phase space and quantization
is completely understood, namely dipole 
halo configurations\cite{deBoer:2008zn,deBoer:2009un}. 
These consists
of $n$-centered configurations with two distinguished centers carrying charges 
$\alpha_1$, $\alpha_2$, with $\alpha_{12}\neq 0$, and $n-2$ centers carrying 
mutually local charges $\alpha_a$, 
such that $\alpha_{1a}=-\alpha_{2a}, \alpha_{ab}=0$. Such configurations
were analyzed in detail in \cite{deBoer:2008zn,deBoer:2009un}, whose notations
we follow. A particular realization of this system, which we shall use to frame our discussion,
is given by a $D6-\overline{D6}$ pair with $n-2$ $D0$-branes orbiting
around it. After reviewing the main features of such configurations in \S\ref{sgen}, 
we describe the 3-centered case in detail in \S\ref{s341} (an example of which
was already analyzed in \S\ref{d6d6d0}), and present a general proof of the
minimal modification hypothesis in \S\ref{srecur}, based on recursion relations
for the equivariant index. Similar recursion relations for the equivariant volume
are presented in \S\ref{srecur2}. Explicit results for 4 and 5 centers can be found
in Appendix \ref{sc}.
In \S\ref{sidcen} we analyze four centered configurations with
two identical centers.

\subsection{Generalities \label{sgen}}

We shall consider a system of $n$-centers,
the first one of which represents a $D6$ brane with
certain U(1) flux, the second one describes a 
$\overline{D6}$-brane 
with opposite $U(1)$ flux and the third one onwards 
corresponds to $D0$-branes carrying
$q_3, \dots, q_n$ units of $D0$-brane charges.
Thus the first and the second centers carry opposite
$D6$ and $D2$-brane charges but the same $D4$ and $D0$-brane
charges. In the language used in \S\ref{d6d6d0} we have:
\be \label{esc1rep}
\begin{split}
&  \tilde\alpha_1 = (1, -U, -\kappa\, 
U^2/2-\cB/24, -\kappa\, U^3/6-\cB U/24), \\
&  \tilde\alpha_2 = (-1, -U,  \kappa\, 
U^2/2 + \cB/24, -\kappa\, U^3/6-\cB U/24), \\
&  \tilde\alpha_a = 
(0,0, 0, q_a), \qquad  \, 
\end{split}
\ee
leading to
\be \label{ech1}
\alpha_{12} = I\equiv
4\kappa U^3/3+\cB U/6, \qquad \alpha_{1a}= - q_a, \qquad
\alpha_{2a}=q_a, \qquad \alpha_{ab}=0, \qquad
3\le a,b\le n\, ,
\ee
where we assume that $I,q_a$ are positive integers. The system described in
\eqref{esc1}, \eqref{esc2} is a special case of this with $n=3$.
We work in the chamber of the moduli space where
\be \label{ecaspec}
c_1=\mu, \qquad c_2=-\mu-\eta\sum_{a=3, \dots , n}q_a, \qquad
c_a=\eta \, q_a \quad \hbox{for $a=3, \dots , n$,}
\ee
where $\mu$ is a positive constant and $\eta$ is a small
positive or negative number.
As described at the end of \S\ref{d6d6d0},
the $\eta=0$ subspace
describes a threshold stability wall on which
a subset of the $D0$-branes can get infinitely separated
from the rest of the system. But the index does not jump across
this wall, since the symplectic product 
of the charge vector of the expelled $D0$-brane charges with the
total charge vector of the $D6-\overline{D6}-D0$ system vanishes.

On the Coulomb branch, the system is described by multi-centered configurations
satisfying  the equilibrium conditions
\be
\label{denefdipole}
\frac{1}{r_{1a}}- \frac{1}{r_{2a}} = \eta\ ,\qquad 
-\frac{I}{r_{12}} + \sum_{a=3, \dots , n} \frac{q_a}{r_{1a}} = -\mu\ . 
\ee
In particular, at $\eta=0$ the $D0$-branes lie either at
infinity or on the plane equidistant to the $D6$
and $\overline{D6}$ branes. As shown in \cite{deBoer:2008zn,deBoer:2009un}, for this type of dipole halo configuration the 
phase space of solutions 
to \eqref{denefdipole} at $\eta=0$ is toric and given by a $T^{n-1}$ bundle over the polytope
\be
\label{defpoly}
\cP(I,\{q_a\}) = \{ (m_a, m) \ :\ 0\leq m_a\leq q_a\ , -j \leq m \leq j\ ,j >  0\}\quad 
\subset \IR^{n-1}\ ,
\ee
where  
\be
\label{defjdip}
j\equiv I/2-\sum_{a=3, \dots , n} m_a
\ee 
is a linear function of the $n-2$ variables $m_a$. These variables parametrize 
the angle $\theta_a$ between $\vec r_{1a}$ and $\vec r_{12}$
via $m_a=q_a \cos\theta_a$, while $m$ parametrizes
the angle $\theta$ between $\vec r_{12}$ and the 
$z$-axis via $m=j\cos\theta$. Physically $m$
represents 
the component of the 
angular momentum along $z$-axis. The coordinates $(m,m_a)$
together with coordinates $(\phi,\phi_a)\in [0,2\pi]^{n-1}$ along the torus fiber 
provide a set of Darboux coordinates on $\cM_n$,  
\be
\label{omdip}
\omega = -\de m\wedge \de\phi 
- \sum_{a=3, \dots , n} \de m_a \wedge \de \phi_a \ .
\ee
Denoting $\phi_a=\tilde\phi_a-\sigma$ where $\sum_a \tilde\phi_a=0$ and using
\eqref{defjdip}, it 
is straightforward to check that \eqref{omdip} agrees with \eqref{omj} with
$\tilde\omega=\sum_a q_a \sin\theta_a \, \de\theta_a\wedge
\de\tilde\phi_a$.  
The equivariant volume \eqref{ephaseint} can be rewritten as
\be \label{echclas}
\begin{split}
g_{\rm classical} (\{\alpha_i\};y)  
&=  (-1)^{I-n+1}\,
 \int_{\substack{ 0\le m_a\le q_a \\  \sum_a m_a \le I/2 }}
\de m_3 \cdots \de m_n\, \int_{-\frac{I}2 + \sum_a m_a}
^{\frac{I}2 - \sum_a m_a} \de m\, 
e^{2\nu m}\, ,
 \\
&=  (-1)^{I-n+1}\,
 \int_{\substack{ 0\le m_a\le q_a \\  \sum_a m_a \le I/2 }}
\de m_3 \cdots \de m_n\, 
\frac{\sinh[(I-2 \sum_{a=3}^n m_a)\nu]}{\nu}\, ,
\quad \nu\equiv \ln y\, .
\end{split}
\ee
For brevity, we shall denote the r.h.s. of \eqref{echclas} 
also by $S(I;\{q_a\}_{a=3, \dots , n};\nu)$.
Moreover, we let $S(I;\{q_a\}_{a=3, \dots , n};\nu)=0$ whenever $I<0$. 

For $\sum_a q_a < I/2$, the upper bound in $\sum m_a\leq I/2$ is never attained,
and the phase space is compact. The equivariant volume can therefore be evaluated
rigorously using localization with respect to $J_3$. For $\eta>0$, the fixed points of $J_3$
are collinear solutions  to \eqref{denefdipole}, where  a (possibly empty)
subset $A\subset \{3, \dots , n\}$  
of the $D0$-branes lie on the segment between the $D6$ and $\overline{D6}$ along 
the $z$ axis, while the $D0$-branes in the complement
 $B=\{3, \dots , n\}\backslash A$  lie 
on the semi-infinite $z$-axis extending from the $D6$-brane to infinity.
In the limit $\eta\to 0$, the centers in $A$ coalesce at the 
mid-point between the $D6$ branes, while the centers in $B$ run off to infinity.
The distance between the $D6$ and $\overline{D6}$ is given 
by $r_{12}=(I-2\sum_{a\in A} q_a)/\mu$, which is positive for any subset $A$.
The angular momentum carried by this configuration is 
$J_3=\left(I - 2\sum_{a\in A} q_a\right)/2$.
In appendix \ref{secsigndip} we show that  
 the sign $s(p)$ evaluates to $(-1)^{n_A}$, 
 where $n_A$ is the cardinality of 
the subset $A$ (the order among the centers inside the clusters $A$ or $B$ is irrelevant
since they carry mutually local charges). Thus, the classical phase
space integral \eqref{echclas} evaluates to   
\be \label{enet1aa}
S(I;\{q_a\}_{a=3, \dots , n};\nu)=  (2\ln y)^{-n+1} \sum_{A} (-1)^{n_A+I-n+1}
\left( y^{I - 2\sum_{a\in A} q_a} + (-1)^{n-1}
y^{-I + 2\sum_{a\in A} q_a}
\right) \, .
\ee
This can of course also be obtained by direct evaluation
of \eqref{echclas}.
Since the fixed points are isolated, 
according to the discussion in \S\ref{secqindex},
the exact refined index is obtained
by replacing  $2\ln y$ by $y-y^{-1}$, leading to
\be \label{enet1ab}
g_{\rm ref} (\{\alpha_i\};y) = (-1)^{I+n-1} (y-y^{-1})^{-n+1} \sum_{A} (-1)^{n_A}
\left( y^{I - 2\sum_{a\in A} q_a} + (-1)^{n-1}
y^{-I + 2\sum_{a\in A} q_a}
\right) \, .
\ee

In contrast, when $I/2\leq \sum_a q_a$, the phase space 
$\cM_n$ is non-compact, 
as it 
has a scaling region represented by the
boundary $\sum_a m_a = I/2$ on which $j$ vanishes, and
all 
centers approach each other at
arbitrarily small distances.\footnote{In this example, 
there are no scaling configurations
where only a subset of the centers scale together, since the total charge
carried by such a subset would be mutually local with respect to the remaining,
non-scaling D0-brane centers.} 
Such configurations 
are therefore invariant under $SO(3)$.
To compactify $\cM_n$, we  include the boundary $j=0$ \i.e.\
supplement the open polytope $\cP$ with the lower-dimensional polytope
\be
\label{defpolyh}
\cQ
(I,\{q_a\}) = \{ (m_a, m) \ :\ 
0\leq m_a\leq q_a\ , \quad 
\sum_{a=3}^n m_a=I/2\ , \quad m=0\}\ .
\ee
Denoting by $\cM_n^{\rm scal}$ the $(2n-4)$-dimensional compact toric manifold 
built over the polytope $\cQ$, with torus fiber parametrized by $\tilde\phi_a$,  we
define the compactification of $\cM_n$ as
\be
\label{hatM}
\hat\cM_n = \cM_n \cup \cM_n^{\rm scal}\ ,
\ee
and require that $SO(3)$ acts trivially on the boundary $\cM_n^{\rm scal}$.

The equivariant volume of $\hat\cM_n$ can in principle be evaluated by localization.
One class of fixed points is given by the same type of collinear configurations as 
described above (\ref{enet1aa}), with the proviso that the subset $A$ must be chosen 
such that $r_{12}>0$, i.e. 
\be \label{eres1}
\sum_{a\in A} q_a < {I\over 2}\, .
\ee
The contribution of these isolated fixed points takes the same form as in \eqref{enet1aa},
with the restriction \eqref{eres1} enforced. We denote this contribution by $S_{\rm \coll}
(I;\{q_a\}_{a=3, \dots , n};\nu)$. In addition to these isolated fixed points,
already present in $\cM_n$, one also expects a contribution from the submanifold 
$\cM_n^{\rm scal}$ inside $\hat\cM_n$. Although this contribution can in principle
be computed using a generalization of the Duistermaat-Heckman formula \cite{MR805808}, 
we shall not attempt to compute it directly, and instead define
\be
\label{corS}
S(I;\{q_a\}_{a=3, \dots , n};\nu)=S_{\rm \coll} (I;\{q_a\}_{a=3, \dots , n};\nu) 
+ \Delta S(I;\{q_a\}_{a=3, \dots , n};\nu)\ ,
\ee
where $\Delta S(I;\{q_a\}_{a=3, \dots , n};\nu)$ is the contribution of $\cM_n^{\rm scal}$.
By a direct computation of the equivariant volume using recursion relations in \S\ref{srecur2}, 
we shall be able to read off $\Delta S(I;\{q_a\}_{a=3, \dots , n};\nu)$. 
Similarly, the exact refined index $g_{\rm ref} (\{\alpha_i\};y)$, which we also denote by 
$\hat S(I; \{q_a\};\nu)$, 
decomposes as 
\be
\label{corhatS}
\hat S(I;\{q_a\}_{a=3, \dots , n};\nu)=\hat S_{\rm \coll}(I;\{q_a\}_{a=3, \dots , n};\nu) 
+ \Delta \hat S(I;\{q_a\}_{a=3, \dots , n};\nu)\ ,
\ee
where $S_{\rm \coll}(I;\{q_a\}_{a=3, \dots , n};\nu)$ denotes the r.h.s. of \eqref{enet1ab}, with
restriction \eqref{eres1} enforced on the subset $A$, and 
$\Delta \hat S(I;\{q_a\}_{a=3, \dots , n};\nu)$ denotes the contribution of the fixed 
submanifold $\cM_n^{\rm scal}$. As explained in \S\ref{scaling},  
the minimal modification hypothesis determines
the correction $\Delta \hat S$ 
uniquely by requiring that $\hat S$ is a $SU(2)$ character and 
$\Delta \hat S\to 0$ as $\nu\to\infty$.

On the other hand,  since
the phase space of $n$-centered solutions is a toric \kahler manifold, it can also be 
quantized exactly. We refer to \cite{deBoer:2008zn,deBoer:2009un} for the details
of this procedure in the dipole halo case, and merely quote the result. When all the
$q_a$'s are distinct, so that the $D0$-branes are distinguishable,
the exact index is given by
\be 
\label{ech2}
g_{\rm ref}(\{\alpha_i\};y)
= (-1)^{I-n+1}\, 
 \sum_{\substack{ m_3, \dots , m_n\\
m_a\in \bZ, 0\le m_a\le q_a-1, \\  \sum_a m_a \le \lfloor (I-n+1)/2\rfloor}}
\sum_{m=-\frac12(I-n+1-2\sum_a m_a)}^{\frac12(I-n+1-2\sum_a m_a)}
y^{2m}\ ,
\ee
where $\lfloor x \rfloor$ denotes the largest integer smaller or equal to $x$,
and the sum over $m\in \frac12 \IZ$ 
is such that\footnote{This condition was not stated
explicitly in \cite{deBoer:2008zn,deBoer:2009un} but is necessary for the consistency
of the $SU(2)$ action.}
 $m - \frac12(I-n+1-2\sum_a m_a)\in \IZ$ .
As already mentioned above \eqref{corhatS}, we denote by 
$\hat S(I; \{q_a\};\nu)$ the r.h.s. 
of \eqref{ech2}, and let $\hat S(I; \{q_a\};\nu)=0$ whenever 
$I< n-1$.
Performing the geometric sum over $m$, we arrive at 
a sum of $SU(2)$ characters
\be
\label{defhatS}
\hat S(I; \{q_a\};\nu)  = (-1)^{I-n+1}\,  \sum_{\substack{m_3, \dots , m_n\\
m_a\in \bZ, 0\le m_a\le q_a-1, \\
 \sum_a m_a \le \lfloor(I-n+1)/2\rfloor}} 
 \frac{\sinh\left[ ( I-n+2 -2\sum_{a=3}^n 
 m_a)\nu\right]}{\sinh \nu}\ .
\ee
If some of the $q_a$'s coincide, and if all $\Omega(\alpha_i)$'s are set to one,
the index is still given by  \eqref{ech2}, \eqref{defhatS}, with the additional restriction
that the $m_a$'s corresponding to identical particles
must be distinct, and the expression must be divided
by a symmetry factor $k!$ for every set of $k$ identical
$q_a$'s. 

In \S\ref{srecur}, we shall give an inductive proof  of the minimal modification prescription described below
\eqref{corhatS}
in this class of dipole halo configurations, by establishing recursion relations for the 
exact index \eqref{defhatS} and comparing them with the recursion relations obeyed
by its regular part $\hat S_{\rm \coll}(I; \{q_a\};\nu)$. Using similar methods, in \S\ref{srecur2}
we shall also demonstrate a variant of this minimal modification
prescription, which allows to recover the exact equivariant index
$\hat S( I; \{q_a\};\nu)$ from the knowledge of the 
equivariant volume  $S(I; \{q_a\};\nu)$. This prescription goes as follows:
\begin{enumerate}
\item In the expression for $S(I; \{q_a\};\nu)$, replace
all $\nu$'s which do not appear as 
arguments of hyperbolic functions, by $\sinh\nu$.
Let us denote the result by $\tilde S(I; \{q_a\};\nu)$.
\item If $\tilde S(I; \{q_a\};\nu)$ 
can be expressed as Laurent
polynomial in $y=e^\nu$, then this is the exact
result for $\hat S(I; \{q_a\};\nu)$. Otherwise we add terms
which vanish as $\nu\to \pm\infty$ to make the expression
into a Laurent
polynomial in $y$. This
gives the exact result for $\hat S(I; \{q_a\};\nu)$.
\end{enumerate}

Before going to the general proof of these statements, 
we shall illustrate
them in the case of  $n=3$ centers in \S\ref{s341}. 
Explicit computations
for $n=4$ and $5$ centers can be found in 
Appendix \ref{sc}.
In \S\ref{sidcen} we shall study four centered configurations
with two identical centers and use the minimal modification
prescription of \S\ref{sidentical} to compute the index.
This is then compared with the known exact results.

\subsection{Three-centered solutions}
\label{s341}

For $n=3$ we have to consider two cases separately.

\subsubsection{Non-scaling case: $q_3< {I\over 2}$}

For $q_3<{I\over 2}$, the sum over collinear configurations \eqref{enet1aa} reduces to
\be \label{esth1A}
\hat S_{\rm \coll} (I;q_3;y) =  {(-1)^{I} \over (y-y^{-1})^2} \, \bigg[
y^I - y^{I-2q_3} 
+ y^{-I} - y^{-I+2q_3} \bigg]\, .
\ee
This can be expressed as a Laurent
polynomial in $y$, and indeed agrees with the 
exact refined index computed from \eqref{ech2}. 

Alternatively we can begin with 
the equivariant volume \eqref{echclas}.
This gives
\be \label{essone}
S(I;q_3;y) = (-1)^{I}  \int_0^{q_3} \de m_3 \, 
\frac{\sinh[(I-2 m_3)\nu]}{\nu} = (-1)^{I}  \frac{\cosh(I\nu)-\cosh[ (I-2q_3)\nu]}{2\nu^2}\, .
\ee
After
the replacement $\nu\to\sinh\nu$ in the denominator,
we get
\be
\tilde S (I;q_3;y) = (-1)^{I}  
\frac{\cosh(I\nu)-\cosh[ (I-2q_3)\nu]}{2\sinh^2\nu}\, .
\ee
It is easy to see that this agrees with \eqref{esth1A}
and hence can be expressed as a Laurent
polynomial in $y$. Thus in this case there is no need to add
any correction terms and $\tilde S(I;q_3;y)$ is the same
as $\hat S(I;q_3;y)$. In particular, the symplectic volume 
$S(I;q_3;y=1)$ agrees with the exact index $\hat S(I;q_3;y=1)$
in the non-scaling regime.

\subsubsection{Scaling case: $q_3\geq  {I\over 2}$}
In this case \eqref{enet1aa}, with the restriction 
\eqref{eres1} on the set $A$,
reduces to
\be \label{esT1}
\hat S_{\rm \coll} (I;q_3;y) =  {(-1)^{I} \over (y-y^{-1})^2} \, \bigg[
y^I  + y^{-I}  \bigg]\, .
\ee
This diverges as $y\to 1$ and hence we must add corrections
described in \eqref{escalprop}. This indeed reproduces the 
exact index computed from \eqref{ech2}, 
\be \label{esth2B}
\begin{split}
\hat S(I;q_3;y) &= {(-1)^{I} \over (y-y^{-1})^2} \, \bigg[
y^I  + y^{-I} - (y+y^{-1}) \bigg]\, ,
\qquad \hbox{for $I$ odd}\\
&= { (-1)^{I}  \over (y-y^{-1})^2} \, \bigg[
y^I  + y^{-I} -  2 \bigg]\, ,
\qquad \hbox{for $I$ even}\, .
\end{split}
\ee

On the other hand the result of the equivariant volume is insensitive to
the parity of $I$, and gives
\be \label{esstwo}
S(I;q_3;y) = (-1)^{I}  
\int_0^{I/2} \de m_3 \, 
\frac{\sinh[(I-2 m_3)\nu]}{\nu} =  (-1)^{I}  \frac{\cosh(I\nu)-1}{2\nu^2}\ .
\ee
The last term in the numerator is recognized as the contribution of the
scaling fixed point $\cQ=\{m_3=I/2,m=0\}$. 
Replacing $\nu\to\sinh\nu$ in the denominator we arrive at 
\be \label{enunu}
\tilde S(I;q_3;y) 
=  (-1)^{I}  \frac{\cosh(I\nu)-1}{2\sinh^2\nu}\ .
\ee

For $I$ even, this agrees with the exact result
\eqref{esth2B}. On the other hand 
if $I$ odd, $\tilde S$ differs from the exact result $\hat S$
by 
\be
\hat S(I;q_3;y) =  \left[ \tilde S(I;q_3;y) 
-  (-1)^{I} \frac{\cosh\nu-1}{2\sinh^2\nu} \right]\qquad \hbox{for $I$ odd}\ .
\ee
Note that the correction term vanishes as $\nu\to\pm\infty$. Thus, 
had we had started with the expression for $\tilde S$ and added 
corrections following the prescription
given below \eqref{defhatS}, we would have arrived at
the correct expression for $\hat S$. It is also worth noting
that when $I$ is odd,  the exact index $\hat S(I;q_3;y=1)$
 differs from the symplectic volume 
$S(I;q_3;y=1)$ by a fraction $(-1)^I/4$, which is necessary to make the result integer. 
Thus the non-renormalization property
which was observed in the non-scaling case breaks down in
this case.  Finally, we note that we have not discussed the 
effect of the regularity conditions \eqref{eex4.5}.
This was discussed in a special
case in \S\ref{d6d6d0}, and we expect this to be satisfied
also for all the solutions described in this section.

\subsection{An inductive proof of the minimal modification hypothesis} \label{srecur}

In this section, we shall prove by induction that $\Delta
\hat S(I;\{q_a\}_{a=3, \dots , n};\nu)$ appearing in
\eqref{corhatS}, giving 
the difference between
the exact index \eqref{ech2} and
the contribution 
$S_{\rm \coll}(I;\{q_a\}_{a=3, \dots , n};\nu)$ 
of regular collinear configurations
given in  \eqref{enet1ab}, 
vanishes as 
$\nu\to\infty$. This proves the validity of the minimal
modification hypothesis for all $n$ for the specific system
under consideration.

First, we establish recursion relations for the equivariant index 
$\hat S$, defined in \eqref{defhatS}. For this purpose it is useful to introduce a 
variant of $\hat S$ defined by
\be
\label{defChat}
\hat C(I; \{q_a\}_{a=3, \dots , n};\nu) = (-1)^{I+n-1} \sum_{\substack{ m_3, \dots , m_n\\
m_a\in \bZ, 0\le m_a\le q_a-1\\ \sum_a m_a \le \lfloor(I-n+1)/2\rfloor}} \frac{\cosh\left[ 
(I-n+2 -2\sum_{a=3, \dots , n} m_a)\nu\right]}{\sinh \nu}\ ,
\ee
and  another quantity
\be
\hat E(I; \{q_a\}_{a=3,\dots n}) =
(-1)^{I+n-1} 
\sum_{\substack{m_3, \dots , m_n\\
m_a\in \bZ, 0\le m_a\le q_a-1,\\ \sum_a m_a = \lfloor(I-n+1)/2\rfloor}} 1\ .
\ee
This is recognized as the number of integer points 
closest to 
the polytope $\cQ$ in \eqref{defpolyh}.
We also define $\hat S$, $\hat C$ and $\hat E$ to be zero
for $I<(n-1)$.
For $n=2$, i.e. in the absence of $D0$-branes, we let
\be
\hat S(I,\nu)=(-1)^{I-1}\frac{\sinh(I\nu)}{\sinh\nu}\ ,\quad
\hat C(I,\nu)=(-1)^{I-1}\frac{\cosh(I\nu)}{\sinh\nu}\  ,\quad
\ee
if $I\geq 0$, and zero otherwise. 
The quantity $\hat E(I; \{q_a\})$ can be evaluated 
from \eqref{defChat} using
\be
\hat E(I; \{q_a\}_{a=3,\dots n}) =
\lim_{\nu\to 0} \left[ \nu\, 
(\hat C(I-1-2q_n; \{q_a\}_{a=3, \dots , n-1};\nu)-\hat C(I-1; \{q_a\}_{a=3, \dots , n-1};\nu) ) \right]\ ,
\ee
valid for $n\geq 3$.

Performing the sum \eqref{defhatS} over the last charge $m_n$ first, we see that the sum is empty
if $\sum_{a=3}^{n-1} 
m_a > \lfloor (I-n+1)/2 \rfloor$, or else runs from $0$ to 
$q-1$ where $q=\min( q_n, \lfloor (I-n+1)/2 \rfloor - 
\sum_{a=3}^{n-1} m_a+1)$.
Using the geometric sum identities
\beq
\sum_{m=0}^{q-1} \sinh(A+2Bm) &=& \frac{\cosh[A+(2q-1)B]-\cosh(A-B)}{2\sinh B}\\
\sum_{m=0}^{q-1} \cosh(A+2Bm) &=& \frac{\sinh[A+(2q-1)B]-\sinh(A-B)}{2\sinh B}
\eeq
with $A=(I-n+2 -2\sum_{a=3}^{n-1} m_a)\nu$ and $B=-\nu$,
we find, for any $n\geq 3$,
\be
\label{Shrec}
\begin{split}
\hat S(I;\{q_a\}_{a=3, \dots , n};\nu) = &  \frac{1}{2\sinh\nu}
\left[ \hat C(I-2q_n;\{q_a\}_{a=3, \dots , n-1};\nu)
- \hat C(I;\{q_a\}_{a=3, \dots , n-1};\nu) \right]\\
&\hspace*{-2cm}
- \frac{\cosh^2\frac{\nu}{2} + (-1)^{I-n} \sinh^2\frac{\nu}{2}}{2\sinh^2\nu}  
 \hat E(I;\{q_a\}_{a=3, \dots , n}) 
\\
\hat C(I;\{q_a\}_{a=3, \dots , n};\nu) = & \frac{1}{2\sinh\nu}
\left[ \hat S(I-2q_n;\{q_a\}_{a=3, \dots , n-1};\nu)
- \hat S(I;\{q_a\}_{a=3, \dots , n-1};\nu) 
\right]  \\
&
- \frac{1+ (-1)^{I-n} }{4\sinh\nu}  \hat E(I;\{q_a\}_{a=3, \dots , n})
\, . 
\end{split}
\ee

Similarly, we can derive a recursion relation for
the contributions of the regular 
collinear solutions to the refined index.
Given a collinear configuration contributing to the
$(n-1)$ particle system with the $D0$-branes carrying
charges $q_3,\dots , q_{n-1}$, we can construct a
collinear configuration contributing to the $n$ particle
system with an additional $D0$-brane with charge
$q_n$ as follows. For simplicity we shall
work in the $\eta\to 0$ limit.
First of all the
additional $D0$-brane can always be placed at infinity
for any collinear configuration of the original $(n-1)$
particle system. 
Also, for any collinear configuration of the
original system, if 
the set $A$ containing the $D0$-branes at the
midpoint between $D6$ and $\overline{D6}$ brane 
satisfies $\sum_{a\in A} q_a< (I-2 q_n)/2$, we can add
the $n$-th $D0$-brane at the midpoint. Taking into account
the various signs appearing in \eqref{enet1ab}
we get 
the recursion relations, for any $n\geq 3$,
\be
\label{Shrecg}
\begin{split}
\hat S_{\rm \coll} (I;\{q_a\}_{a=3, \dots , n};\nu) = & \frac{1}{2\sinh \nu}
\left[ \hat C_{\rm \coll} (I-2 q_n;\{q_a\}_{a=3, \dots , n-1};\nu)
- \hat C_{\rm \coll} (I;\{q_a\}_{a=3, \dots , n-1};\nu) \right]\\
\hat C_{\rm \coll}  (I;\{q_a\}_{a=3, \dots , n};\nu) = & \frac{1}{2\sinh \nu}
\left[ \hat S_{\rm \coll} (I-2 q_n;\{q_a\}_{a=3, \dots , n-1};\nu)
-\hat S_{\rm \coll} (I;\{q_a\}_{a=3, \dots , n-1};\nu) 
\right]  
\end{split}
\ee
where $\hat C_{\rm \coll}$  is defined with an opposite sign compared to 
$\hat S_{\rm \coll}$ in 
\eqref{enet1ab}, 
\be
\hat C_{\rm \coll} (I;\{q_a\}_{a=3, \dots , n};\nu) \equiv 
 (y-y^{-1})^{-n+1} \sum_{A} (-1)^{n_A+I+n-1}
\left( y^{I - 2\sum_{a\in A} q_a} - (-1)^{n-1}
y^{-I + 2\sum_{a\in A} q_a}
\right) \, .
\ee
Subtracting \eqref{Shrecg} from \eqref{Shrec}, we arrive at 
\be
\label{Shrecd}
\begin{split}
\Delta \hat S(I;\{q_a\}_{a=3, \dots , n};\nu) = &  \frac{1}{2\sinh\nu}
\left[ \Delta \hat C(I-2 q_n;\{q_a\}_{a=3, \dots , n-1};\nu)
- \Delta \hat C(I;\{q_a\}_{a=3, \dots , n-1};\nu) \right]\\
&\hspace*{-2cm}
- \frac{\cosh^2\frac{\nu}{2} + (-1)^{I-n} \sinh^2\frac{\nu}{2}}{2\sinh^2\nu}  
\hat E(I;\{q_a\}_{a=3, \dots , n}) 
\\
\Delta \hat C(I;\{q_a\}_{a=3, \dots , n};\nu) = &  \frac{1}{2\sinh\nu}
\left[ \Delta \hat S(I-2 q_n;\{q_a\}_{a=3, \dots , n-1};\nu)
- \Delta \hat S(I;\{q_a\}_{a=3, \dots , n-1};\nu) 
\right]  \\
&- \frac{1+ (-1)^{I-n} }{4\sinh\nu} \, \hat E(I;\{q_a\}_{a=3, \dots , n}) 
\end{split}
\ee
Assuming that $\Delta \hat S,\Delta \hat C\to 0$ as $y\to \infty$ for $n-1$ particles, it immediately
follows from \eqref{Shrecd} that the same statement holds for $n$ particles. 
The validity of the assumption at $n=3$ is easily checked using the explicit results in the
previous subsection. Thus, the minimal modification prescription is proved
for this class of multi-centered configurations. It would be interesting to 
compute $ \Delta \hat S$ directly by using the formula \eqref{ephaseqloc}.

\subsection{Recursion relations for the equivariant volume\label{srecur2}}

In this section we shall derive recursion relations
for the equivariant volume 
\eqref{echclas} similar to the ones given in \S\ref{srecur} , and then use them 
to prove the prescription given below \eqref{defhatS}. 
 We first define a variant of  the equivariant volume \eqref{echclas},
\be \label{ech2clasC}
\begin{split}
C(I;\{m_a\}_{a=3, \dots , n};\nu)
& =(-1)^{I+n-1} \int_{\substack{0\le m_a\le q_a, \\\sum_a m_a \le I/2}}
\de m_3 \cdots \de m_n\, 
\frac{\cosh[(I-2 \sum_{a=3}^n m_a)\nu]}{\nu}
\end{split}
\ee
and
\be
\label{defE}
 E(I;\{q_a\}_{a=3, \dots , n})  =
(-1)^{I+n-1}  \int_{\substack{0\le m_a\le q_a}}
\de m_3 \cdots \de m_n\, \delta\Big(\sum_{a=3}^n m_a -I/2
\Big)\ .
\ee
This last expression is recognized  as the symplectic volume  of 
the submanifold of fixed points based over the polytope $\cQ$.
For $n=2$, i.e. in the absence of $D0$-branes, we set
\be
S(I,\nu)=(-1)^{I-1}\frac{\sinh(I\nu)}{\nu}\ ,\qquad
C(I,\nu)=(-1)^{I-1}\frac{\cosh(I\nu)}{\nu}\
\ee
if $I\geq 0$, and zero otherwise. 
$E$ can be evaluated
in terms of \eqref{ech2clasC} using 
\be
 E(I;\{q_a\}_{a=3, \dots , n})  =   \lim_{\nu\to 0} \left[ \nu\, 
 (C(I-2q_n;\{q_a\}_{a=3, \dots , n-1};\nu)-C(I;\{q_a\}_{a=3, \dots , n-1};\nu)) \right] \ ,
\ee
or directly from its Fourier representation,
\be
E(I;\{q_a\}_{a=3, \dots , n})   
=(-1)^{I+n-1}  \int_{\IR} \frac{\de t}{2\pi t^{n-2}}
e^{-i(I-\sum_{a=3, \dots , n} q_a)t/2}
\prod_{a=3}^{n}  2\sin \frac{q_a}{2} t\ .
\ee
In particular, it is a piecewise polynomial in $I,q_a$.

By integrating first with respect to the last charge $m_n$
in \eqref{echclas}, \eqref{ech2clasC},  
it is straightforward to establish the following relations, valid for $n\geq 3$:
\be
\label{Srec}
\begin{split}
S(I;\{q_a\}_{a=3, \dots , n};\nu) = &  \frac{1}{2\nu}
\left[ C(I-2 q_n;\{q_a\}_{a=3, \dots , n-1};\nu)
- C(I;\{q_a\}_{a=3, \dots , n-1};\nu) \right]\\
&
- \frac{1}{2\nu^2} \, E(I;\{q_a\}_{a=3, \dots , n}) 
\end{split}
\ee
\be
\label{Crec}
C(I;\{q_a\}_{a=3, \dots , n};\nu) = \frac{1}{2\nu}
\left[ S(I-2 q_n;\{q_a\}_{a=3, \dots , n-1};\nu)
- S(I;\{q_a\}_{a=3, \dots , n-1};\nu) 
\right] 
\ee
Similarly as in \eqref{Shrecg}, the contribution to $S(I;\{q_a\})$ from the regular, collinear
fixed points satisfies
\be
\label{Srecg}
\begin{split}
S_{\rm \coll} (I;\{q_a\}_{a=3, \dots , n};\nu) = & \frac{1}{2\nu}
\left[ C_{\rm \coll} (I-2 q_n;\{q_a\}_{a=3, \dots , n-1};\nu)
- C_{\rm \coll} (I;\{q_a\}_{a=3, \dots , n-1};\nu) \right]\\
C_{\rm \coll}  (I;\{q_a\}_{a=3, \dots , n};\nu) = & \frac{1}{2 \nu}
\left[ S_{\rm \coll} (I-2 q_n;\{q_a\}_{a=3, \dots , n-1};\nu)
- S_{\rm \coll} (I;\{q_a\}_{a=3, \dots , n-1};\nu) 
\right]  
\end{split}
\ee
where $C_{\rm \coll}$  is defined with an opposite sign compared to 
$S_{\rm \coll}$ in \eqref{enet1aa}. Applying \eqref{Srec},\eqref{Srecg}
recursively, the difference $\Delta S=S-S_{\rm \coll}$ evaluates to
\be
\label{delSrec}
\Delta S (I;\{q_a\}_{a=3, \dots , n};\nu) =2 \sum_{m=3\atop
n-m\in 2\bZ}^n
\sum_{B} 
\frac{(-1)^{n_B+1}}{(2\nu)^{n-m+2}}  E(I-2\sum_{b\in B} q_b;\{q_a\}_{a=3, \dots , m}) 
\ee
where for given $m$,
the sum runs 
over subsets $B\subset \{m+1,\dots n\}$ subject to the restriction 
that $\sum_{b\in B} q_b<I/2$. This difference can be understood as 
the  contribution of the fixed submanifold $\cM_n^{\rm scal}$ built over the polytope $\cQ$.
In particular, the first term with $m=n$ in the sum \eqref{delSrec} is just the symplectic
volume $E(I,\{q_a\}_{a=3, \dots , n})$ of  $\cM_n^{\rm scal}$, rescaled by a 
factor $-1/(2\nu^2)$. The remaining terms in \eqref{delSrec}  should originate from
the Euler class of the normal bundle to $\cM_n^{\rm scal}$ which determines
the integration mesure on the fixed submanifold, as in \eqref{ephaseqloc}.
We give evidence for this claim in two specific
examples with $n=4$ and $n=5$ in \S\ref{sec_meq}.

We now turn to the proof of the prescription given below \eqref{defhatS} for obtaining
the exact equivariant index from the equivariant volume.
Recall that $\tilde S$ was defined as the result of replacing in $S$
all the $\nu$'s outside the argument of
hyperbolic functions by $\sinh\nu$. 
We define $\tilde C$ as the result of a similar replacement
in $C$.
Then the 
recursion relations for $\tilde S$ and $\tilde C$ 
can be obtained by modifying \eqref{Srec}, \eqref{Crec}
to:
\be\label{Srectilde}
\begin{split}
\tilde S(I;\{q_a\}_{a=3, \dots , n};\nu) = &  \frac{1}{2
\sinh\nu}
\left[ \tilde C(I-2 q_n;\{q_a\}_{a=3, \dots , n-1};\nu)
- \tilde C(I;\{q_a\}_{a=3, \dots , n-1};\nu) \right]\, ,\\
&
- \frac{1}{2\sinh^2\nu} \, E(I;\{q_a\}_{a=3, \dots , n})  \\
\tilde C(I;\{q_a\}_{a=3, \dots , n};\nu) = &  \frac{1}{2\sinh\nu}
\left[ \tilde S(I-2 q_n;\{q_a\}_{a=3, \dots , n-1};\nu)
- \tilde S(I;\{q_a\}_{a=3, \dots , n-1};\nu) 
\right] \, .
\end{split}
\ee
Subtracting \eqref{Shrecg} from \eqref{Srectilde}, 
and defining $\Delta \tilde S=\tilde S -\hat S_{\rm \coll}$,
$\Delta \tilde C=\tilde C -\hat C_{\rm \coll}$, we arrive at 
\be
\label{DSrec}
\begin{split}
\Delta \tilde S(I;\{q_a\}_{a=3, \dots , n};\nu) = &   \frac{1}{2
\sinh\nu}
\left[ \Delta \tilde C(I-2 q_n;\{q_a\}_{a=3, \dots , n-1};\nu)
- \Delta \tilde C(I;\{q_a\}_{a=3, \dots , n-1};\nu) \right]\\
&
- \frac{1}{2\sinh^2\nu} \, E(I;\{q_a\}_{a=3, \dots , n}) 
\\
\Delta \tilde C(I;\{q_a\}_{a=3, \dots , n};\nu) =& \frac{1}{2
\sinh\nu}
\left[ \Delta \tilde S(I-2 q_n;\{q_a\}_{a=3, \dots , n-1};\nu)
- \Delta \tilde S(I;\{q_a\}_{a=3, \dots , n-1};\nu) 
\right] 
\end{split}
\ee
Assuming that $\Delta \tilde S,\Delta \tilde 
C\to 0$ as $\nu\to \infty$ for $n-1$ particles, it immediately
follows from \eqref{DSrec} that the same statement holds for $n$ particles.
The validity of the assumption is easily checked for $n=3$.
On the other hand we have already seen in \S\ref{srecur}
that $\hat S - \hat S_{\rm \coll}$ and $\hat C - \hat C_{\rm
\coll}$ vanishes as $\nu\to\infty$. This shows that
$\hat S - \tilde S$ and $\hat C - \tilde C$ vanishes as
$\nu\to\infty$, thereby confirming the prescription described
below  \eqref{defhatS}.

\subsection{Four-centered case with two identical centers} \label{sidcen}

So far we have only considered the cases where the centers
carry distinct charges. For the dipole halo configuration
analyzed here, the first case of identical charges arise
for four centered configurations in which $q_3$ and $q_4$
coincide.
We shall now examine this case and compare the
results obtained using the minimal modification
hypothesis with the exact results. In this case our formula 
\eqref{enn1full}, \eqref{efullp2}
for the total index
associated with this configuration is:
\be \label{eids1}
\begin{split}
& {1\over 2}
g_{\rm ref} (\alpha_1, \alpha_2, \alpha_3, \alpha_3;y)
\Omega^{\one}_{\rm ref}(\alpha_1,y) \, \Omega^{\one}_{\rm ref}(\alpha_2,y)\,
\Omega^{\one}_{\rm ref}(\alpha_3, y)^2 \\
& + {1\over 2}\, 
g_{\rm ref}(\alpha_1, \alpha_2, 2 \alpha_3;y)
{y-y^{-1}\over y^2 - y^{-2}}\, 
\Omega^{\one}_{\rm ref}(\alpha_1,y) \, \Omega^{\one}_{\rm ref}(\alpha_2,y)\, 
\Omega^{\one}_{\rm ref}(\alpha_3, y^2)\\
&
+ h(\alpha_1, \alpha_2, r\beta;  s^{(1)}_1=1,
s^{(1)}_2=1,s^{(1)}_3=2; y)
\Omega^{\one}_{\rm ref}(\alpha_1,y) \, \Omega^{\one}_{\rm ref}(\alpha_2,y)\,
\Omega^{\one}_{\rm ref}(\alpha_3, y^2)\, .
\end{split}
\ee
We shall determine the function $h$ using the
prescription of \S\ref{sidentical}.
Following the convention of \cite{deBoer:2008fk} and appendix
\ref{sc} we shall consider three separate cases
{\bf A, B} and {\bf D}.
The case {\bf C}, for which $q_3<I/2<q_4$
is not relevant here since we have $q_3=q_4$.

\subsubsection*{Case A}
 In this case we have $2q_3< I/2$.
Thus according to \eqref{esth1A} we have
\be \label{eids2}
g_{\rm ref}(\alpha_1,\alpha_2, 2\alpha_3;y)
= { (-1)^{I}  \over (y-y^{-1})^2} \, \bigg[
y^I - y^{I-4q_3} 
+ y^{-I} - y^{-I+4q_3} \bigg]
= { (-1)^{I} (1 - y^{- 4 q_3}) (y^I - y^{-I+4q_3}) \over (y-y^{-1})^2}\, . 
\ee
Since the first factor has a factor of $(1+y^2)$ our prescription
of \S\ref{sidentical} tells 
us that we have $h=0$. 
Thus using \eqref{eids1},
\eqref{ech3a} and \eqref{eids2} we see that
the full index
is given by
\be \label{eids3}
\begin{split}
& {1\over 2} {(-1)^{I-1} \over (y-y^{-1})^3} \, \bigg[
y^I - 2 y^{I-2q_3}  + y^{I - 4 q_3}
- y^{-I} + 2 y^{-I+2q_3}
- y^{-I+4q_3} \bigg]
\Omega^{\one}_{\rm ref}(\alpha_1,y) \, \Omega^{\one}_{\rm ref}(\alpha_2,y)
\, \Omega^{\one}_{\rm ref}(\alpha_3, y)^2 \\
&+{1\over 2}  {(-1)^{I}  \over (y-y^{-1})(y^2 - y^{-2})} \, 
\bigg[
y^I - y^{I-4q_3} 
+ y^{-I} - y^{-I+4q_3} \bigg]\Omega^{\one}_{\rm ref}(\alpha_1,y) \,
\Omega^{\one}_{\rm ref}(\alpha_2,y)\, 
\Omega^{\one}_{\rm ref}(\alpha_3, y^2)\, .
\end{split}
\ee
It is straightforward to check that Eq. \eqref{eids3} with
$\Omega^{\one}_{\rm ref}(\alpha,y)=1$ agrees
with the exact refined index  computed from \eqref{ech2}, 
with the additional restriction $m_3<m_4$
on the sum to account for the identity of the particles.

\subsubsection*{Case B}

In this case we have $q_3\le I/2$, $2q_3 \ge I/2$, 
and hence according to \eqref{esth2B} we have
\be \label{eids4}
\begin{split}
g_{\rm ref}(\alpha_1,\alpha_2, 2\alpha_3;y)
&={ (-1)^{I} \over (y-y^{-1})^2} \, \bigg[
y^I  + y^{-I} - (y+y^{-1}) \bigg]\, ,
\qquad \hbox{for $I$ odd}\\
&=  { (-1)^{I} \,  \over (y-y^{-1})^2} \, \bigg[
y^I  + y^{-I} -  2 \bigg]\, ,
\qquad \hbox{for $I$ even}\, .
\end{split}
\ee
This is the same as in \eqref{enet4} with $\alpha_{12}$
replaced by $I$. Since $h$ is determined from
$g_{\rm ref}(\alpha_1,\alpha_2, 2\alpha_3;y)$, it will
be given by \eqref{egref4}:
\be\label{eids5}
\begin{split}
h(\alpha_1, \alpha_2, r\beta;  s^{(1)}_1=1,
s^{(1)}_2=1,s^{(1)}_3=2; y)
& = 0 \quad \hbox{for} \quad I\in 2\bZ+1\\
& = 0 \quad \hbox{for} \quad I\in 4\bZ\\
& = {1\over 2}\,
(-1)^{I +3}   (y+y^{-1})^{-1}
\quad \hbox{for} \quad I\in 4\bZ+2\, .
\end{split}
\ee
Using \eqref{eids1},
\eqref{ech4b}, \eqref{eids4} and \eqref{eids5}
we now get the total contribution to the index from
this configuration to be
\be \label{eids6}
\begin{split}
&  {1\over 2}  {(-1)^{I-1} \over (y-y^{-1})^3} \, \bigg[
y^I - 2y^{I-2q_3} - y^{-I} + 2 y^{-I+2q_3} - (y-y^{-1})(4q_3  - I) \bigg]
\\
& \qquad \qquad 
\Omega^{\one}_{\rm ref}(\alpha_1,y) \, \Omega^{\one}_{\rm ref}(\alpha_2,y)\, 
\Omega^{\one}_{\rm ref}(\alpha_3, y)^2\\
& + {1\over 2}
 {(-1)^{I} \, \over (y-y^{-1}) (y^2 - y^{-2})} \, \bigg[
y^I  + y^{-I} - (y+y^{-1}) \bigg]
\Omega^{\one}_{\rm ref}(\alpha_1,y)\,  \Omega^{\one}_{\rm ref}(\alpha_2,y)\, 
\Omega^{\one}_{\rm ref}(\alpha_3, y^2)
\\ & \qquad \qquad \qquad \qquad\qquad \qquad
\qquad \qquad\qquad \qquad\qquad \qquad\qquad \qquad
\qquad \qquad \hbox{for $I\in 2\bZ+1$}
\end{split}
\ee
\be \label{eids6a}
\begin{split}
& {1\over 2}
 \, {(-1)^{I-1} \over (y-y^{-1})^3} \, \bigg[
y^I - 2 y^{I-2q_3} - y^{-I} + 2 y^{-I+2q_3}  - {1\over 2}
(y^2-y^{-2})(4q_3  - I) \bigg]\
\\
& \qquad \qquad \Omega^{\one}_{\rm ref}(\alpha_1,y) \, \Omega^{\one}_{\rm ref}(\alpha_2,y)
\, \Omega^{\one}_{\rm ref}(\alpha_3, y)^2\\
& + {1\over 2} 
{(-1)^{I} \over (y-y^{-1}) (y^2 - y^{-2})} \, \bigg[
y^I  + y^{-I} -  2 \bigg]
\Omega^{\one}_{\rm ref}(\alpha_1,y)\,  \Omega^{\one}_{\rm ref}(\alpha_2,y)\,
\Omega^{\one}_{\rm ref}(\alpha_3, y^2)
\\ & \qquad \qquad \qquad \qquad\qquad \qquad
\qquad \qquad\qquad \qquad\qquad \qquad\qquad \qquad
\qquad \qquad\hbox{for $I\in 4\bZ$} 
\end{split}
\ee
\be \label{eids6b}
\begin{split}
& {1\over 2}
\, {(-1)^{I-1}  \over (y-y^{-1})^3} \, \bigg[
y^I - 2 y^{I-2q_3} - y^{-I} + 2 y^{-I+2q_3}- {1\over 2}
(y^2-y^{-2})(4q_3 - I) \bigg]
\\
& \qquad \qquad \Omega^{\one}_{\rm ref}(\alpha_1,y) \, \Omega^{\one}_{\rm ref}(\alpha_2,y)
\, \Omega^{\one}_{\rm ref}(\alpha_3, y)^2\\
& + {1\over 2} 
\, {(-1)^{I}  \over (y-y^{-1}) (y^2 - y^{-2})} \, \bigg[
y^I  + y^{-I} -  2 \bigg]
\Omega^{\one}_{\rm ref}(\alpha_1,y) \, \Omega^{\one}_{\rm ref}(\alpha_2,y)\, 
\Omega^{\one}_{\rm ref}(\alpha_3, y^2)\\
& +
{1\over 2}\,
(-1)^{I +3}   (y+y^{-1})^{-1} 
\Omega^{\one}_{\rm ref}(\alpha_1,y) \, \Omega^{\one}_{\rm ref}(\alpha_2,y)\, 
\Omega^{\one}_{\rm ref}(\alpha_3, y^2)\, ,
\\ & \qquad \qquad \qquad \qquad\qquad \qquad
\qquad \qquad\qquad \qquad\qquad \qquad\qquad \qquad
\qquad \qquad \hbox{for $I\in 4\bZ+2$}\, .
\end{split}
\ee
Again, one may check that Eq. \eqref{eids6}-\eqref{eids6b} with
$\Omega^{\one}_{\rm ref}(\alpha,y)=1$ agree
with the exact refined index  computed from \eqref{ech2}
with the restriction $m_3<m_4$.

\subsubsection*{Case D}
In this case we have $q_3 >I/2$
and as a result $g_{\rm ref}(\alpha_1,\alpha_2, 2\alpha_3,y)$
and hence $h$ will be given by \eqref{eids4} and
\eqref{eids5} respectively. Thus the only difference from
case {\bf B} is in the expression for
$g_{\rm ref}(\alpha_1, \alpha_2, \alpha_3, \alpha_3;y)$
given in \eqref{ech5b}. Thus using 
\eqref{eids1},
\eqref{ech5b}, \eqref{eids4} and \eqref{eids5}
we now get the total contribution to the index from
this configuration to be
\be \label{eids7}
\begin{split}
&  {1\over 2}\, {(-1)^{I-1}  \over (y-y^{-1})^3} \, \bigg[
y^I - y^{-I}  - I\, (y-y^{-1}) \bigg]
\Omega^{\one}_{\rm ref}(\alpha_1,y) \Omega^{\one}_{\rm ref}(\alpha_2,y)
\Omega^{\one}_{\rm ref}(\alpha_3, y)^2\\
& + {1\over 2}
 {(-1)^{I} \, \over (y-y^{-1}) (y^2 - y^{-2})} \, \bigg[
y^I  + y^{-I} - (y+y^{-1}) \bigg]
\Omega^{\one}_{\rm ref}(\alpha_1,y) \Omega^{\one}_{\rm ref}(\alpha_2,y)
\Omega^{\one}_{\rm ref}(\alpha_3, y^2)
\\ & \qquad \qquad \qquad \qquad\qquad \qquad
\qquad \qquad\qquad \qquad\qquad \qquad\qquad \qquad
\qquad \qquad \hbox{for $I\in 2\bZ+1$}
\end{split}
\ee
\be \label{eids7a}
\begin{split}
& {1\over 2}
 \, {(-1)^{I-1} \over (y-y^{-1})^3} \, \bigg[
y^I - y^{-I}  - {I\over 2}\, (y^2-y^{-2}) \bigg]
\Omega^{\one}_{\rm ref}(\alpha_1,y) \Omega^{\one}_{\rm ref}(\alpha_2,y)
\Omega^{\one}_{\rm ref}(\alpha_3, y)^2\\
& + {1\over 2} 
\, {(-1)^{I}  \over (y-y^{-1}) (y^2 - y^{-2})} \, \bigg[
y^I  + y^{-I} -  2 \bigg]
\Omega^{\one}_{\rm ref}(\alpha_1,y) \Omega^{\one}_{\rm ref}(\alpha_2,y)
\Omega^{\one}_{\rm ref}(\alpha_3, y^2)
\\ & \qquad \qquad \qquad \qquad\qquad \qquad
\qquad \qquad\qquad \qquad\qquad \qquad\qquad \qquad
\qquad \qquad\hbox{for $I\in 4\bZ$} 
\end{split}
\ee
\be \label{eids7b}
\begin{split}
& {1\over 2}
 \, {(-1)^{I-1} \over (y-y^{-1})^3} \, \bigg[
y^I - y^{-I}  - {I\over 2}\, (y^2-y^{-2}) \bigg]
\Omega^{\one}_{\rm ref}(\alpha_1,y) \Omega^{\one}_{\rm ref}(\alpha_2,y)
\Omega^{\one}_{\rm ref}(\alpha_3, y)^2\\
& + {1\over 2} 
 \, {(-1)^{I} \over (y-y^{-1}) (y^2 - y^{-2})} \, \bigg[
y^I  + y^{-I} -  2 \bigg]
\Omega^{\one}_{\rm ref}(\alpha_1,y) \Omega^{\one}_{\rm ref}(\alpha_2,y)
\Omega^{\one}_{\rm ref}(\alpha_3, y^2)\\
& +
{1\over 2}\,
(-1)^{I +3}   (y+y^{-1})^{-1} 
\Omega^{\one}_{\rm ref}(\alpha_1,y) \Omega^{\one}_{\rm ref}(\alpha_2,y)
\Omega^{\one}_{\rm ref}(\alpha_3, y^2)\, ,
\\ & \qquad \qquad \qquad \qquad\qquad \qquad
\qquad \qquad\qquad \qquad\qquad \qquad\qquad \qquad
\qquad \qquad \hbox{for $I\in 4\bZ+2$}\, .
\end{split}
\ee
Again, one may check that Eq. \eqref{eids7}-\eqref{eids7b} with
$\Omega^{\one}_{\rm ref}(\alpha,y)=1$ agree
with the exact refined index  computed from \eqref{ech2}
with the restriction $m_3<m_4$.

\acknowledgments

We are grateful to I.~Bena, M.~Berkooz, F.~Denef, S.~El Showk and 
M.~Vergne 
for valuable discussions. J.M. thanks the LPTHE, 
IHES, Zhejiang University and Kyoto University for hospitality during parts of this work. 
The work of J.M. is partially supported by ANR grant BLAN06-3-137168. B.P. and A.S. wish to thank
the organizers of the ISM 2011 workshop in Puri for their generous hospitality during the 
last stage of this project.
A.S. acknowledges the support of 
the J. C.  Bose fellowship of the Department of
Science and Technology, India and of the
project 11-R\& D-HRI-5.02-0304.

\appendix

\section{Sign rules for collinear fixed points} \label{ssignrule}

In this appendix, we provide some details on the computation 
of the sign  of  the contributions
from collinear fixed points near a wall of marginal stability (\S\ref{swall})
and for dipole halo configurations (\S\ref{sgen}). Recall that $s(p)$ is
given by \eqref{esigneq}, where $\hat M(p)$ is the Hessian of the 
`superpotential' \eqref{defhatW} at the critical point $p$.

\subsection{Sign rules
near a wall of marginal stability} \label{sd}

First we shall compute the sign $s(p)$
associated with a collinear configuration $p$
near a wall of
marginal stability, if the collinear configuration breaks up
into two widely separated clusters as we approach the wall.
Such a configuration has been described in \S\ref{swall}
where the two sets into which the centers split have been
called $A$ and $B$. We follow the notation of \S\ref{swall},
and work with the configuration for which $z_B>z_A$,
\i.e.\ $z_j - z_i>0$ for $i\in A$, $j\in B$.
Near $R\equiv |z_A-z_B|\to \infty$, 
the superpotential \eqref{defhatW}
 decomposes as 
\be
\hat W(z_i,\lambda) \sim \hat W_A\left(z_{i\in A}, \lambda_A=\frac{n_A}{n}\lambda \right) 
+\hat W_B\left(z_{j\in B}, \lambda_B=\frac{n_B}{n}\lambda \right) 
-\sum_{i\in A, j\in B}\alpha_{ij} \ln | z_i - z_j |\ .
\ee
For such a configuration
 the Hessian of $\hat W$ with respect to $\lambda,z_{i\in A},z_{j\in B}$ 
takes the form
\be
\hat M= \hat M_0 + \hat M_1\, ,
\ee
where $\hat M_0$ is the Hessian in the strict $R\to \infty$ limit and
$M_1$ is of order $1/R^2$. We have
\be
\hat M_0=\begin{pmatrix} 
0 & \frac{n_A}{n} u_A^T &  \frac{n_B}{n} u_B^T \\
\frac{n_A}{n} u_A &  M_A &0 \\
 \frac{n_B}{n} u_B &  0& M_B
 \end{pmatrix}
\ee
where $M_{A}$ is the Hessian of $W_{A}$ with respect to $z_{i\in A}$, $u_A$ is the $n_A$-dimensional column vector with entries $1/n_A$, and similarly for $M_B, u_B$.  In
particular, we note that $M_A\, u_A=M_B\, u_B=0$. 
$\hat M_1$ is given by
\be \label{ehatm1}
\begin{split}
& (\hat M_1)_{00} = (\hat M_1)_{0i}= (\hat M_1)_{i0}=0
\quad \hbox{for $1\le i\le n$}, \\
& (\hat M_1)_{ij} = 
\frac{1}{R^2}
\begin{cases}  \delta_{ij} 
\sum_{k\in B}
{\alpha_{ik}
} \quad \hbox{for $i,j\in A$}\\
  \delta_{ij} \sum_{k\in A} {\alpha_{ki}
} \quad \hbox{for $i,j\in B$}\\
- {\alpha_{ij}
} \quad  \hbox{for $i\in A$, $j\in B$}\\
-  {\alpha_{ji}
} \quad  \hbox{for $i\in B$, $j\in A$}
\end{cases}
\end{split}
\ee
On the other
hand, the Hessian of $\hat W_A$ (respectively, $\hat W_B$) with respect to $\lambda_A,z_a$ ($a\in A$)
(respectively, $\lambda_B, z_b$, $b\in B$) is given by 
\be
\hat M_A= 
\begin{pmatrix} 
0 & u_A^T \\
u_A & M_A
\end{pmatrix}\ ,\qquad
\hat M_B= 
\begin{pmatrix} 
0 & u_B^T \\
u_B & M_B
\end{pmatrix}\, . 
\ee
To compare the signs of $\det \hat M$ and 
$\det \hat M_A\, \det \hat M_B$, we shall construct
an eigensystem of $\hat M$ in terms of the eigensystems of $\hat M_A$ and $\hat M_B$, in the limit 
$R\to \infty$. First, we note that an eigensystem  of 
$\hat M_A$ is given by the $n_A+1$
(eigenvectors,eigenvalues)
\be
\left( [\pm \frac{1}{\sqrt{n_A}} ,u_A], \pm\frac{1}{\sqrt{n_A}} \right)\ ,\quad 
\left([0,v_A^{(i)}],   \lambda_A^{(i)} \right)
\ee
where $v_A^{(i)}$, $i=1, \dots , n_A-1$ are eigenvectors of $M_A$ in the subspace orthogonal
to the null eigenvector $u_A$. Similarly, let $v_B^{(j)}$ be a system of eigenvectors of 
$M_B$ in the subspace orthogonal
to the null eigenvector $u_B$, with eigenvalues $\lambda_B^{(j)}$. 
In the strict $R\to \infty$ limit, $\hat M$ reduces to
$\hat M_0$ and an eigensystem of $\hat M_0$
is  given by the $n_A+n_B+1$ 
(eigenvectors,eigenvalues)
\be
\Big( [\pm \frac{1}{\sqrt{n}},\frac{n_A}{n}u_A, \frac{n_B}{n}u_B],\pm \frac{1}{\sqrt{n}} \Big),\quad
\Big([0,v_A^{(i)},0_B],   \lambda_A^{(i)})\ ,\quad ([0,0_A,v_B^{(j)}],   \lambda_B^{(j)}\Big),\quad
\Big([0,u_A, -u_B], 0\Big)
\ee
In particular, the last eigenvector, corresponding to a change in the relative separation
between the two clusters keeping their inner structure fixed, yields a zero-mode of $\hat M_0$. Since the eigenvalues 
$\lambda_A^{(i)}, \lambda_B^{(j)}$ 
are generically distinct and non-zero, the structure of the spectrum will retain its 
form at  large but finite $R$, except that the last eigenvalue will be lifted to a non-zero
eigenvalue $\lambda(R)$. 
To compute $\lambda(R)$, it suffices to use non-degenerate perturbation theory
and calculate the expectation value of
$\hat M_1$ on the unperturbed eigenvector 
$U=[0,u_A, -u_B]$. This gives
\be
\lambda(R) \sim  \frac{U^T\cdot \hat M_1 \cdot U}{U^T \cdot U} \sim                                      
\frac{\gamma_{AB} (n_A+n_B)}{\, n_A n_B R^2}\, .
\ee
As a result, 
\be
\label{hdet3}
\det \hat M \sim - \frac{1}{n} \lambda(R) \prod_i  \lambda_A^{(i)})  
\prod_j  \lambda_B^{(j)} 
 \sim - \frac{1} {R^2}
\, \gamma_{AB}\, \det \hat M_A\, \det \hat M_B 
\ee
Using \eqref{esigneq}, we arrive at \eqref{esigndiv} for
$z_B>z_A$. The result for $z_B<z_A$ follows by
exchanging $A$ and $B$ in the above analysis.

\subsection{Sign rules for collinear dipole halos \label{secsigndip}}

In this section, we establish the sign rule 
$s(p)=(-1)^{n_A}$ used in \eqref{enet1aa}
for collinear dipole halo configurations in the non-scaling regime. 
For small positive $\eta$, 
the solutions described above \eqref{enet1aa} satisfy
\be
z_a\to \frac12(z_1+z_2) - {(r_{12})^3\over 8 R^2}
\quad  \forall a\in A\ ,\quad z_b \to \frac12(z_1+z_2)-R 
\quad \forall b\in B\ ,
\ee
where $R\simeq \sqrt{r_{12}/\eta}$. 
In this limit, the Hessian $\hat M$ 
takes the following form to order $R^{-3}$:
\be \label{eqhatM}
\begin{split}
& \hat M_{00} = 0, \quad \hat M_{0i}=\hat M_{i0} = {1\over n}
\quad \hbox{for $1\le i,j\le n$},  \quad \hat M_{12} =\hat M_{21} = - {I \over (r_{12})^2}\, ,  \\
& \hat M_{11} = {I-4\sum_{a\in A} q_a  \over r_{12}^2} + 
\frac{1}{R^2} ( \sum_{b\in B} {q_b} - 2  \sum_{a\in A} q_a) 
+ \frac{ r_{12} }{R^{3}}\sum_{b\in B} q_b\, ,\\
& \hat M_{22} = 
{I - 4\sum_{a\in A} q_a   \over r_{12}^2} 
+ \frac{1}{R^2} ( 2 \sum_{a\in A} q_a -\sum_{b\in B} {q_b} )
+  \frac{ r_{12} }{R^{3}} \sum_{b\in B} q_b  
\, , \\
&\hat M_{1a} = \hat M_{a1} = {4\over (r_{12})^2} q_a + \frac{2}{R^2} 
q_a \quad \hbox{for $a\in A$}, \quad
\hat M_{1b} = \hat M_{b1}
= - \frac{1}{R^{2}} q_b -  \frac{r_{12} }{R^{3}}  q_b  
\quad \hbox{for $b\in B$}\, ,
 \\ 
&  \hat M_{2a} = \hat 
M_{a2} = {4\over (r_{12})^2} q_a - \frac{2}{R^2} q_a
\quad \hbox{for $a\in A$}, \quad \hat M_{2b} = \hat M_{b2} 
= \frac{1}{R^2} q_b - \frac{r_{12} }{R^{3}} q_b \quad 
\hbox{for $b\in B$}\, ,
\\
& \hat M_{aa'} = -{8 \over (r_{12})^2} \, q_a \, \delta_{aa'} \quad
\hbox{for $a, a'\in A$}, 
\quad \hat M_{bb'}  = 2 \, \frac{r_{12}}{R^{3}} \, q_b\, \delta_{bb'}
\quad \hbox{for $b,b'\in B$}, \\ & 
\hat M_{ab}=\hat M_{ba}
=0 \quad  \hbox{for $a\in A$, $b\in B$}\, .
\end{split}
\ee
To evaluate the determinant of this matrix in the large
$R$ limit we proceed as follows.
First we define a new matrix $\tilde M$ by dropping the
first two rows and columns of $\hat M$; we have seen earlier
that $\det \hat M = -\widehat\det M = -\det \tilde M$.
To evaluate $\det \tilde M$, we can add half of the
the second to last row
of the matrix to the first row and then add 
half of the second to last
columns of the matrix to the first column. This does not
change the determinant but simplifies the matrix. 
Let us denote the resulting matrix by
$\tilde M_0 + \tilde M_1$ where $\tilde M_0$ is the 
limit of the matrix as
$R\to\infty$ and $\tilde M_1$ is the remainder.  
It is straightforward to check that $\tilde M_0$ is
diagonal and has eigenvalues
\be \label{etilM1}
{(I-2\sum_{a\in A} q_a)\over r_{12}^2}, \quad
\left\{- {8 q_a\over
(r_{12})^2}\right\}, \quad \{0_{n_B}\}\, ,
\ee 
where $0_{n_B}$ denotes that the eigenvalue 0 is
repeated $n_B$ times.
When we take into account the effect of $\tilde M_1$, the
non-zero eigenvalues are not affected appreciably but the
zero eigenvalues are lifted and can be obtained using
first order degenerate perturbation theory. This gives the
approximate eigenvalues of $\tilde M_0+\tilde M_1$
to be:
\be \label{etilM2}
{(I-2\sum_{a\in A} q_a)\over r_{12}^2}, \quad
\left\{- {8 q_a\over
(r_{12})^2}\right\}, \quad
\left\{\left({2 r_{12}\over R^3}\right)_{n_B}\right\} \, .
\ee 
To leading order in the limit $R\to \infty$, 
$\det\hat M=-\det \tilde M$ is therefore   given by
\be
\begin{split}
\det \hat M \simeq 2^{n+2n_A-2} \, (-1)^{n_A+1}
\frac{(I-2 \sum _{a\in A} q_a)\, (\prod_{a\in A} q_a )\, 
(\prod_{b\in B} q_b) }
{R^{3(n-2-n_A)} r_{12}^{3n_A-n+4}}\, .
\end{split}
\ee
In particular, the sign of $\det \hat M$ is $(-1)^{n_A+1}$. Note that to obtain 
this result, it is important to keep all subleading terms through order $1/R^3$ as 
indicated in \eqref{eqhatM}, since the determinant at lower order vanishes. Using \eqref{esigneq} we now arrive at
\be \label{essigma}
s(p) = (-1)^{n_A}\, .
\ee

\section{Laurent polynomial property in absence of scaling
solutions} \label{sinteg}

Eq.\eqref{enn1}, \eqref{enn2} (or equivalently
\eqref{eform22new}) gives the index associated with a multi-centered black
hole solution when the charges of the components do not allow for scaling solutions.
Let us restrict to the case where
$\gamma$ is primitive so that \eqref{enn1} directly gives the
refined index $\tot_{\rm ref}(\gamma, y)$ rather than its rational counterpart $\bar\tot_{\rm ref}(\gamma, y)$.
In this case the right hand side of \eqref{enn1} must 
be a Laurent polynomial in $y$ (i.e. a finite linear combination of $y^{\pm m}$)
whenever the $\Omega^{\one}_{\rm ref}(\alpha_i, y)$'s are
since otherwise the result cannot be interpreted as
an $SU(2)$ character.
Our goal in this appendix will be to prove this 
property of \eqref{enn1}.\footnote{In order to prove that
the result is indeed an $SU(2)$ character, one 
must also show that the coefficients of $y^{\pm m}$ are
integers. This is indeed true, but we shall omit
the proof.}

Clearly this will not be true for an arbitrary
choice of the functions $g^{}_{\rm ref}$, but 
we shall use
\eqref{esplit1} -- valid when at the attractor point only single
centered black holes contribute to the index -- to restrict the form
of $g_{\rm ref}$.
It  is clear from this
equation that the right hand side of \eqref{esplit1} will
be a Laurent
polynomial in $y$
if the $g^{}_{\rm ref}$'s
appearing on the right hand side have this
property.
Thus by iterative application of \eqref{esplit1} we can
establish that $g_{\rm ref}(\alpha_1,\dots , \alpha_n, y)$
is given by a Laurent
polynomial in $y$. 
It now follows from \eqref{enn1} that
when the $\alpha_i$'s are   all primitive
then the right hand side of
this equation is a Laurent
polynomial in $y$
since the individual $\bOm^{\one}_{\rm ref}(\alpha_i,y)
=\Omega^{\one}_{\rm ref}(\alpha_i,y)$'s and
the $g_{\rm ref}(\alpha_1,\dots ,\alpha_n, y)$ 
have this property. This argument fails when 
some of the $\alpha_i$'s are not primitive:
in this case
$\bOm^{\one}_{\rm ref}(\alpha_i,y)$ defined in \eqref{enn2}
have extra factors of $m(y^m-y^{-m})$ in the denominator,
which must cancel in order that the  final expression is
a Laurent polynomial in $y$. 
We shall now
demonstrate that
this cancellation does take place.

By iterative application of \eqref{esplit1} we can express
$g_{\rm ref}(\alpha_1,\dots ,\alpha_n)$ as a sum over
attractor flow trees in which a total charge $\gamma$
decays via an appropriate tree to the charges $\alpha_1$,
$\dots ,\alpha_n$. 
At a vertex at which a charge $\beta_1$ decays into two clusters
of charge $\beta_2$ and  $\beta_3$, we get
a multiplicative factor of
\be 
(-1)^{\langle \beta_2, \beta_3\rangle +1}
\sign\langle \beta_2, \beta_3\rangle 
{\sinh  (\langle \beta_2, \beta_3\rangle \nu)\over \sinh\nu}
\, .
\ee
Let us focus on the vertex at which a non-primitive
charge $\alpha_i$ gets attached to the tree.
Suppose at this vertex an internal line 
carrying charge $\beta+\alpha_i$
decays into another internal line of charge $\beta$
and the external line of charge $\alpha_i$.
Our goal will be to show that the contribution from the
vertex factor cancels the $(y^m - y^{-m})$ factors in
the denominator appearing in \eqref{enn2}.
The product of the vertex factor and the
$\bar \Omega^{\one}(\alpha_i, y)$ factor is given by
\be \label{ec.1}
\begin{split}
& (-1)^{\langle \beta, \alpha_i\rangle +1}
\sign\langle \beta, \alpha_i\rangle 
{\sinh  (\langle \beta, \alpha_i\rangle \nu)\over \sinh\nu}
\, \bar \Omega^{\one}(\alpha_i,y)\\
=& (-1)^{\langle \beta, \alpha_i\rangle +1}
\sign\langle \beta, \alpha_i\rangle \sum_{m|\alpha_i}
m^{-1} 
{\sinh  (\langle \beta, \alpha_i\rangle \nu)\over \sinh(m\nu)}
\, \Omega^{\one}(\alpha_i/m,y^m) \, .
\end{split}
\ee
Since $m|\alpha_i$, 
$\langle \beta, \alpha_i\rangle$ is an integral 
multiple of $m$. In
this case the unwanted denominator factors
cancel and ${\sinh  (\langle \beta, \alpha_i\rangle \nu)/ \sinh(m\nu)}$
is a Laurent polynomial in $y$.

This leaves the end vertices of the attractor flow
tree, at which 
a charge $\alpha_i+\alpha_j$ decays into
a pair of charges $\alpha_i$ and $\alpha_j$. The
contribution from such a vertex will be of the form
\be \label{eckt1}
\begin{split}
& (-1)^{\langle \alpha_i, \alpha_j\rangle +1}
\sign\langle \alpha_i, \alpha_j\rangle 
{\sinh  (\langle \alpha_i, \alpha_j\rangle \nu)\over \sinh\nu}
\, \bar \Omega^{\one}(\alpha_i,y) 
\bar \Omega^{\one}(\alpha_j,y)\\
=& (-1)^{\langle \alpha_i, \alpha_j\rangle +1}
\sign\langle \alpha_i, \alpha_j\rangle \sum_{m|\alpha_i}
\sum_{p|\alpha_j}
m^{-1} p^{-1}
{\sinh  (\langle \alpha_i, \alpha_j\rangle \nu)
\sinh\nu\over \sinh(m\nu)
\sinh (p\nu)}
\, \Omega^{\one}(\alpha_i/m,y^m) 
\Omega^{\one}(\alpha_j/p,y^p)\, .
\end{split}
\ee
Let us define $z=y^2 = e^{2\nu}$.
Since $m|\alpha_i$ and $p|\alpha_j$, 
$\langle \alpha_i, \alpha_j\rangle$ is an integral
multiple of $mp$. 
In this case $\sinh(\langle \alpha_i, \alpha_j\rangle \nu)$
will have zeroes at $z^{mp}=1$. On the other hand the
denominator $\sinh(m\nu) \sinh(p\nu)$ has zeroes
at $z^m=1$ and also at $z^p=1$. If $m$ and $p$ are
relatively prime then the locations of these zeroes
are distinct except for a common zero of both
factors at $z=1$. Furthermore they
coincide with the zeroes of the numerator
at $z^{mp}=1$. Thus all the
zeroes of the denominator 
$\sinh(m\nu) \sinh(p\nu)$ cancel against the zeroes
of the numerator $\sinh  (\langle \alpha_i, \alpha_j\rangle \nu)\, \sinh\nu$.
Thus as long as $m$ and $p$ are relatively prime, \eqref{eckt1}
is a Laurent polynomial in $y$. 

Now suppose that $q\equiv {\rm gcd}(m,p)>1$. 
Then $\sinh(m\nu)\sinh(p\nu)$ will have
double zeroes at each of the $q$ solutions to
$z^q=1$. In contrast the $\sinh  (\langle \alpha_i, \alpha_j\rangle \nu)$ factor in the numerator will generically have only
single zeroes at each solution of $z^q=1$.  
Combining this with the extra
factor of $\sinh\nu$ in the numerator, we see that
there is effectively a left-over factor of 
$\sinh\nu/\sinh(q\nu)$ from the vertex, 
multiplied by factors which are Laurent
polynomials in $y$. We now show that the
factor  $\sinh\nu/\sinh(q\nu)$ is cancelled
by other vertex factors in the same tree.

To see this, note that since $q$ divides both $\alpha_i$, $\alpha_j$, it
divides  their sum $\alpha_i+\alpha_j$.
We can now repeat the analysis at
the next vertex where say a line carrying charge 
$\alpha_i+\alpha_j +\beta$ splits into charges
$\beta$ and $\alpha_i+\alpha_j$. As long
as $\beta$ and $q$ do not have a common factor, 
the analysis of the previous paragraph shows that
the vertex factor will cancel all the unwanted denominators,
including the left-over factor of $\sinh(q\nu)/\sinh\nu$ from
the previous vertex.
If on the other hand $\beta$ and $q$ have a common factor
$s$ then we shall have a left-over factor of $\sinh\nu/
\sinh(s\nu)$ besides factors containing Laurent
polynomials in $y$. Furthermore 
$\alpha_i+\alpha_j+\beta$ will have the
same common factor $s$.
The analysis can now be repeated for the next vertex.
Proceeding this way, and using the fact that the initial
charge $\gamma$ is taken to be primitive, one can prove that
at the end all the denominator factors cancel, and we
are left with a Laurent polynomial in $y$, proving the
desired result.

\section{Equivariant volumes and indices in dipole halos: $n=4,5$} \label{sc}

In this section, we provide explicit results for the equivariant volume and equivariant
index of the moduli space of $n$-centered dipole halo configurations with 4 and 5 
centers. This serves as a check on our minimal modification hypothesis and  
on the recursion relations derived in \S\ref{srecur},\S\ref{srecur2}, and provides useful
insight of the fixed points responsible for these contributions.

\subsection{Four-centered case with distinct centers}
\label{s342}

For $n=4$ we have to treat several different cases, depending on the value of $I/2$
relative to $q_3$, $q_4$ and $q_3+q_4$. We label the four possible cases 
as in \cite{deBoer:2009un}, see Figure \ref{figpoly4}.

\FIGURE{\centerline{\includegraphics[totalheight=14cm]{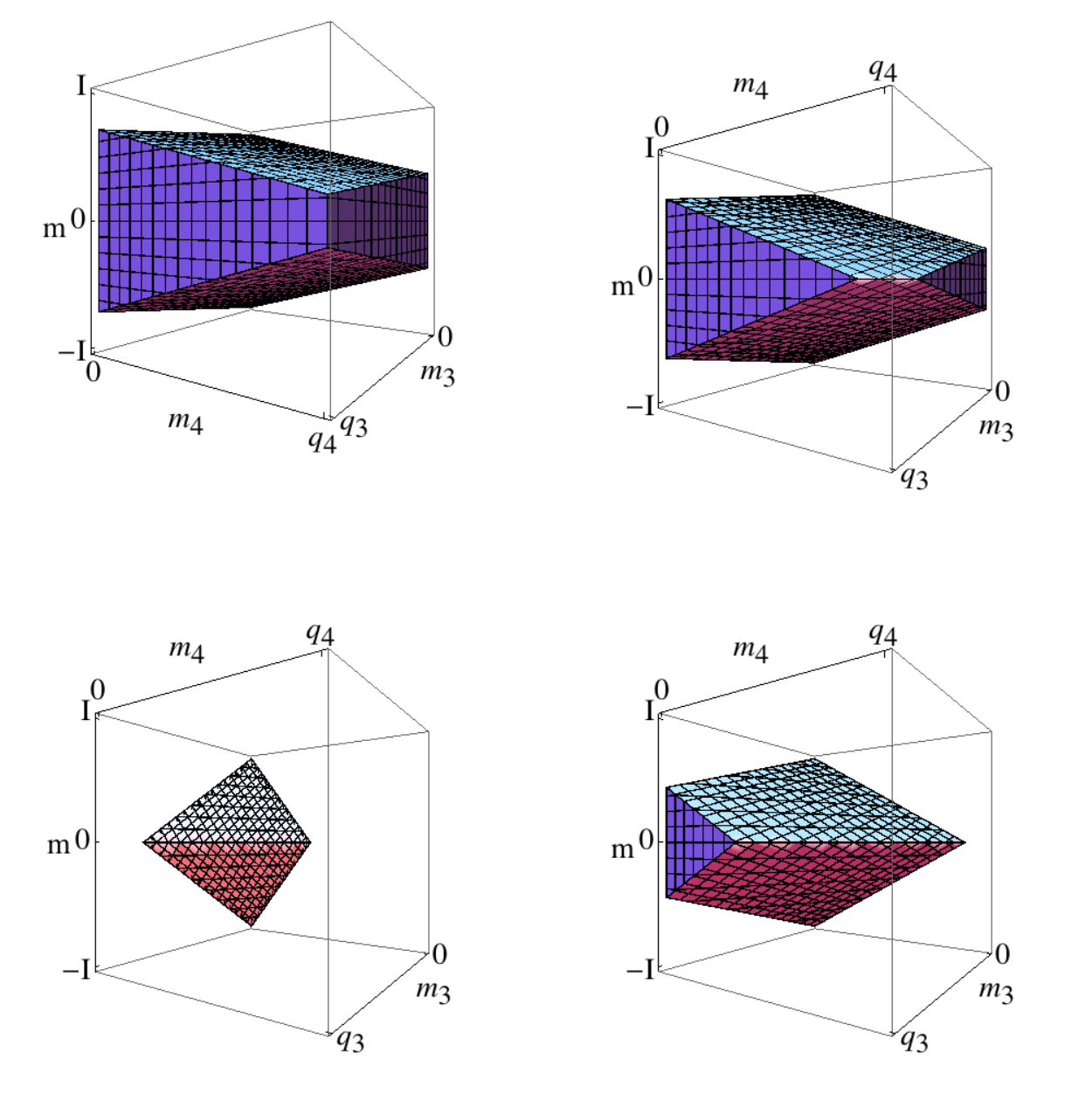}}
\caption{Polytopes associated to 4-center dipole halos in charge regimes $A,B,C,D$ 
respectively,  in clockwise order starting from top-left corner. \label{figpoly4}}
}

\subsubsection*{Case A: $q_3+q_4 < {I\over 2}$}

 In this case, the contribution  to 
 $g_{\rm ref}(\alpha_1,\dots, \alpha_4;y)$ from
collinear solutions,
given in \eqref{enet1ab},
is:
\be \label{ech3a}
\hat S_{\rm \coll} (I;q_3,q_4;y)=
 {(-1)^{I-1} \over (y-y^{-1})^3} \, \bigg[
y^I - y^{I-2q_3} - y^{I-2q_4} + y^{I - 2 q_3 - 2q_4}
- y^{-I} + y^{-I+2q_3} + y^{-I+2q_4} - y^{-I+2q_3+2q_4}
\bigg]\, .
\ee
This has finite $y\to 1$ limit and hence, according to our
proposal in \S\ref{scaling}, should be  the complete answer.
Indeed, it can be checked that \eqref{ech3a} agrees with the 
exact result \eqref{ech2}.  

On the other hand, the equivariant volume 
evaluates to 
\be
\begin{split}
S(I;q_3,q_4;\nu) =& (-1)^{I-1}  \, \int_0^{q_3} \de m_3 \,  \int_0^{q_4} \de m_4\, 
\frac{\sinh[(I-2(m_3+m_4)\nu]}{\nu}  \\ =
\frac{(-1)^{I-1} }{4\nu^3}&
\left( \sinh(I\nu)+\sinh[(2q_3-I)\nu ]+\sinh[(2q_4-I)\nu ]
+\sinh[(I-2q_3-2q_4)\nu ] \right)\, .
\end{split}
\ee
Replacing $\nu\to\sinh\nu$ in the denominator we get
\be \label{ewget}
\begin{split}
\tilde S(I;q_3,q_4;\nu)  =
\frac{(-1)^{I-1} }{4\sinh^3 \nu} &
\left\{ \sinh(I\nu)+\sinh[(2q_3-I)\nu ]+\sinh[(2q_4-I)\nu ] \right.\\
&\left.+\sinh[(I-2q_3-2q_4)\nu ] \right\}\, .
\end{split}
\ee
Thus already agrees with the exact result $\hat S$ given in
\eqref{ech3a}. Hence the prescription below
\eqref{defhatS} and the minimal modification
hypothesis gives the same result.

\subsubsection*{Case B: $q_3, q_4\leq {I\over 2}\leq q_3+q_4$}

 In this case \eqref{enet1ab}
 leads to the following contribution to
 $g_{\rm ref}(\alpha_1,\dots, \alpha_4;y)$:
\be \label{ech4a}
\hat S_{\rm \coll} (I;q_3,q_4;y)=
 {(-1)^{I-1} \over (y-y^{-1})^3} \, \bigg[
y^I - y^{I-2q_3} - y^{I-2q_4} - y^{-I} + y^{-I+2q_3} + y^{-I+2q_4}
 \bigg]\, .
\ee
This diverges in the $y\to 1$ limit. Hence we must add
extra contributions of the form given in \eqref{escalprop}
to have a Laurent polynomial. This leads to
\be \label{ech4b}
\begin{split}
\hat S (I;q_3,q_4;y)=
&=  {(-1)^{I-1} \over (y-y^{-1})^3} \, \bigg[
y^I - y^{I-2q_3} - y^{I-2q_4} - y^{-I} + y^{-I+2q_3} + y^{-I+2q_4}
\\
& \qquad \qquad - (y-y^{-1})(2q_3 + 2 q_4 - I) \bigg]\, ,
\qquad \hbox{for $I$ odd}\\
&={ (-1)^{I-1}  \over (y-y^{-1})^3} \, \bigg[
y^I - y^{I-2q_3} - y^{I-2q_4} - y^{-I} + y^{-I+2q_3} + y^{-I+2q_4}
\\
& \qquad \qquad - {1\over 2}
(y^2-y^{-2})(2q_3 + 2 q_4 - I) \bigg]\, ,
\qquad \hbox{for $I$ even}\, .
\end{split}
\ee
The need for adding correction terms shows  that in this
case there are scaling solutions. 
It can be checked that \eqref{ech4b} indeed agrees with the 
exact result \eqref{ech2}.  

On the other hand, the equivariant volume 
evaluates to 
\be
\begin{split}
S(I;q_3,q_4;\nu) =
&
\frac{ (-1)^{I-1} }{4\nu^3}
\left\{ \sinh(I\nu)  +\sinh[(2q_3-I)\nu ]+\sinh[(2q_4-I)\nu ]
-(2q_3+2q_4-I) \nu  \right\} \ .
\end{split}
\ee
The last term is recognized as the contribution of the submanifold 
of fixed points $\cM_n^{\rm scal}$ in \eqref{hatM}, 
with symplectic volume $E(I,q_3,q_4)=(-1)^I (q_3+q_4-I/2)$. 
Replacing $\nu$ by $\sinh\nu$ in the denominator and in the
last term of the numerator, we arrive at 
\be
\begin{split}
\tilde S(I;q_3,q_4;\nu) =&
\frac{ (-1)^{I-1} }{4\sinh^3\nu} \Big\{
\sinh(I\nu)  +\sinh[(2q_3-I)\nu ]+\sinh[(2q_4-I)\nu ]
\\ & \qquad \qquad \qquad
-(2q_3+2q_4-I) \sinh\nu  \Big\}\, .
\end{split}
\ee
Again, this differs from $\hat S$ given in \eqref{ech4b}
by a term that vanishes as $\nu\to\pm\infty$:
\be
\hat S(I;q_3,q_4;y) = \tilde S(I;q_3,q_4;y) 
+(-1)^{I-1}   \frac{2q_3+2q_4-I}{4\sinh^3\nu} 
\left\{ \begin{matrix} 0 \ :\ \hbox{$I$ odd} \\
\sinh\nu-\frac12\sinh2\nu \ :\  \hbox{ $I$ even} 
\end{matrix}\right\} \, .
\ee
Thus the minimal modification hypothesis and the
prescription given below \eqref{defhatS} agree with
each other and the exact result.

\subsubsection*{ Case C: $q_3\le {I\over 2} \le q_4$}
In this case \eqref{enet1ab}
 leads to the following contribution to
 $g_{\rm ref}(\alpha_1,\dots, \alpha_4;y)$:
\be \label{ech7a}
\hat S_{\rm \coll} (I;q_3,q_4;y)= { (-1)^{I-1}  \over (y-y^{-1})^3} \, \bigg[
y^I - y^{I-2q_3}  - y^{-I} + y^{-I+2q_3} 
 \bigg]\, .
\ee
This diverges in the $y\to 1$ limit. Hence we must add
extra contributions of the form given in \eqref{escalprop}
to have a Laurent polynomial. This leads to
\be \label{ech7b}
\begin{split}
\hat S (I;q_3,q_4;y)
&={ (-1)^{I-1} \over (y-y^{-1})^3} \, \bigg[
y^I - y^{I-2q_3} - y^{-I} + y^{-I+2q_3} 
 - 2\, (y-y^{-1})q_3 \bigg]\, ,
\\
& \qquad \qquad\qquad \hbox{for $I$ odd}\\
&=   {(-1)^{I-1} \over (y-y^{-1})^3} \, \bigg[
y^I - y^{I-2q_3}  - y^{-I} + y^{-I+2q_3} 
 - 
(y^2-y^{-2}) q_3  \bigg]\, ,
\\
& \qquad \qquad\qquad \hbox{for $I$ even}\, ,
\end{split}
\ee
in agreement with the exact result \eqref{ech2}. 

On the other hand,
the equivariant volume leads to 
\be
\label{s34c}
\begin{split}
S(I;q_3,q_4;\nu) = &
\frac{(-1)^{I-1}}{4\nu^3}
\left\{ \sinh(I\nu)  +\sinh[(2q_3-I)\nu ]
-2 q_3 \nu\right\}\, .
\end{split}
\ee
Again, 
the last term in \eqref{s34c} is recognized as the contribution
from the fixed submanifold $\cM_n^{\rm scal}$, with symplectic 
volume $E(I,q_3,q_4)=(-1)^I q_3$.
Following the rules described earlier, we get
\be
\begin{split}
\tilde S(I;q_3,q_4;\nu) = &
\frac{(-1)^{I-1}}{4\sinh^3\nu}
\left\{ \sinh(I\nu)  +\sinh[(2q_3-I)\nu ]
-2 q_3 \sinh\nu\right\}\, .
\end{split}
\ee
It is easy to see that this differs from $\hat S$ given in
\eqref{ech7b} by
terms which vanish as $\nu\to \pm\infty$. 

\subsubsection*{Case D: $q_3, q_4\ge {I\over 2}$} 
In this case \eqref{enet1ab}
 leads to the following contribution to
 $g_{\rm ref}(\alpha_1,\dots, \alpha_4;y)$:
 \be \label{ech5a}
\hat S_{\rm \coll} (I;q_3,q_4;y)=   {(-1)^{I-1}  \over (y-y^{-1})^3} \, \bigg[
y^I - y^{-I}  \bigg]\, .
\ee
This reduces to the case discussed in \S\ref{sillustrate}.
Therefore
Eq. \eqref{enet2} gives
\be \label{ech5b}
\begin{split}
\hat S(I;q_3,q_4;y) = 
&=  {(-1)^{I-1} \over (y-y^{-1})^3} \, \bigg[ 
y^I - y^{-I}  - I\, (y-y^{-1}) \bigg]\, ,
\qquad \hbox{for $I$ odd}\\
&=   {(-1)^{I-1} \over (y-y^{-1})^3} \, \bigg[
y^I - y^{-I}  - {I\over 2}\, (y^2-y^{-2}) \bigg]\, ,
\qquad \hbox{for $I$ even}\, .
\end{split}
\ee
Again, the results agree with the exact refined index 
\eqref{ech2}.

The equivariant volume gives
\be
\label{s34d}
\begin{split}
 S(I;q_3,q_4;\nu) =&
\frac{(-1)^{I-1} }{4\nu^3}
\left( \sinh(I\nu) - I \nu  \right) = S_{\rm \coll}(I;q_3,q_4;\nu) - \frac{(-1)^{I-1} I }{4\nu^2}\, ,
\end{split}
\ee
and hence
\be
\begin{split}
\tilde S(I;q_3,q_4;\nu) =&
\frac{(-1)^{I-1} }{4\sinh^3\nu}
\left( \sinh(I\nu) - I \, \sinh \nu \right) \, .
\end{split}
\ee
Again this differs from the exact result \eqref{ech5b}
by a term that vanishes as $\nu\to\pm\infty$.
The last term in \eqref{s34d} is moreover recognized as the contribution
from the  fixed submanifold $\cM_n^{\rm scal}$, with symplectic 
volume $E(I,q_3,q_4)=(-1)^I I/2$. We further comment on this contribution
in \S\ref{sec_meq}.

\subsection{Five-centered case with distinct centers}
In this section we  compute the equivariant volume
for 5 distinct centers. Without loss of generality, we assume that $q_3<q_4<q_5$. 
If in addition $q_3+q_4\geq q_5$, then we have the ordering 
\be
i)\qquad 
0\leq q_3 \leq q_4 \leq  q_5 \leq   q_3+q_4 \leq  q_3+q_5\leq  q_4+q_5 \leq   q_3+q_4+q_5
\ee
If instead $q_5\geq q_3+q_4$, then 
\be
ii)\qquad 
0\leq q_3 \leq q_4 \leq  q_3+q_4 \leq q_5 \leq  q_3+q_5\leq  q_4+q_5 \leq   q_3+q_4+q_5
\ee
We now split the discussion according to the position of $I/2$ relative to these values,
starting with case i) and then discussing the appropriate change in case ii). The polytopes
arising in  case i) are depicted in Figure \ref{figpoly5}.

\FIGURE{\centerline{\includegraphics[totalheight=7cm]{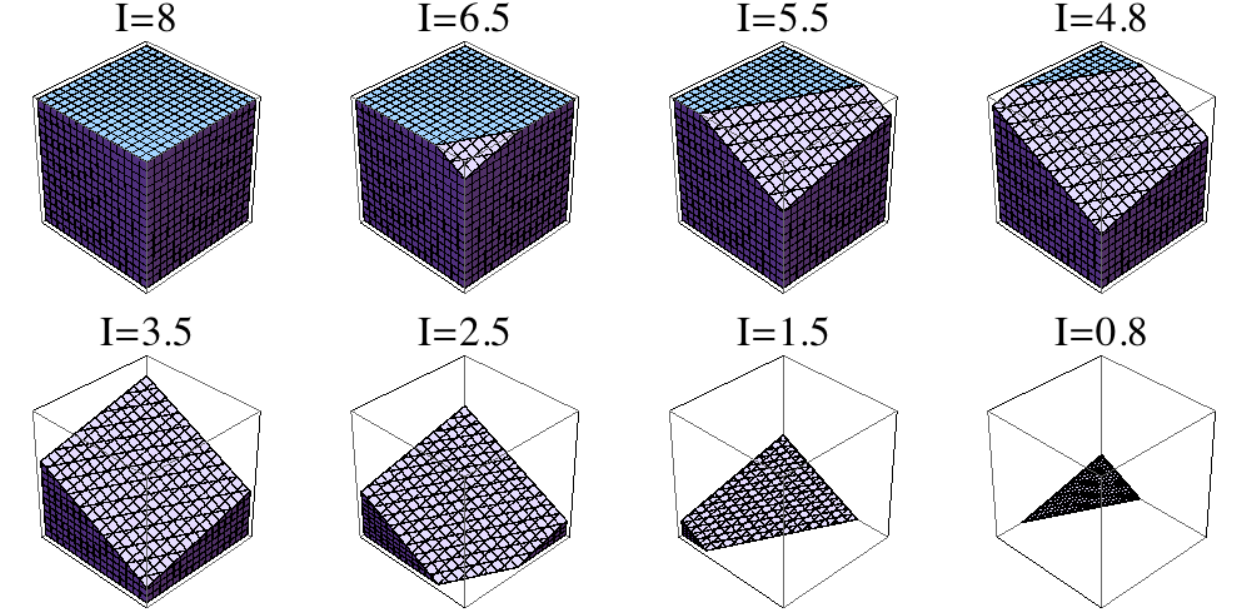}}
\caption{Polytopes associated to 5-center dipole halos with 
$(q_3,q_4,q_5)=(1,2,4)$ and varying values of $I$. The vertical, left and right axes
correspond to $m_5, m_3,m_4$. The dimension
associated to $m\in [-j,j]$ with $j=\frac12I-\sum_a m_a$ is suppressed.\label{figpoly5}}
}

For $q_3+q_4+q_5<I/2$,
\be
\label{S51}
\begin{split}
S(I;q_3,q_4,q_5;\nu) =& \frac{(-1)^I}{8\nu^4}\left( \cosh(I\nu)
-\cosh[(I-2q_3)\nu ]-\cosh[(I-2q_4)\nu ]-\cosh[(I-2q_5)\nu ]\right.\\
&\left.
+\cosh[(I-2q_3-2q_4)\nu ]+\cosh[(I-2q_3-2q_5)\nu ]\right.\\
&\left. +\cosh[(I-2q_4-2q_5)\nu ] -\cosh[(I-2q_3-2q_4-2q_5)\nu ] \right)\, .
\end{split}
\ee
For $q_4+q_5<I/2<q_3+q_4+q_5$, 
\be
\label{S52}
\begin{split}
S(I;q_3,q_4,q_5;\nu) =& \frac{(-1)^I}{8\nu^4}\left( \cosh(I\nu)
-\cosh[(I-2q_3)\nu ]-\cosh[(I-2q_4)\nu ]-\cosh[(I-2q_5)\nu ]\right.\\
&\left.
+\cosh[(I-2q_3-2q_4)\nu ]+\cosh[(I-2q_3-2q_5)\nu ]\right.\\
&\left. +\cosh[(I-2q_4-2q_5)\nu ] \right)
-\frac{(-1)^I}{16\nu^2}(I-2q_3-2q_4-2q_5)^2-\frac{(-1)^I}{8\nu^4}
\, .
\end{split}
\ee
For $q_3+q_5<I/2<q_4+q_5$,
\be
\label{S53}
\begin{split}
S(I;q_3,q_4,q_5;\nu) =& \frac{(-1)^I}{8\nu^4}\left( \cosh(I\nu)
-\cosh[(I-2q_3)\nu ]-\cosh[(I-2q_4)\nu ]-\cosh[(I-2q_5)\nu ]\right.\\
&\left.
+\cosh[(I-2q_3-2q_4)\nu ]+\cosh[(I-2q_3-2q_5)\nu ]\right]\\
&+\frac{(-1)^I}{4\nu^2} q_3(I-q_3-2q_4-2q_5) 
\end{split}
\ee
For $q_3+q_4<I/2<q_3+q_5$, 
\be
\label{S54}
\begin{split}
S(I;q_3,q_4,q_5;\nu) =& \frac{(-1)^I}{8\nu^4}\left( \cosh(I\nu)
-\cosh[(I-2q_3)\nu ]-\cosh[(I-2q_4)\nu ]-\cosh[(I-2q_5)\nu ]\right.\\
&\left.
+\cosh[(I-2q_3-2q_4)\nu ]\right]+\frac{(-1)^I}{16\nu^2}(I^2-8q_3 q_4 -4 I q_5 + 4 q_5^2)
+\frac{(-1)^I}{8\nu^4}\, .
\end{split}
\ee
For $q_5<I/2<q_3+q_4$, 
\be
\label{S55}
\begin{split}
S(I;q_3,q_4,q_5;\nu) =& \frac{(-1)^I}{8\nu^4}\left( \cosh(I\nu)
-\cosh[(I-2q_3)\nu ]-\cosh[(I-2q_4)\nu ]-\cosh[(I-2q_5)\nu ]\right)\\
&+\frac{(-1)^I}{8\nu^2}(I^2-2I(q_3+q_4+q_5)+2(q_3^2+q_4^2+q_5^2))
+\frac{(-1)^I}{4\nu^4}\, .
\end{split}
\ee
For $q_4<I/2<q_5$, 
\be
\label{S56}
\begin{split}
S(I;q_3,q_4,q_5;\nu) =& \frac{(-1)^I}{8\nu^4}\left( \cosh(I\nu)
-\cosh[(I-2q_3)\nu ]-\cosh[(I-2q_4)\nu ]\right)\\
&+\frac{(-1)^I}{16\nu^2}(I^2-4I(q_3+q_4)+4(q_3^2+q_4^2))
+\frac{(-1)^I}{8\nu^4}\, .
\end{split}
\ee
For $q_3<I/2<q_4$,
\be
\label{S57}
\begin{split}
S(I;q_3,q_4,q_5;\nu) =& \frac{(-1)^I}{8\nu^4}\left( \cosh(I\nu)
-\cosh[(I-2q_3)\nu ] \right)- \frac{(-1)^I}{4\nu^2} q_3(I-q_3) 
\, .
\end{split}
\ee
For $0\leq I/2\leq q_3$,
\be
\label{S58}
S(I; q_3,q_4,q_5;\nu)= \frac{(-1)^I}{16\nu^4}
\left[ 2\cosh(I\nu)  - \nu^2 I^2 - 2\right]\, .
\ee
This case is further discussed in \S\ref{sec_meq}.

In case (ii), the region  $q_4<I/2<q_3+q_5$ instead splits in three regions:
$q_5<I/2<q_3+q_5$, where $S$ is still given by \eqref{S54}, $q_3+q_4<I/2<q_5$, where
\be
\label{S55b}
\begin{split}
S(I;q_3,q_4,q_5;\nu) =& \frac{(-1)^I}{8\nu^4}\left( \cosh(I\nu)
-\cosh[(I-2q_3)\nu ]-\cosh[(I-2q_4)\nu  ]\right.\\
&\left.
+\cosh[(I-2q_3-2q_4)\nu ] \right) - \frac{(-1)^I}{2\nu^2} q_3 q_4,
\end{split}
\ee
and $q_4<I/2<q_3+q_4$, where $S$ is still given by \eqref{S56}.

\subsection{Multi-equivariant volumes \label{sec_meq}}

To interpret the above results as a sum over isolated and non-isolated fixed points, 
it is useful to compute the equivariant volume for the most general torus action on $\cM_n$,
\be \label{echclasgen}
\begin{split}
S (I,\{ q_a\};\nu,\{\nu_a\})  =  (-1)^{I-n+1}\,
 \int_{\substack{ 0\le m_a\le q_a \\  \sum_a m_a \le I/2 }}
\de m_3 \cdots \de m_n\, e^{2\sum\nu_a m_a}
\frac{\sinh[(I-2 \sum_{a=3}^n m_a)\nu]}{\nu}\, .
\end{split}
\ee
 and compare it to the corresponding equivariant 
volume of the fixed submanifold $\cM^{\rm scal}_n$
\be \label{echclasgenQ}
\begin{split}
E(I,\{ q_a\};\{\nu_a\})  =  (-1)^{I-n+1}\,
 \int_{\substack{ 0\le m_a\le q_a  }}
\de m_3 \cdots \de m_n\, e^{2\sum\nu_a m_a} 
\delta\left(\sum_a m_a - {I\over 2}\right)\, .
\end{split}
\ee
We shall refer to \eqref{echclasgen} and \eqref{echclasgenQ} as the
multi-equivariant' volume of $\cM_n$ and $\cM_n^{\rm scal}$, respectively.
We shall compute these equivariant volumes in two simple cases with
$n=4$ and $n=5$ centers, which demonstrate that  the non-isolated fixed 
point contribution to $S (I,\{ q_a\};\nu;\{\nu_a\})$ is closely related to 
$E(I,\{ q_a\};\nu;\{\nu_a\})$, though not identical.

For $n=4$ and $I/2<q_3,q_4$, Eq. \eqref{echclasgen}  evaluates to 
\be
\label{s34gen}
\begin{split}
S (I,q_3,q_4;\nu,\nu_3,\nu_4)  =& (-1)^{I-1} \left( 
\frac{e^{\nu I}}{8\nu(\nu-\nu_3)(\nu-\nu_4)}
-\frac{e^{-\nu I}}{8\nu(\nu+\nu_3)(\nu+\nu_4)} \right. \\
&\left. +\frac{e^{\nu_3 I}}{4(\nu^2-\nu_3^2)(\nu_4-\nu_3)}
+\frac{e^{\nu_4 I}}{4(\nu^2-\nu_4^2)(\nu_3-\nu_4)} \right)\ .
\end{split}
\ee
These four contributions correspond to the four vertices of the polytope $\cP$ (see 
Fig.\ref{figpoly4}, bottom-left graph): the first
two arise from collinear configurations ($m_{3,4}=0, m=\pm I/2$) while the last 
two are of scaling type, with $j=0$ ($m_3=I/2, m_4=m=0$ or $m_4=I/2, m_3=m=0$). 
Rescaling $\nu_a$ by a common factor $\epsilon$ and taking the 
limit $\epsilon\to 0$ (as we shall always do when taking the limit $\nu_a\to 0$), 
Eq. \eqref{s34gen} reduces to \eqref{s34d}. 
On the other hand, the multi-equivariant volume of $\cM_n^{\rm scal}$ is given by 
\be
\label{s34genE}
E (I,q_3,q_4;\nu_3,\nu_4)  =(-1)^{I-1}\, \left( 
\frac{e^{\nu_3 I}}{2(\nu_3-\nu_4)} -\frac{e^{\nu_4 I}}{2(\nu_3-\nu_4)}\right) \ .
\ee
This differs from the second line in \eqref{s34gen} for general values of $\nu_3,\nu_4$,
although it agree with it  in the limit  $\nu_3,\nu_4\to 0$, after rescaling by $-1/(2\nu^2)$.  
The difference between \eqref{s34genE} and the second line of \eqref{s34gen} should
originate from the Euler class of the normal bundle of $\cM_n^{\rm scal}$ inside 
$\cM_n$, which appears in the denominator of the localization formula. The comparison
of the two formulae shows that this Euler class should contribute a factor
of $\nu^2/(\nu^2-\nu_a^2)$ at each of the fixed points of the toric action.

To see that such corrections can be important even in the limit $\nu\to 0$, let
us consider the case $n=5$ and $I/2<q_3,q_4,q_5$, and identify the fixed points contributing
to the equivariant volume computed by direct integration in \eqref{S58}. In this case,
Eq. \eqref{echclasgen}  evaluates to 
\be
\label{S58gen}
\begin{split}
S (I,q_3,q_4,q_5;\nu;\nu_3,\nu_4,\nu_5)  =&(-1)^I\, \left( 
\frac{e^{\nu I}}{16\nu(\nu-\nu_3)(\nu-\nu_4)(\nu-\nu_5)} \right. \\
&+\frac{e^{-\nu I}}{16\nu(\nu+\nu_3)(\nu+\nu_4)(\nu+\nu_5)} 
-\frac{e^{\nu_3 I}}{8(\nu^2-\nu_3^2)(\nu_3-\nu_4)(\nu_3-\nu_5)}\\
&\left.  -\frac{e^{\nu_4 I}}{8(\nu^2-\nu_4^2)(\nu_4-\nu_3)(\nu_4-\nu_5)}
 -\frac{e^{\nu_5 I}}{8(\nu^2-\nu_5^2)(\nu_5-\nu_3)(\nu_5-\nu_4)} \right)
\end{split}
\ee
These five contributions correspond to the five vertices of the polytope $\cP$,
displayed on the bottom-right corner of Fig.\ref{figpoly5} (after restoring 
the direction along $m$): the first
two arise from collinear configurations ($m_{3,4,5}=0, m=\pm I/2$) while the last 
three are of scaling type, with $j=0$ ($m_3=I/2, m_4=m_5=m=0$ and permutations
thereof). Rescaling $\nu_a$ by a common factor $\epsilon$ and taking the 
limit $\epsilon\to 0$, \eqref{S58gen} reduces to \eqref{S58}. 
In particular, the first two terms in  \eqref{S58gen} have a smooth limit at $\nu_a\to 0$
and reproduce the first term in \eqref{S58}. The second,  $\cO(I^2/\nu^2)$ term 
in  \eqref{S58} arises by expanding $e^{\nu_a I}$
to second order in $\nu_a$, while the last, $\cO(1/\nu^4)$ term in  \eqref{S58} 
arises by expanding $1/(\nu^2-\nu_a^2)$ to second order in $\nu_a$. 
In contrast,  the multi-equivariant 
volume of $\cM_n^{\rm scal}$ is given by
\be
\begin{split}
E (I,q_3,q_4,q_5;\nu_3,\nu_4,\nu_5) =& (-1)^{I}\\
&\left(
\frac{e^{\nu_3 I}}{4(\nu_3-\nu_4)(\nu_3-\nu_5)}+
\frac{e^{\nu_4 I}}{4(\nu_4-\nu_3)(\nu_4-\nu_5)}+
\frac{e^{\nu_5 I}}{4(\nu_5-\nu_3)(\nu_3-\nu_4)}\right)\ .
\end{split}
\ee
This reduces to $E (I,q_3,q_4,q_5)=(-1)^{I} I^2/8$ in the limit $\nu_a\to 0$. 
Thus, after rescaling by a factor $-1/(2\nu^2)$, the 
multi-equivariant volume 
$ -E (I,q_3,q_4,q_5;\nu_3,\nu_4,\nu_5)/(2\nu^2)$ correctly accounts for 
the $\cO(I^2/\nu^2)$ term in \eqref{S58}, but fails to reproduce the $\cO(1/\nu^4)$.
Again, this indicates that the Euler class of the normal bundle of $\cM_n^{\rm scal}$
which appears in the denominator of the localization formula should
produce an additional  factor $\nu^2/(\nu^2-\nu_a^2)$ at each fixed point. 
 
 More generally, we expect that  the linear combination of equivariant volumes $E(I,\{q_a\})$ 
 appearing in \eqref{delSrec}  can be interpreted as the integral 
  of ${\rm Ch}(\L,\nu)/$ ${\rm Eu}({ N  \cM_n^{\rm scal}} )$
  over  the fixed submanifold  $\cM_n^{\rm scal}$. Similarly, we expect that 
 the analog linear combination of equivariant indices 
 $\hat E(I,\{q_a\})$ which would appear in a similar formula for $\Delta\hat S$
corresponds to the equivariant integral \eqref{ephaseqloc}.
It would be interesting to 
 carry this out in detail.


\providecommand{\href}[2]{#2}\begingroup\raggedright\endgroup

\end{document}